\documentclass[10pt,aps,twocolumn,pre,floatfix]{revtex4-2}
\usepackage{placeins}

\usepackage{float}
\usepackage{hyperref}
\usepackage{xr-hyper}
\usepackage{amssymb}
\usepackage{amsmath}

\usepackage{graphicx}
\usepackage{amsmath,mathtools,esint}
\usepackage{color,ulem,soul}
\usepackage{booktabs}
\usepackage{physics}

\begin{document}
	
	\title{Cotunneling-Assisted Redistribution of Nonpassivity in a Photosynthetic Junction}

	\author{Manjet Rajak$^{1}$}
	\author{Trishna Kalita$^{1}$}
	\author{Debasish Sharma$^{1.2}$}
	\author{Himangshu Prabal Goswami$^{1}$}
	\email{hpg@gauhati.ac.in}
	\affiliation{$^{1}$QuAinT Research Group, Department of Chemistry, Gauhati University, Jalukbari, Guwahati-781014, Assam, India}{}
	\affiliation{$^{2}$Department of Inorganic and Physical Chemistry, Indian Institute of Science, Bengaluru 560012, Karnataka, India}
	\date{\today}
	
	\begin{abstract}
		
		We investigate how cotunneling affects passive-state ordering in a Photosystem II reaction center modeled as a nonequilibrium molecular junction. We use a full pigment excitation manifold, with excitonic and charge-transfer rates computed within a Nakajima-Zwanzig formalism using realistic spectral densities. Cotunneling is introduced through a microscopically constructed doubly reduced acceptor state obtained from relevant quinone-side two-electron configurations, while photoexcitation is allowed into the full excitonic manifold. Passive-state permutations are quantified through ergotropy and energetic capacitance. We find that thermodynamic smoothness in ergotropy and energetic capacitance can conceal microscopic state rearrangements. Cotunneling redistributes the extractable nonequilibrium energy among the quinone acceptor, the donor-pheophytin radical pair, and excitonic states with strong contributions from pigments on the inactive branch.  At high Coulomb interaction, cotunneling produces no passive-state reordering. Piecewise-linear relations among ergotropy, energetic capacitance, and mean energy further reveal the population reorderings generated by cotunneling.
	\end{abstract}
	
	\maketitle

	\section{Introduction}
	Passivity is related to how state populations are ordered with respect to their energies. In isolated systems, a state is passive when its populations decrease with increasing energy. Lower-energy states then have populations equal to or larger than those of higher-energy states. Such an ordering cannot be rearranged to lower the mean energy and extract work. A state that violates this ordering is nonpassive or active. Its population distribution contains extractable work because an appropriate rearrangement can lower the mean energy. Such rearrangements can be generated through cyclic unitary operations. \cite{Allahverdyan2004} At thermal equilibrium, the populations follow the Boltzmann distribution at a fixed temperature and define a Gibbs state. A Gibbs state has a thermal population ordering and is completely passive. However, a general passive state need not be Gibbsian. Passivity requires the appropriate ordering of populations with energy but does not require a thermal distribution as observed in open quantum dynamics.~\cite{Allahverdyan2004,breuer2002theory}  Pumping, relaxation, particle exchange, and internal transfer continuously redistribute populations among the energy levels. The resulting steady state need not be Gibbsian even when its populations remain passively ordered with respect to the system Hamiltonian. It may therefore be passive but nonthermal. Alternatively, the dynamics can produce a nonpassive population arrangement that contains extractable work. \cite{Pusz1978,Allahverdyan2004} The system can also move between different passive configurations as its populations are reorganized. \cite{Allahverdyan2004,Kalita2026}
	
	The photosystem II reaction center (PSIIRC) provides a natural molecular setting in which this distinction between energy and population ordering becomes nontrivial. Charge separation proceeds through an interacting network of excitonic states and charge-transfer configurations rather than through independent hopping events.~\cite{Novoderezhkin2007,Diner2002} Electronic coupling, Coulomb interaction, optical excitation, vibrational relaxation, and environmental charge exchange collectively redistribute populations across these states. Primary and secondary charge-transfer processes occur on picosecond timescales, while the associated electronic dynamics extend from hundreds of femtoseconds to nanoseconds.~\cite{Romero2010,Jha2024,Sirohiwal2020} The protein environment further modifies state energies and transfer pathways and contributes to processes such as triplet localization and photoprotection of the active D1 branch.~\cite{Sirohiwal2020,Bhattacharjee2023,Sharma2025,Hayase2023} 
	
	PSIIRC is therefore not only a charge-transfer network but also a finite nonequilibrium system in which competing processes continuously redistribute populations across the many-body energy levels. This redistribution can generate active states when higher-energy excitonic or charge-transfer states acquire populations that violate the passive energy ordering. These active population configurations provide the extractable part of the PSIIRC energy, since their populations can be reordered toward a lower-energy passive configuration. Our previous work established that such nonequilibrium population structure in the PSIIRC can modify the amount of extractable work associated with the steady state during sequential electronic transfer.~\cite{Kalita2026}  In a general PSIIRC system, Coulomb interaction and many-body level alignment suppress or compete with ordinary single-electron transfer, correlated higher-order processes such as cotunneling become relevant in estimating transport and thermodynamic observables. ~\cite{Averin1990,Sharma2025,carmi01,Gustavsson2008,Emary2009,Roch2011} 
	
	Cotunneling introduces a correlated many-electron transport sector in PSIIRC whose occupation is controlled by Coulomb interaction and by the surrounding charge-transfer network. Its significance therefore extends beyond its effect on conventional observables such as current and power.\cite{Sharma2025} By opening an additional population-transfer pathway, cotunneling can reorganize the steady state populations across the many-body energy levels and thereby alter the passive reference state and the configurations responsible for nonpassivity. Cotunneling and passivity thus describe complementary aspects of the same nonequilibrium dynamics. Cotunneling specifies the additional many-body transport mechanism, while passivity determines how the resulting population redistribution changes the energetic organization of PSIIRC. Consequently, steady states with similar mean energies can exhibit different population orderings and different degrees of nonpassivity. A thermodynamic description must therefore account for the population ordering generated by the complete kinetic network rather than rely solely on conventional transport observables. Ergotropy and energetic capacitance are used here to quantify two distinct consequences of this passivity structure.
	Ergotropy measures the energetic separation between the actual nonequilibrium steady state and its passive counterpart and therefore quantifies the extractable work associated with nonpassivity.~\cite{Allahverdyan2004,Kalita2026} Energetic capacitance instead measures the full energetic range between the passive and anti-passive arrangements compatible with the same population spectrum.~\cite{yang2023battery,zhang2024local} These quantities are therefore not independent performance measures, but complementary measures arising from the same population-ordering structure. The central aim of the present work is to analyse how Coulomb interaction, through cotunneling, reorganizes the population ordering and passivity of PSIIRC as it is driven into a nonequilibrium steady state.
	
	We employ a microscopic source-system-drain description that uses relevant information at the microscopic pigment level. The reduced dynamics are formulated using the Nakajima-Zwanzig projection-operator framework \cite{Nakajima1958,Zwanzig1960,breuer2002theory}, with second-order kernels for optical, vibrational, charge-transfer, and sequential processes and fourth-order kernels for cotunneling. From the resulting steady state populations, we identify changes in the passive state configurations and evaluate the associated ergotropy, energetic capacitance, mean energy, and a transition-defined work. The paper is organized as follows. We first construct the PSIIRC molecular-junction model and identify the molecular states and virtual channels involved in sequential and correlated transport. We then formulate the reduced population dynamics and the passivity-based thermodynamic description. Finally, we analyze how the transport parameters and Coulomb interaction reorganize the steady state population ordering and the corresponding energetic quantities.

	\begin{figure}[ht]
		\centering
		\includegraphics[width=1.0\linewidth]{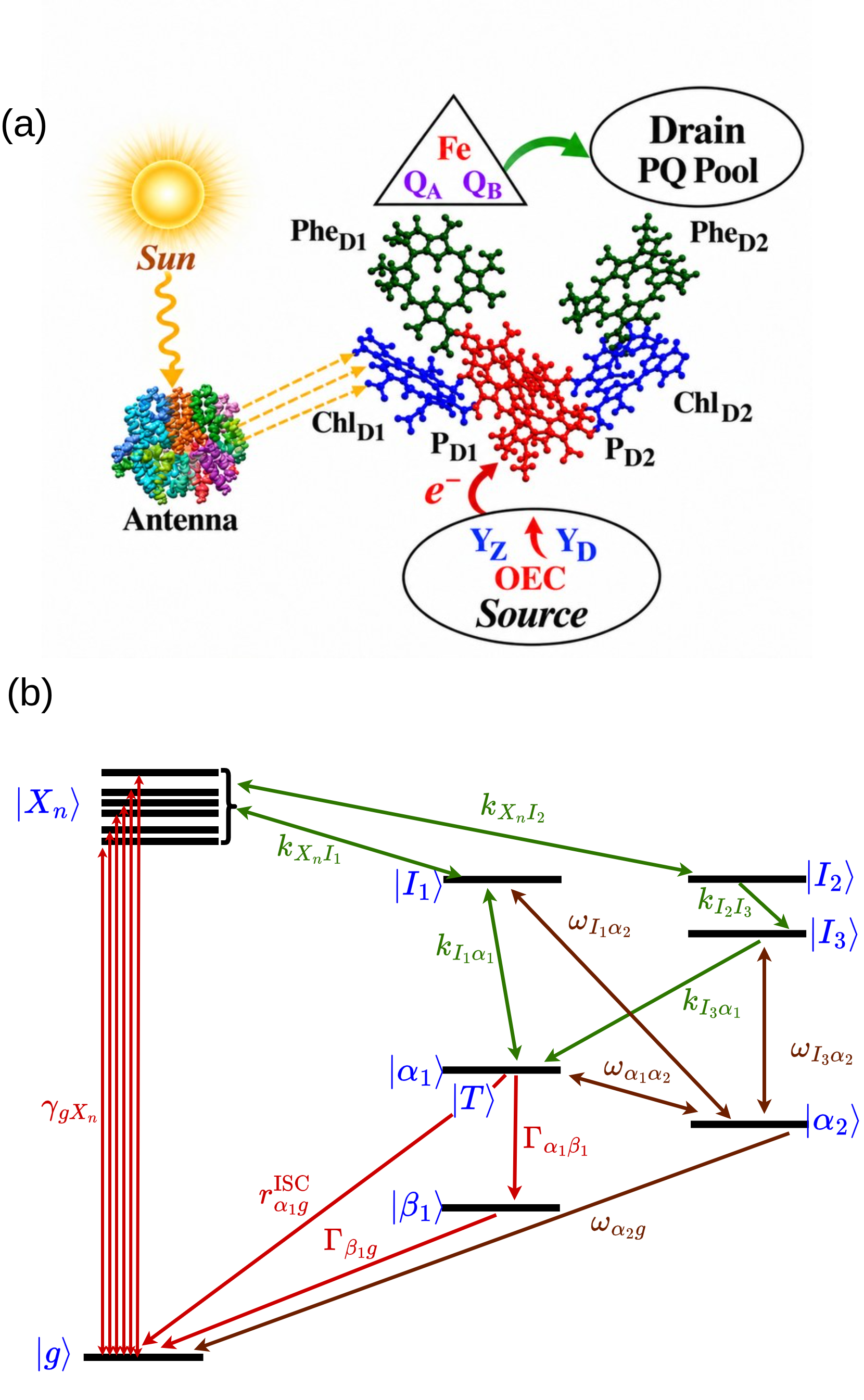}
		\caption{Schematic of the PSIIRC modeled as a nonequilibrium quantum junction. (a) Reaction-center organization showing antenna excitation, the OEC as the electronic source, and the acceptor side coupled to the downstream PQ pool. (b) Thirteen-state kinetic model containing the ground, excitonic, charge-transfer, sequential, and cotunneling-assisted states. Optical, internal transfer, sequential transfer, and cotunneling processes are denoted by $\gamma_{gX_n}$, $k_{xy}$, $\Gamma_{ab}$, and $\omega_{ab}$, respectively. The vibrational redistribution rates ($r_{mn}$) between $X_m\leftrightarrow X_n$ are not shown but included in calculations.  $\ket{T}$ represents a triplet state and involves an intersystem crossing recombination.}
		\label{fig:model}
	\end{figure}
	
	\section{PSIIRC Model and Dynamics}
	\label{sec:methodology}
	
	The PSIIRC is modeled as an open nonequilibrium quantum junction in which the oxygen-evolving complex (OEC) acts as the source and the downstream plastoquinone (PQ) pool acts as the drain. ~\cite{Sharma2025,KernRenger2007} Photoexcitation of the reaction center is mediated by the surrounding light-harvesting antenna, which is represented as an effective nonequilibrium bosonic environment rather than as an equilibrium free-space radiation bath.~\cite{breuer2002theory,Dorfman2013,creatore2013efficient,Nitzan2006} The antenna couples the ground state to the complete six-state excitonic manifold, with the relative pumping strengths determined by the corresponding excitonic transition-dipole moments.~\cite{Gelzinis2013,stones2017performance} Continuous photoexcitation pumps population into the excited states. The source-drain gradient along with the pumping drives the reaction center away from equilibrium. This finite quantum system contains the ground state $\ket{g}$, six delocalized excitonic states $\ket{X_n}, n = 1\ldots 6$, and three intermediate charge-transfer states,
	$\ket{I_1}\equiv\ket{\mathrm{Chl}_{D1}^{+}\mathrm{Phe}_{D1}^{-}}$,
	$\ket{I_2}\equiv\ket{P_{D2}^{+}P_{D1}^{-}}$, and
	$\ket{I_3}\equiv\ket{P_{D1}^{+}\mathrm{Chl}_{D1}^{-}}$. \cite{stones2017performance,Jha2024}
	The six excitonic states originate from coherent mixing of the pigment basis states $P_{D1}$, $P_{D2}$, $\mathrm{Chl}_{D1}$, $\mathrm{Chl}_{D2}$, $\mathrm{Phe}_{D1}$, and $\mathrm{Phe}_{D2}$, consistent with the excitonic description commonly employed for coupled pigment manifolds in photosynthetic reaction centers~\cite{kizmann2024intraband}. The charge-transfer states, in contrast, retain the electronic couplings, electrostatic shifts, and reorganization energies of their microscopic configurations~\cite{Kalita2026,Sharma2025}. In contrast to descriptions in which optical excitation is restricted predominantly to $\ket{g}\rightarrow\ket{X_1}$, we retain excitation into all six delocalized excitonic states. The quinone configurations required to construct the doubly reduced acceptor sector are included in the finite system, whereas only the remaining downstream degrees of freedom constitute the effective drain.~\cite{DeCausmaecker2019,Sharma2025} Taking $N$ as the reference electron number, the radical-pair state $\ket{\alpha_1}\equiv\ket{P_{D1}^{+}\mathrm{Phe}_{D1}^{-}}$, the terminal cationic $(N-1)$ state $\ket{\beta_1}\equiv\ket{P_{D1}^{+}\mathrm{Phe}_{D1}}$, and the cotunneling-assisted $(N+2)$ state $\ket{\alpha_2}$ form a thirteen-state finite system.

	The doubly anionic state in the $(N+2)$ sector entering the proposed cotunneling pathway, $\ket{\alpha_2}$, is constructed separately within a reduced electronic subspace of the PSIIRC acceptor side coupled to the PQ-side drain. The introduction of this sector should not be interpreted as replacing the established sequential mechanism of physiological PSII electron transfer. Following primary charge separation, electron transfer proceeds through the acceptor-side sequence $\mathrm{Pheo}_{D1}\rightarrow Q_A\rightarrow Q_B$, in which $\mathrm{Pheo}_{D1}^{-}$ reduces the tightly bound quinone $Q_A$, and $Q_A^{-}$ subsequently transfers the electron to the exchangeable terminal quinone $Q_B$. \cite{Diner2002,Muh2012,DeCausmaecker2019} Whereas $Q_A$ functions predominantly as a one-electron relay, $Q_B$ constitutes the terminal two-electron gate of the acceptor branch. {~\cite{Diner2002,Muh2012}} It receives two reducing equivalents in successive photochemical turnovers, accompanied by proton uptake, before the resulting plastoquinol $Q_BH_2$ leaves the binding site and enters the membrane PQ pool. \cite{Diner2002,Muh2012,DeCausmaecker2019} This charge-storage functionality provides the physical motivation for considering an interacting $(N+2)$ sector, but does not by itself imply cotunneling. In particular, the ability of the acceptor side to accommodate a second reducing equivalent means that its many-electron description contains both singly and doubly reduced charge sectors. Transfer of the second electron then takes place in the presence of an already reduced acceptor, so that electron-electron interactions enter naturally in a many-body description. The $(N+1)$ states act as intermediate configurations. The interacting $(N+2)$ sector then allows higher-order transitions to the doubly reduced state through virtual $(N+1)$ configurations, in addition to ordinary sequential transfer. Such terms provide the microscopic basis for testing a cotunneling contribution, in which the intermediate one-electron configurations participate virtually in the correlated transition rather than being required to form independently populated kinetic states. Thus, the present construction does not assume that physiological PSII bypasses its established sequential $\mathrm{Pheo}_{D1}\rightarrow Q_A\rightarrow Q_B$ pathway. Instead, it examines whether an additional correlated two-electron channel is permitted when the same acceptor-side system is treated within an open many-body junction description. This interpretation is consistent with the formulation of Sharma \textit{et al.}, in which two interacting virtual $(N+1)$ channels mediate a fourth-order simultaneous two-electron contribution involving a negatively charged many-body state. \cite{Sharma2025}
	
	The acceptor-side topology is then used to define a minimal electronic subspace in which the corresponding $(N+2)$ sector can be represented. Since transfer toward the PQ-side drain proceeds through the terminal quinone $Q_B$ \cite{Diner2002,Bhattacharjee2023}, we restrict this sector to two-electron configurations containing $Q_B$ occupation and allow the second excess electron to occupy $Q_B$ itself or either of the upstream redox-active acceptors belonging to the same functional chain, $Q_A$ and $\mathrm{Pheo}_{D1}$. This gives the localized occupations $(Q_B,Q_B)$, $(Q_A,Q_B)$, and $(\mathrm{Pheo}_{D1},Q_B)$ and hence the diabatic basis $\{\ket{Q_B^{2-}},\ket{Q_A^-Q_B^-},\ket{\mathrm{Pheo}_{D1}^-Q_B^-}\}$. The first configuration represents the purely electronic double occupation of the terminal quinone, the second distributes the two excess electrons over the quinone pair, and the third extends the same charge sector to the upstream primary acceptor. The basis therefore retains the distinct charge-localization patterns available along the functional $\mathrm{Pheo}_{D1}$-$Q_A$-$Q_B$ branch while keeping at least one excess electron in the terminal $Q_B$ sector through which charge is ultimately transferred toward the PQ pool.
	
	These configurations are localized diabatic electronic states and should not be identified individually with experimentally isolated intermediates of the physiological $Q_B$ cycle. In particular, inclusion of $\ket{Q_B^{2-}}$ does not imply physiological accumulation of a bare, long-lived $Q_B^{2-}$ species. Under physiological conditions, acquisition of the second reducing equivalent by $Q_B$ is coupled to proton uptake and ultimately produces $Q_BH_2$. \cite{Muh2012,DeCausmaecker2019} Likewise, $\ket{\mathrm{Pheo}_{D1}^-Q_B^-}$ is a model-defined charge-localized configuration within the reduced electronic Hilbert space rather than an assignment of a known spectroscopic intermediate. The three-state construction should therefore be regarded as an effective $Q_B$-occupied $(N+2)$ active space, rather than a complete two-electron Hilbert space of PSII ~\cite{Diner2002,Muh2012}, or a microscopic description of the full proton-coupled $Q_B$ cycle. Configurations without $Q_B$ occupation, those involving other reaction-center pigments, and explicit protonated quinone states lie outside the reduced electronic sector considered here.
	
	No individual diabatic configuration is identified a priori with $\ket{\alpha_2}$. Electronic coupling among the three localized configurations is introduced through the effective $(N+2)$, whose diagonalization yields three correlated eigenstates. The state retained as $\ket{\alpha_2}$ is identified by its connection to the terminal two-electron branch represented within the chosen active space. Upon continuously introducing the off-diagonal electronic couplings, it is the eigenstate adiabatically connected to the diabatic double-occupation configuration $\ket{Q_B^{2-}}$. For the adopted parametrization, this eigenstate occurs near $1.50~\mathrm{eV}$ and is found after diagonalization to retain overwhelmingly $Q_B^{2-}$ character, with only weak contributions from $\ket{Q_A^-Q_B^-}$ and $\ket{\mathrm{Pheo}_{D1}^-Q_B^-}$. We therefore denote this correlated terminal-acceptor branch by $\ket{\alpha_2}$. Importantly, its dominant $Q_B^{2-}$ character is an outcome of the Hamiltonian and the continuity of the corresponding eigenstate, rather than a criterion imposed on its eigenvector before diagonalization or a consequence of energetic proximity to $\ket{\alpha_1}$. The other two $(N+2)$ eigenstates remain valid eigenstates of the reduced Hamiltonian but correspond continuously to different charge-localization branches and are not retained as the terminal double-occupation state in the present cotunneling description. The explicit Hamiltonian, eigenstate composition and effective parametrization are provided in the SI, Sec.~A.
	
	The reduced dynamics is constructed by partitioning the total Hamiltonian into a finite PSIIRC Hamiltonian, the vibrational and electronic reservoir Hamiltonians, and their respective interaction terms. A Nakajima-Zwanzig projection onto the populations of the thirteen many-body states is then used to eliminate the environmental degrees of freedom.\cite{Nakajima1958,Zwanzig1960,breuer2002theory} The resulting population dynamics contains distinct classes of processes according to the physical interaction responsible for each transition, as shown in Fig.~\ref{fig:model}(b). These comprise optical excitation and relaxation $\ket{g}\leftrightarrow\ket{X_n}$, vibrationally induced redistribution $\ket{X_n}\leftrightarrow\ket{X_m}$ within the excitonic manifold, internal transfer pathways $\ket{X_n}\leftrightarrow\ket{I_1}$, $\ket{X_n}\leftrightarrow\ket{I_2}$, $\ket{I_2}\leftrightarrow\ket{I_3}$, and $\ket{I_{1,3}}\leftrightarrow\ket{\alpha_1}$, sequential ejection through $\ket{\alpha_1}\rightarrow\ket{\beta_1}\rightarrow\ket{g}$, and the cotunneling pathways $\ket{I_{1,3}}\leftrightarrow\ket{\alpha_2}$, $\ket{\alpha_1}\leftrightarrow\ket{\alpha_2}$, and $\ket{\alpha_2}\rightarrow\ket{g}$. These processes are not represented by a single universal rate expression, and each rate is obtained from the corresponding system-environment interaction in its appropriate coupling regime, as detailed in the SI, Sec.~B. We summarize below only the principal ingredients required for the present analysis.
	
	Continuous optical excitation populates all six excitonic states from the ground state, while the corresponding reverse processes provide relaxation back toward $\ket{g}$. These optical transitions are described by rates $\gamma_{gX_n}$ and $\gamma_{X_n g}$, with $n=1,\ldots,6$. Vibrationally assisted redistribution within the excitonic manifold is described using a modified-Redfield treatment \cite{Novoderezhkin2007,Renger2001,Yang2002}, with $r_{qp}$ denoting population transfer from $\ket{m}$ to $\ket{n}$ for $m,n=1,\ldots,6$ and $m\neq n$ (not shown in the schematic figure for brevity). Transfer between the excitonic and charge-transfer manifolds, together with transfer among the weakly coupled charge-transfer configurations, is treated in the Förster regime \cite{Renger2001,Nitzan2006,may2023charge} and denoted by $k_{x y}, x,y\in X_n,I_1,I_2,I_3,\alpha_1$. The  Förster network consists of $\ket{X_n}\leftrightarrow\ket{I_1}$, $\ket{X_n}\leftrightarrow\ket{I_2}$, $\ket{I_2}\leftrightarrow\ket{I_3}$, $\ket{I_1}\leftrightarrow\ket{\alpha_1}$, and $\ket{I_3}\leftrightarrow\ket{\alpha_1}$. These rates incorporate the spectral overlap between the corresponding donor and acceptor configurations and describe the  internal electron-transfer pathways connecting the excitonic and charge-separated manifolds. Direct transfer from $\ket{X_n}$ to $\ket{I_3}$ is neglected because the associated electronic coupling and spectral overlap are known to be negligible. \cite{stones2017performance} Note that these microscopic rates incorporate experimentally characterized vibrational structure, reorganization energies, line-broadening functions, and realistic absorption and fluorescence line shapes. The spectral description includes information from the experimentally resolved vibrational modes used in the PSIIRC parametrization.\cite{stones2017performance,Novoderezhkin2005} The present master-equation treatment is applied in the parameter regime where its dynamics have previously been shown to agree with those obtained from the Hierarchical Equations of Motion approach.~\cite{stones2017performance} In particular, a hybrid spectral density containing the prominent vibrational modes near \(342~\mathrm{cm}^{-1}\) and \(742~\mathrm{cm}^{-1}\), together with a Drude background, was found to yield converged dynamics in both descriptions.~\cite{Novoderezhkin2005,stones2017performance}  
	A time-nonlocal formulation would provide an alternative treatment in which memory effects are retained explicitly. In the present setting, however, this would require construction and propagation of a memory kernel using the full vibrational spectral density, thereby converting the calculation into a substantially different and more demanding numerical problem. The spectral densities, overlap integrals, density-of-states contributions, reorganization energies, and complete rate expressions are provided in the SI, Sec.~D. 
	
	Coupling to the electronic source and drain is treated separately from the vibrational dynamics. Sequential one-electron exchange with the source and drain reservoirs is obtained to second order in the molecule-reservoir coupling \cite{Harbola2006,Timm2008} and is denoted by $\Gamma_{a'b'}, a'b'\in \alpha_1,\beta_1,g $. In Fig.~\ref{fig:model}(b), the relevant rates are $\Gamma_{\alpha_1\beta_1}$ and $\Gamma_{\beta_1 g}$. Correlated two-electron processes are retained to fourth order \cite{Averin1990,Timm2008,Konig1996} and are described by the cotunneling rates $\omega_{x y}$. The corresponding pathways are $\omega_{I_1\alpha_2}$, $\omega_{\alpha_2 I_1}$, $\omega_{I_3\alpha_2}$, $\omega_{\alpha_2 I_3}$, $\omega_{\alpha_1\alpha_2}$, $\omega_{\alpha_2\alpha_1}$, and the direct two-electron rate $\omega_{\alpha_2 g}$, represented by the corresponding arrows in Fig.~\ref{fig:model}(b). The cotunneling contributions are  regularized effective fourth-order rates that retain the intermediate charge configurations entering the correlated-transfer propagators together with their dependence on the Coulomb interaction energy $U$, allowing the cotunneling contribution to be distinguished from sequential one-electron transport. All explicit calculations are provided in the SI, Sec.~D.
	
	We also examine recombination of the secondary radical-pair state $\ket{\alpha_1}$ to the ground-state manifold through an intermediate triplet sector. Experimental and computational studies indicate that triplet configurations associated with the primary donor $P_{680}$ can contribute to charge recombination, particularly when forward electron transfer is suppressed or thermally activated back reactions become favorable~\cite{stones2017performance,Bhattacharjee2023}. These configurations are also associated with the photoprotective response of the reaction center under excess excitation.~\cite{Bhattacharjee2023} The corresponding microscopic pathway is represented as $\ket{\alpha_1}\longrightarrow\ket{T}\longrightarrow\ket{g}$, where $\ket{T}$ denotes an intermediate triplet configuration.
	The triplet manifold is used only as an intermediate microscopic subspace and is not added as a separate population sector of the thirteen-state kinetic model, but treated within a Schrieffer-Wolff reduction. It generates an effective $\alpha_1$-$g$ recombination coupling.~\cite{Schrieffer1966,Bravyi2011} The lowest triplet state of the D1 branch is a local excitation
	$^{3}\mathrm{Chl}_{D1}$ near $1.22~\mathrm{eV}$, whereas higher-lying
	charge-transfer triplets include
	${}^{3}[P_{D1}^{\delta+}\mathrm{Pheo}_{D1}^{\delta-}]$ near
	$1.55~\mathrm{eV}$ and
	${}^{3}[\mathrm{Chl}_{D1}^{\delta+}\mathrm{Pheo}_{D1}^{\delta-}]$ near
	$1.71~\mathrm{eV}$.~\cite{Bhattacharjee2023} Among these, the latter lies
	closest to the $\alpha_1$ energy and therefore provides the most restrictive
	perturbative construction, including its range of validity, is given in the SI, Sec.~D.

	The resulting description is a population-level Liouvillian in which the different microscopic processes retain their respective perturbative orders and environmental correlation functions. The population vector is ordered as
	$
	\bra{\hat{\rho}_q}
	=
	\{
	\rho_g,\allowbreak
	\rho_{X_1},\allowbreak
	\rho_{X_2},\allowbreak
	\rho_{X_3},\allowbreak
	\rho_{X_4},\allowbreak
	\rho_{X_5},\allowbreak
	\rho_{X_6},\allowbreak
	\rho_{I_1},\allowbreak
	\rho_{I_2},\allowbreak
	\rho_{I_3},\allowbreak
	\rho_{\alpha_1},\allowbreak
	\rho_{\alpha_2},\allowbreak
	\rho_{\beta_1}
	\}
	$. The corresponding quantum master equation is
	$
	\ket{\dot{\hat{\rho}}_q}
	=
	\breve{\mathcal L}
	\ket{\hat{\rho}_q},
	$
	where $\breve{\mathcal L}$ is the population Liouvillian superoperator  (see Eq. 165 of SI for the complete analytical form). It is decomposed as
	$
	\breve{\mathcal L}
	=
	\breve{\mathcal L}_{\gamma}^{(2)}
	+
	\breve{\mathcal L}_{r}^{(2)}
	+
	\breve{\mathcal L}_{k}^{(2)}
	+
	\breve{\mathcal L}_{\Gamma}^{(2)}
	+
	\breve{\mathcal L}_{\omega}^{(4)}.
	$
	Here, $\breve{\mathcal L}_{p}^{(o)}$ denotes the contribution of perturbative order $o$ associated with the rate family $p$, where $p=\gamma,r,k,\Gamma,\omega$ corresponds respectively to optical excitation and relaxation, modified-Redfield excitonic redistribution, Förster electron transfer, sequential reservoir exchange, and cotunneling. The exact forms of these contributions are provided in the SI, Sec.~D. The final Liouvillian provides the starting point for calculating the steady state populations.  
	
	The master-equation framework used here  evaluates the relevant environmental correlation functions with realistic spectral profiles while remaining computationally tractable for the repeated steady state calculations required for the ergotropy and energetic-capacitance analysis. Its use therefore defines the scope of the present treatment. Environmental spectral structure and microscopic transfer processes are retained, whereas an explicit analysis of non-Markovian memory effects using a time-nonlocal kernel lies beyond the current framework.
	The use of a population-based kinetic description does not imply that coherent electronic mixing is discarded at the microscopic level. The reduced nonequilibrium dynamics is formulated in terms of populations of the states obtained after the microscopic PSIIRC construction described. Only the coherences between distinct states of the final 13-state many-body basis, corresponding to off-diagonal elements of the $13\times13$ density matrix, are neglected in the final kinetic propagation under the rotating-wave approximation. \cite{Sharma2025,Kalita2026,Harbola2006,Timm2008,stones2017performance} The electronic mixing already incorporated into the construction and composition of the retained states remains encoded in their energies, transition matrix elements, and resulting microscopic rates. We introduced controlled perturbation in energies and rates (SI,Fig. 2.) during the numerical pipeline as an illustrative sensitivity test of the ensemble mean under 5\% energy perturbations and did not find any substantial changes that alter the interpretation of our results. 
	
	Recent two-dimensional electronic spectroscopy studies have reported updated
	site energies (a few of the thirteen states) and electronic couplings for the chlorophyll and pheophytin
	cofactors of the PSIIRC.~\cite{Jha2024,zhou2024unraveling} We had examined the sensitivity of the thermodynamic response to
	this alternative pigment parametrization and found that the qualitative ergotropy
	behavior and its physical interpretation remained unchanged during sequential transfer. ~\cite{Kalita2026} 
	The conclusions drawn in this work are therefore expected to be robust against
	reasonable variations ($<5\%$) in the underlying pigment-site parametrization. 
	
	\begin{figure}[htbp]
		\centering
		\includegraphics[width=1.0\linewidth]{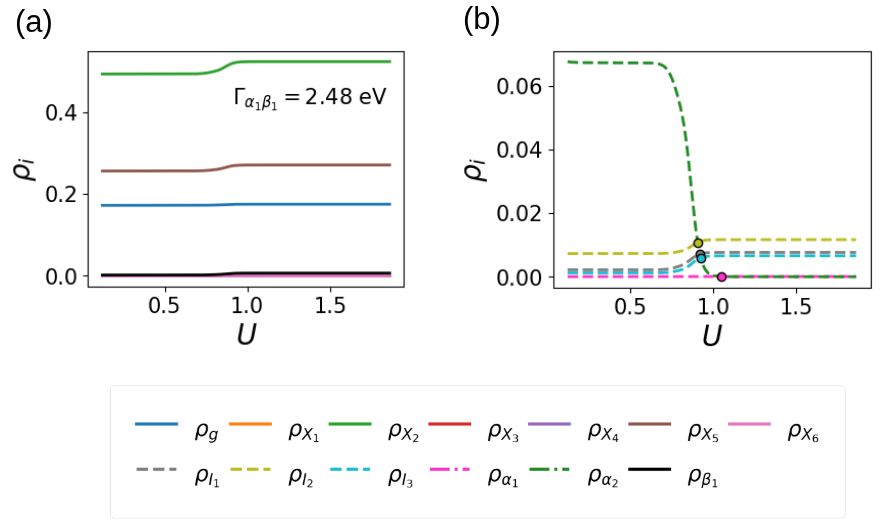}
		\caption{Steady-state populations $\rho_i$ as functions of the Coulomb interaction energy $U$ for $\Gamma_{\alpha_1\beta_1}=2.48~\mathrm{eV}$. (a) Populations of all states. (b) Enlarged view of the weakly populated states near the population redistribution region. All numerical parameters are provided in the SI. }
		\label{fig:2}
	\end{figure}

	\section{Passivity and Its Energetic Quantification}
	
	The thermodynamic characterization used in this work is based on the
	population ordering of the PSIIRC nonequilibrium steady state. For the bare PSIIRC
	Hamiltonian
	$
	\hat{H}_q
	=
	\sum_{k=1}^{13}
	E_k
	\ket{k}\bra{k},
	$ 
	with energies ordered as
	$
	E_1\leq E_2\leq \cdots \leq E_{13},
	$
	a state is passive when its eigenvalues are arranged in the opposite order,
	$
	P_1\geq P_2\geq \cdots \geq P_n.
	$
	The largest population then occupies the lowest-energy state, the next
	largest population occupies the next-lowest state, and so forth. No cyclic
	unitary transformation can lower the energy of such a state further.
	Conversely, any departure from this ordering identifies a nonpassive state
	from which energy can, in principle, be extracted through an
	entropy-preserving unitary rearrangement.~\cite{Allahverdyan2004}In the present population-level description of the PSIIRC, the
	nonequilibrium steady state is diagonal,
	$
	\hat{\rho}_{\mathrm{ss}}
	=
	\sum_{k=1}^{13}
	\rho_k^{ }
	\ket{k}\bra{k},
	$
	where $\rho_k^{ }$ are obtained from the stationary solution of the
	master equation. The mean energy of this state is $
	\langle E\rangle
	=
	\sum_{k=1}^{n}
	E_k\rho_k^{ }.
	\label{eq:avg_energy}
	$
	The passive state $\hat{\rho}_p$ is constructed from the same set of
	steady state populations by rearranging them in decreasing order over the
	increasing energy spectrum. Its energy,
	$
	E_{\mathrm{pas}}
	=
	\mathrm{Tr}
	\left(
	\hat{H}_q\hat{\rho}_p
	\right),
	$ 
	is the minimum energy accessible by unitary rearrangement of the given
	population spectrum. 
	The ergotropy follows directly from this construction as the energetic
	distance between the actual steady state and its passive counterpart,
	\begin{equation}
		\mathcal{E}
		=
		\mathrm{Tr}
		\left(
		\hat{H}_q\hat{\rho}_{\mathrm{ss}}
		\right)
		-
		\mathrm{Tr}
		\left(
		\hat{H}_q\hat{\rho}_p
		\right).
		\label{eq:ergotropy}
	\end{equation}
	For the diagonal steady states considered here,
	\begin{equation}
		\mathcal{E}
		=
		\sum_{k=1}^{13}
		E_k
		\left(
		\rho_k^{ }-P_k
		\right),
		\label{eq:ergotropy_diag}
	\end{equation}
	where $P_k$ denotes the passively reordered populations. The mean energy
	can therefore be decomposed as
	\begin{equation}
		\langle E\rangle
		=
		E_{\mathrm{pas}}
		+
		\mathcal{E}.
		\label{eq:energy_decomp}
	\end{equation}
	Thus, ergotropy is not an independent energetic quantity introduced
	alongside passivity. It is the energetic consequence of the departure of
	the actual steady state from passive ordering. A passive state has
	$\mathcal{E}=0$, whereas $\mathcal{E}>0$ quantifies the work content
	associated with its nonpassivity. 
	The same population spectrum also defines an opposite extremal
	configuration. The anti-passive state $\hat{\rho}_{\mathrm{ap}}$ is
	constructed by assigning the largest populations to the highest-energy
	levels and therefore has the maximum energy attainable by unitary
	rearrangement of the state spectrum. The energetic capacitance is defined
	as the energy interval between these two extremal arrangements,
	\begin{equation}
		\mathcal{C}
		=
		\mathrm{Tr}
		\left(
		\hat{H}_q\hat{\rho}_{\mathrm{ap}}
		\right)
		-
		\mathrm{Tr}
		\left(
		\hat{H}_q\hat{\rho}_p
		\right),
		\label{eq:capacitance}
	\end{equation}
	which reduces, for the diagonal steady state, to
	
	\begin{equation}
		\mathcal{C}
		=
		\sum_{k=1}^{n}
		E_k
		\left(
		P_k^{\mathrm{ap}}-P_k
		\right),
		\label{eq:capacitance_diag}
	\end{equation}
	where $P_k^{\mathrm{ap}}$ and $P_k$ denote the anti-passive and passive
	rearrangements of the same steady state populations, respectively. Since
	these two states are the maximum- and minimum-energy configurations on the
	same unitary orbit, $\mathcal{C}$ is necessarily nonnegative. It was originally introduced in the context of quantum batteries.~\cite{yang2023battery,zhang2024local} Here, this quantity can be used  without interpreting the PSIIRC as a quantum battery. Its role is instead to characterize the energetic range over which the finite molecular system can be placed through population rearrangements that preserve the state spectrum.
	Ergotropy and energetic capacitance therefore quantify two different
	consequences of the same passivity structure. The ergotropy determines
	where the actual PSIIRC steady state lies relative to its passive limit,
	whereas the energetic capacitance specifies the full energetic interval
	allowed by rearrangement of the same population spectrum. Therefore the ergotropy and capacitance are ideal measures based on unrestricted rearrangement of the thirteen populations.

	\section{Results and Discussion}

	\begin{figure}[htbp]
		\centering
		\includegraphics[width=1\linewidth]{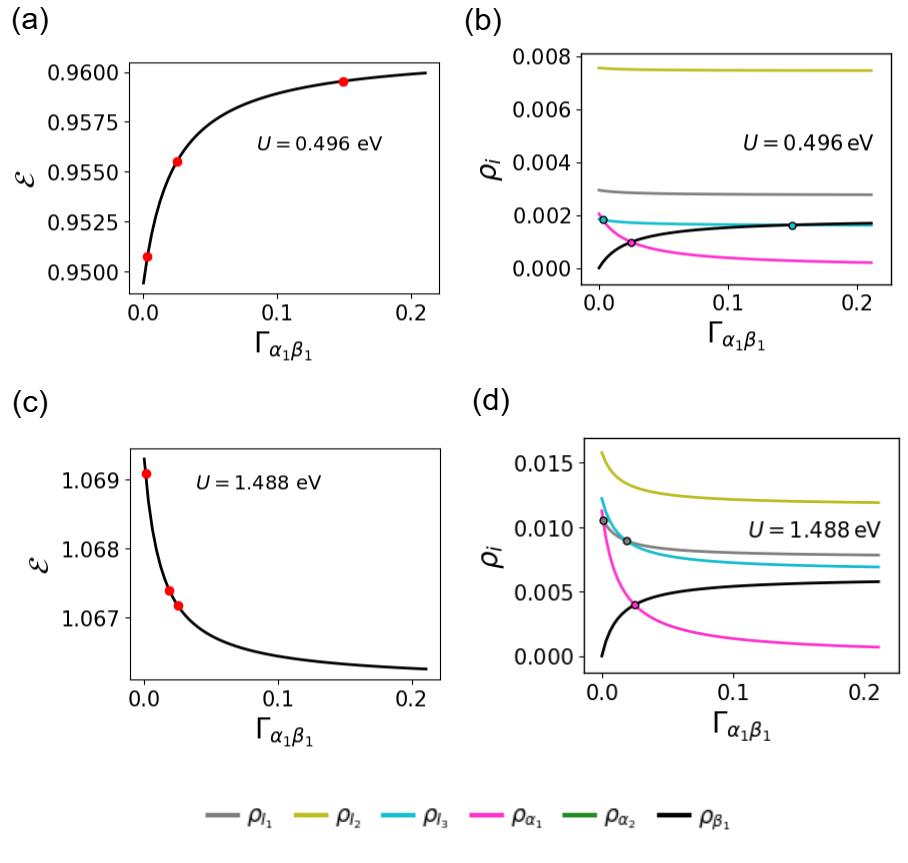}
		\caption{Dependence of the ergotropy $\mathcal{E}$ and relevant steady state populations $\rho_i$ on the electron-ejection rate $\Gamma_{\alpha_1\beta_1}$ for two Coulomb interaction energies. Panel (a) shows the ergotropy at $U=0.496~\mathrm{eV}$, while panel (b) shows the corresponding selected steady state populations. Panel (c) shows the ergotropy at $U=1.488~\mathrm{eV}$, and panel (d) shows the corresponding selected steady state populations. The marked points indicate the population-crossing positions associated with changes in the passive state ordering. }
		\label{fig:3}
	\end{figure}
	
	In Figs.~\ref{fig:2}(a) and \ref{fig:2}(b), we first examine how the steady state population hierarchy changes with the Coulomb interaction over $0<U<1.860~\mathrm{eV}$ at fixed $\Gamma_{\alpha_1\beta_1}$. Based on the energy basis, if we arrange the populations just by following the label of the energy basis i.e $\{\rho_{g},\rho_{\beta},\rho_{\alpha_2},\rho_{\alpha_1},\rho_{I_3},\rho_{I_2},\rho_{I_1},\rho_{X_1},\rho_{X_2},\rho_{X_3},\rho_{X_4},\rho_{X_5},\rho_{X_6}\}$  is not a Gibbs or passive state, in contrast to isolated systems. The passive state needs to be identified from the steadystate values and arranging these from highest population to lowest population which keeps changing as $U$ is swept.  For example, in the interval $U\lesssim0.9~\mathrm{eV}$, the populations vary weakly. The passive state is  
	$\{\rho_{X_2},\allowbreak$$
	\rho_{X_5},\allowbreak$$
	\rho_g,\allowbreak$$
	\rho_{\alpha_2},\allowbreak$$
	\rho_{I_2},\allowbreak$$
	\rho_{I_1},\allowbreak$$
	\rho_{\beta_1},\allowbreak$$
	\rho_{I_3},\allowbreak$$
	\rho_{\alpha_1},\allowbreak$$
	\rho_{X_6},\allowbreak$$
	\rho_{X_1},\allowbreak$$
	\rho_{X_4},\allowbreak$$
	\rho_{X_3}\}.$  A qualitatively different state emerges near $U\simeq0.913~\mathrm{eV}$. Beyond this point, $\rho_{\alpha_2}$ decreases sharply, whereas the remaining populations increase gradually. Between approximately $0.913$ and $1.051~\mathrm{eV}$, the resulting crossings among $\rho_{\alpha_2}$, $\rho_{I_1}$, $\rho_{I_3}$, and $\rho_{\beta_1}$ successively reorganize the passive state ordering.  At $U = 0.913eV$, the passive state is$\{\rho_{X_2},\allowbreak$
	$\rho_{X_5},\allowbreak$
	$\rho_g,\allowbreak$
	$\rho_{\alpha_2},\allowbreak$
	$\rho_{I_2},\allowbreak$
	$\rho_{I_1},\allowbreak$
	$\rho_{\beta_1},\allowbreak$
	$\rho_{I_3},\allowbreak$
	$\rho_{\alpha_1},\allowbreak$
	$\rho_{X_6},\allowbreak$
	$\rho_{X_1},\allowbreak$
	$\rho_{X_4},\allowbreak$
	$\rho_{X_3}\}$,which continues to $U = 0.916eV$. Then the passive state changes to $\rho_{X_2}$, $\rho_{X_5}$, $\rho_g$, $\rho_{I_2}$, $\rho_{I_1}$, $\rho_{\beta_1}$, $\rho_{\alpha_2}$, $\rho_{I_3}$, $\rho_{\alpha_1}$, $\rho_{X_6}$, $\rho_{X_1}$, $\rho_{X_4}$, $\rho_{X_3}$. At $U\ge0.934eV$ the passive state changes to$\{\rho_{X_2},\allowbreak
	\rho_{X_5},\allowbreak
	\rho_g,\allowbreak
	\rho_{I_2},\allowbreak
	\rho_{I_1},\allowbreak
	\rho_{I_3},\allowbreak
	\rho_{\beta_1},\allowbreak
	\rho_{\alpha_2},\allowbreak
	\rho_{\alpha_1},\allowbreak
	\rho_{X_6},\allowbreak
	\rho_{X_1},\allowbreak
	\rho_{X_4},\allowbreak
	\rho_{X_3}\}$ which continues to $U = 1.051eV$. and after the $U = 1.051eV$ it changes to $\rho_{X_2}$, $\rho_{X_5}$, $\rho_g$, $\rho_{I_2}$, $\rho_{I_1}$, $\rho_{I_3}$, $\rho_{\beta_1}$, $\rho_{\alpha_1}$, $\rho_{X_6}$, $\rho_{X_1}$, $\rho_{X_4}$, $\rho_{X_3}$, $\rho_{\alpha_2}$ . Overall, within a small range of high $U$, we observe a total of five different passive states.

	Although the $U$ sweep separates the population structure into five regions, we focus on two regimes of $U$ based on the relative value of the population of the $\alpha_2$ state. One where $\rho_{\alpha_2}$ is large, it corresponds to a low $U$ regime and a high $U$ regime in which $\alpha_2$ remains strongly depleted. The  labeling of low and high is  not associated with a single population crossing, but with a finite interaction interval over which the ordering of the steady state populations is reconstructed.  The distinction between low and high $U$ is associated with the occupation of the interacting $\alpha_2$ sector. $\rho_{\alpha_2}$ remains appreciable at low $U$ but is strongly depleted at high $U$. The same visual separation between low- and high-interaction population structures persists when $\Gamma_{\alpha_1\beta_1}$ is varied, as shown by the population maps in Fig.~1 of the SI. For our numerics, we use $U=0.496~\mathrm{eV}$ and $U=1.488~\mathrm{eV}$ as representative low- and high-interaction regimes, respectively.
	
	In Fig.~\ref{fig:3}(a), we show the ergotropy evaluated using Eq.~\eqref{eq:ergotropy_diag}, while the corresponding steady state populations are shown in Fig.~\ref{fig:3}(b). The passive state is identified independently at each value of the electron-ejection rate $\Gamma_{\alpha_1\beta_1}$ from the numerically obtained steady state populations. At low $U$, the ergotropy increases smoothly but nonlinearly with $\Gamma_{\alpha_1\beta_1}$, although three population crossings divide the investigated range into four distinct passive state configurations. For $0<\Gamma_{\alpha_1\beta_1}<0.00293~\mathrm{eV}$, the passive state ordering is $\{\rho_{X_2},\allowbreak\rho_{X_5},\allowbreak\rho_g,\allowbreak\rho_{\alpha_2},\allowbreak\rho_{I_2},\allowbreak\rho_{I_1},\allowbreak\rho_{\alpha_1},\allowbreak\rho_{I_3},\allowbreak\rho_{\beta_1},\allowbreak\rho_{X_6},\allowbreak\rho_{X_1},\allowbreak\rho_{X_4},\allowbreak\rho_{X_3}\}$. Although the complete ergotropy contains thirteen energy weighted population differences, its magnitude in this interval is dominated by four contributions and can be written as $\mathcal{E}\simeq E_{\alpha_2}(\rho_{\alpha_2}-\rho_g)+\allowbreak E_{\alpha_1}(\rho_{\alpha_1}-\rho_{\alpha_2})+\allowbreak E_{X_2}(\rho_{X_2}-\rho_{\beta_1})+\allowbreak E_{X_5}(\rho_{X_5}-\rho_{X_4})$. These dominant contributions were identified by seperately analysing each weighted contribution and checking which contributions reproduce the ergotropy curve.  The dominant energy weighted population contributions for the different passive state regions at low $U$ are shown in Fig.~4. of the SI.  For $0.00293~\mathrm{eV}<\Gamma_{\alpha_1\beta_1}<0.02502~\mathrm{eV}$, the passive state ordering changes to $\{\rho_{X_2},\allowbreak\rho_{X_5},\allowbreak\rho_g,\allowbreak\rho_{\alpha_2},\allowbreak\rho_{I_2},\allowbreak\rho_{I_1},\allowbreak\rho_{I_3},\allowbreak\rho_{\alpha_1},\allowbreak\rho_{\beta_1},\allowbreak\rho_{X_6},\allowbreak\rho_{X_1},\allowbreak\rho_{X_4},\allowbreak\rho_{X_3}\}$. Despite this rearrangement, the dominant contribution remains $\mathcal{E}\simeq E_{\alpha_2}(\rho_{\alpha_2}-\rho_g)+\allowbreak E_{\alpha_1}(\rho_{\alpha_1}-\rho_{\alpha_2})+\allowbreak E_{X_2}(\rho_{X_2}-\rho_{\beta_1})+\allowbreak E_{X_5}(\rho_{X_5}-\rho_{X_4})$. For $0.02502~\mathrm{eV}<\Gamma_{\alpha_1\beta_1}<0.1493~\mathrm{eV}$, the passive state again changes, now to $\{\rho_{X_2},\allowbreak\rho_{X_5},\allowbreak\rho_g,\allowbreak\rho_{\alpha_2},\allowbreak\rho_{I_2},\allowbreak\rho_{I_1},\allowbreak\rho_{I_3},\allowbreak\rho_{\beta_1},\allowbreak\rho_{\alpha_1},\allowbreak\rho_{X_6},\allowbreak\rho_{X_1},\allowbreak\rho_{X_4},\allowbreak\rho_{X_3}\}$. The ergotropy nevertheless continues to be governed by the same four dominant energy weighted population differences, $\mathcal{E}\simeq E_{\alpha_2}(\rho_{\alpha_2}-\rho_g)+\allowbreak E_{\alpha_1}(\rho_{\alpha_1}-\rho_{\alpha_2})+\allowbreak E_{X_2}(\rho_{X_2}-\rho_{\beta_1})+\allowbreak E_{X_5}(\rho_{X_5}-\rho_{X_4})$. Finally, for $0.1493~\mathrm{eV}<\Gamma_{\alpha_1\beta_1}<0.217~\mathrm{eV}$, the passive state ordering becomes $\{\rho_{X_2},\allowbreak\rho_{X_5},\allowbreak\rho_g,\allowbreak\rho_{\alpha_2},\allowbreak\rho_{I_2},\allowbreak\rho_{I_1},\allowbreak\rho_{\beta_1},\allowbreak\rho_{I_3},\allowbreak\rho_{\alpha_1},\allowbreak\rho_{X_6},\allowbreak\rho_{X_1},\allowbreak\rho_{X_4},\allowbreak\rho_{X_3}\}$. The same dominant ergotropic structure is retained, $\mathcal{E}\simeq E_{\alpha_2}(\rho_{\alpha_2}-\rho_g)+\allowbreak E_{\alpha_1}(\rho_{\alpha_1}-\rho_{\alpha_2})+\allowbreak E_{X_2}(\rho_{X_2}-\rho_{\beta_1})+\allowbreak E_{X_5}(\rho_{X_5}-\rho_{X_4})$.
	
	Thus, although three population crossings involving $\rho_{I_3}$, $\rho_{\alpha_1}$, and $\rho_{\beta_1}$ divide the low $U$ sweep into four distinct passive state regions, none of these rearrangements replaces the four dominant energy weighted population differences controlling the ergotropy. A change in passive state ordering therefore does not necessarily produce an abrupt thermodynamic signature. In the present case, the dominant $\alpha_2$, $\alpha_1$, $X_2$, and $X_5$ associated population differences evolve continuously such that their combined energy weighted contribution increases with $\Gamma_{\alpha_1\beta_1}$. The ergotropy consequently increases smoothly despite the successive changes in passive configuration.
	
	The corresponding high $U$ behavior is shown in Figs.~\ref{fig:3}(c) and \ref{fig:3}(d).  In this regime, the ergotropy decreases smoothly and nonlinearly with increasing $\Gamma_{\alpha_1\beta_1}$. Three population crossings again generate four distinct passive state configurations. Unlike the low $U$ case, the high $U$ passive state rearrangements additionally involve $\rho_{I_1}$. For $0<\Gamma_{\alpha_1\beta_1}<0.00106~\mathrm{eV}$, the passive state ordering is $\{\rho_{X_2},\allowbreak\rho_{X_5},\allowbreak\rho_g,\allowbreak\rho_{I_2},\allowbreak\rho_{I_3},\allowbreak\rho_{\alpha_1},\allowbreak\rho_{I_1},\allowbreak\rho_{\beta_1},\allowbreak\rho_{X_1},\allowbreak\rho_{X_6},\allowbreak\rho_{X_3},\allowbreak\rho_{X_4},\allowbreak\rho_{\alpha_2}\}$. Although the complete ergotropy again contains contributions from all thirteen states, only three terms dominate at high $U$, giving $\mathcal{E}\simeq E_{\alpha_2}(\rho_{\alpha_2}-\rho_g)+\allowbreak E_{X_2}(\rho_{X_2}-\rho_{X_1})+\allowbreak E_{X_5}(\rho_{X_5}-\rho_{X_4})$. In contrast to the low $U$ regime, the $\alpha_1$-associated contribution is no longer appreciable. For $0.00106~\mathrm{eV}<\Gamma_{\alpha_1\beta_1}<0.01870~\mathrm{eV}$, the passive state ordering changes to $\{\rho_{X_2},\allowbreak\rho_{X_5},\allowbreak\rho_g,\allowbreak\rho_{I_2},\allowbreak\rho_{I_3},\allowbreak\rho_{I_1},\allowbreak\rho_{\alpha_1},\allowbreak\rho_{\beta_1},\allowbreak\rho_{X_1},\allowbreak\rho_{X_6},\allowbreak\rho_{X_3},\allowbreak\rho_{X_4},\allowbreak\rho_{\alpha_2}\}$, while the dominant contribution remains $\mathcal{E}\simeq E_{\alpha_2}(\rho_{\alpha_2}-\rho_g)+\allowbreak E_{X_2}(\rho_{X_2}-\rho_{X_1})+\allowbreak E_{X_5}(\rho_{X_5}-\rho_{X_4})$. For $0.01870~\mathrm{eV}<\Gamma_{\alpha_1\beta_1}<0.02503~\mathrm{eV}$, the passive state ordering becomes $\{\rho_{X_2},\allowbreak\rho_{X_5},\allowbreak\rho_g,\allowbreak\rho_{I_2},\allowbreak\rho_{I_1},\allowbreak\rho_{I_3},\allowbreak\rho_{\alpha_1},\allowbreak\rho_{\beta_1},\allowbreak\rho_{X_1},\allowbreak\rho_{X_6},\allowbreak\rho_{X_3},\allowbreak\rho_{X_4},\allowbreak\rho_{\alpha_2}\}$. The ergotropy continues to be governed by the three dominant energy weighted population differences, $\mathcal{E}\simeq E_{\alpha_2}(\rho_{\alpha_2}-\rho_g)+\allowbreak E_{X_2}(\rho_{X_2}-\rho_{X_1})+\allowbreak E_{X_5}(\rho_{X_5}-\rho_{X_4})$. Finally, for $0.02503~\mathrm{eV}<\Gamma_{\alpha_1\beta_1}<0.217~\mathrm{eV}$, the passive state ordering shifts to $\{\rho_{X_2},\allowbreak\rho_{X_5},\allowbreak\rho_g,\allowbreak\rho_{I_2},\allowbreak\rho_{I_1},\allowbreak\rho_{I_3},\allowbreak\rho_{\beta_1},\allowbreak\rho_{\alpha_1},\allowbreak\rho_{X_1},\allowbreak\rho_{X_6},\allowbreak\rho_{X_3},\allowbreak\rho_{X_4},\allowbreak\rho_{\alpha_2}\}$, while the dominant ergotropic structure remains $\mathcal{E}\simeq E_{\alpha_2}(\rho_{\alpha_2}-\rho_g)+\allowbreak E_{X_2}(\rho_{X_2}-\rho_{X_1})+\allowbreak E_{X_5}(\rho_{X_5}-\rho_{X_4})$. The plots of dominant energy weighted population contributions for the different passive state regions are shown in Fig.~5. of the SI.
	
	Although the high $U$ regime contains three successive passive state rearrangements involving $\rho_{I_1}$, $\rho_{I_3}$, $\rho_{\alpha_1}$, and $\rho_{\beta_1}$, the ergotropy remains controlled by the only three dominant energy weighted population differences throughout the investigated range. Their combined contribution decreases with increasing $\Gamma_{\alpha_1\beta_1}$, producing the smooth decrease of the ergotropy at high $U$. Relative to the low $U$ regime, the principal change in the ergotropic structure is the loss of the $\alpha_1$-associated contribution.
	
	The comparison of the two interaction regimes  reveals two different levels of passivity response. Population crossings can modify the passive state permutation without appreciably changing the dominant energetic imbalance, in which case the ergotropy remains smooth across the crossing. By contrast, changing the interaction regime modifies which energy weighted population differences dominate the nonpassive energy itself, leading to opposite ergotropic responses. A smooth ergotropy curve should therefore not be interpreted as evidence for a single passive configuration.  Several microscopic passive state rearrangements can occur while the dominant contribution to the extractable energy remains structurally unchanged. More generally, thermodynamic smoothness can conceal microscopic state rearrangements.

	\begin{figure}[htbp]
		\centering
		\includegraphics[width=1.0\linewidth]{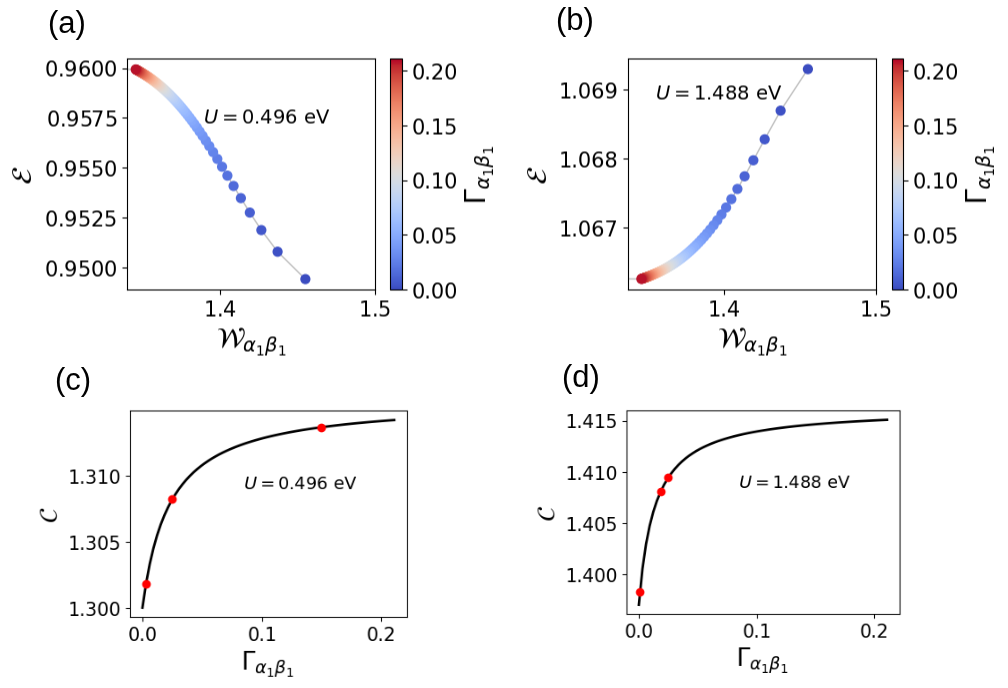}
		\caption{Energetic response of the system under variation of  $\Gamma_{\alpha_1\beta_1}$ for two Coulomb interaction strengths. Panels (a) and (b) show the ergotropy $\mathcal{E}$ as a function of work $\mathcal{W}_{\alpha_1\beta_1}$ for $U=0.496$ eV and $U=1.488$ eV, respectively, with the color scale indicating $\Gamma_{\alpha_1\beta_1}$. Panels (c) and (d) show the corresponding energetic capacitance $\mathcal{C}$ as a function of $\Gamma_{\alpha_1\beta_1}$. The red markers denote the population-crossing points along the $\Gamma_{\alpha_1\beta_1}$ sweep.}
		\label{fig:4}
	\end{figure}
	
	The dominant ergotropic contributions reveal that nonpassivity in PSIIRC is distributed between molecular sectors with distinct functional roles. At low $U$, the contribution associated with $\ket{\alpha_1}\equiv\ket{P_{D1}^{+}Phe_{D1}^{-}}$ connects part of the nonpassive energy to the productive D1 charge-separation, while the $\alpha_2$ contribution introduces the interacting terminal acceptor sector. A different role emerges from the persistent excitonic contributions. The pigment decomposition shows that $X_2$ is strongly weighted toward $Phe_{D2}$, whereas $X_5$ is dominated by $Chl_{D2}$ with additional $Phe_{D2}$ character. The occurrence of two distinct excitonic states with strong D2 character among the leading ergotropic terms therefore points to a D2-side contribution rather than an isolated effect of a single delocalized eigenstate. These D2-weighted states need not participate directly in electron ejection to be thermodynamically important. Their steady state populations remain displaced from the populations assigned to the corresponding energies in the passive state and thereby sustain part of the nonpassive energy. This distinction becomes clear at high $U$, where the $\alpha_1$ contribution becomes subdominant despite its continued role in transport, whereas the $\alpha_2$, $X_2$, and $X_5$ contributions remain dominant. Coulomb interaction therefore shifts the ergotropic balance away from the terminal D1 radical pair while preserving population imbalances in the acceptor-side and D2-weighted excitonic sectors. The D1 and D2 branches consequently acquire different thermodynamic roles in the steady state. The D1 branch primarily supports productive charge separation and extraction, whereas the D2-weighted excitonic manifold provides a persistent population-retention sector for nonpassivity. This interpretation does not assign independently extractable work to individual pigments, since ergotropy remains a property of the global population rearrangement, but it identifies the molecular regions of PSIIRC that most strongly sustain its nonequilibrium work content.
	
	At high $U$, the dominant ergotropic structure becomes
	$E_{\alpha_2}(\rho_{\alpha_2}-\rho_g)+\allowbreak
	E_{X_2}(\rho_{X_2}-\rho_{X_1})+\allowbreak
	E_{X_5}(\rho_{X_5}-\rho_{X_4})$.
	The most important change is the loss of the dominant $\alpha_1$ contribution, even though $\ket{\alpha_1}$ remains part of the  transport pathway. Increasing the Coulomb interaction therefore weakens the thermodynamic role of the terminal D1 radical pair without removing its kinetic role in electron transfer. In contrast, the $\alpha_2$, $X_2$, and $X_5$ contributions persist. The D2-weighted excitonic contribution is thus robust against this change in interaction strength, although the passive population paired with $X_2$ changes from $\rho_{\beta_1}$ at low $U$ to $\rho_{X_1}$ at high $U$. This change reflects a reorganization of the passive state rather than a change in the molecular identity of $X_2$.
	
	Coulomb interaction therefore does more than rescale the total ergotropy. It redistributes the thermodynamic importance of different molecular sectors. At low $U$, the nonpassive energy involves the $P_{D1}^{+}Phe_{D1}^{-}$ radical pair together with the interacting acceptor-side state and the D2-weighted excitons. At high $U$, the D1 radical-pair contribution becomes subdominant, while the acceptor-side $\alpha_2$ state and the $Phe_{D2}$- and $Chl_{D2}$-weighted excitons remain among the leading contributions. The interaction therefore shifts the ergotropic balance away from the productive D1 radical-pair sector while preserving nonpassivity associated with the terminal acceptor and D2 excitonic manifold. Importantly, thermodynamic significance is not determined by population magnitude alone. The population of $\alpha_2$ is only about $6\%$ and decreases further with increasing $U$, yet its contribution remains among the dominant terms. Its importance instead follows from the energy weighted difference between its actual population and the population assigned to the corresponding energy in the passive state. Conversely, a highly populated state can contribute little when its occupation is already close to its passive assignment. The ergotropic analysis therefore distinguishes three quantities that need not coincide, viz. kinetic participation, steady state population, and thermodynamic importance. In PSIIRC, Coulomb interaction reorganizes these relationships and reveals which molecular population imbalances sustain nonpassivity even when their associated states are neither the most populated nor the dominant      carriers of electron transport.
	
	To examine how the energetic content of the nonequilibrium steady state is related to the work associated with the electron ejection terminal ($\alpha_1\to \beta_1$), we evaluate an effective  work measure which is a transition-defined energy indicator (equivalent to a load bias) \cite{Dorfman2013},
	\begin{equation}
		\mathcal{W}_{\alpha_1\beta_1}
		=
		eV
		=
		E_{\alpha_1}
		-
		E_{\beta_1}
		+
		k_BT
		\ln\!\left(
		\frac{\rho_{\alpha_1}^{ }}
		{\rho_{\beta_1}^{ }}
		\right),
		\label{eq:work}
	\end{equation}
	and compare it with the ergotropy while varying the electron-ejection rate $\Gamma_{\alpha_1\beta_1}$ at fixed Coulomb interaction $U$. Each value of $\Gamma_{\alpha_1\beta_1}$ produces a different steady state population distribution and therefore a corresponding pair $\left(\mathcal{W}_{\alpha_1\beta_1},\mathcal{E}\right)$ in Fig.~\ref{fig:4}(a,b).
	At low $U$ [Fig.~\ref{fig:4}(a)], the ergotropy decreases with increasing $\mathcal{W}_{\alpha_1\beta_1}$. Equivalently, increasing $\Gamma_{\alpha_1\beta_1}$ lowers the transition-defined work while increasing the ergotropy. The decrease in $\mathcal{W}_{\alpha_1\beta_1}$ follows primarily from the increase of the terminal cationic population with increasing rate [Fig.~\ref{fig:3}(b)], which reduces the logarithmic population term in Eq.~\ref{eq:work}. The ergotropy, however, responds in the opposite direction because the same change in rate redistributes population throughout the coupled reaction-center network and increases the dominant energy weighted population differences that determine the deviation from the passive state.

	The behavior changes qualitatively at high $U$ [Fig.~\ref{fig:4}(b)]. The work decreases with increasing electron-ejection rate, as in the low $U$ regime, but the ergotropy now decreases as well. The change from a negative to a positive $\mathcal{E}$-$\mathcal{W}_{\alpha_1\beta_1}$ relation therefore originates from a reversal of the ergotropic response rather than from a change in the response of the terminal work. At low $U$, this redistribution increases the dominant nonpassive population imbalances. In addition to the $P_{D1}^{+}Phe_{D1}^{-}$ radical pair, the leading ergotropic structure contains the correlated $\alpha_2$ acceptor-side state and the $X_2$ and $X_5$ excitons. The latter predominantly sample $Phe_{D2}$ and $Chl_{D2}$, respectively. Thus, although increasing the ejection rate reduces the work associated with the terminal D1 transition, the resulting population redistribution enhances nonpassivity in molecular sectors extending beyond. The increase in ergotropy at low $U$ can therefore be understood as a network response that compensates for the decrease in terminal work.
	
	At high $U$, this compensation is lost. The $P_{D1}^{+}Phe_{D1}^{-}$ contribution becomes subdominant, while the $\alpha_2$, $X_2$, and $X_5$ sectors remain in the leading ergotropic structure. Their combined contribution decreases. The persistence of $X_2$ and $X_5$ is particularly informative because the process itself acts on the D1 side. Their continued thermodynamic importance shows that the response to local D1 extraction remains distributed across the reaction center, including the D2 weighted excitonic manifold. 
	
	This distinction is also visible in the passive state assignments. At low $U$, the $X_2$ contribution contains $E_{X_2}(\rho_{X_2}-\rho_{\beta_1})$, whereas at high $U$ it becomes $E_{X_2}(\rho_{X_2}-\rho_{X_1})$. The $X_5$ contribution, $E_{X_5}(\rho_{X_5}-\rho_{X_4})$, remains among the dominant terms in both regimes. The molecular identity of the D2-weighted excitonic contribution therefore persists even though the passive configuration against which it is measured changes. Increasing $U$ reorganizes not only the magnitude of the population differences but also their thermodynamic placement within the passive state.
	The opposite $\mathcal{E}$-$\mathcal{W}_{\alpha_1\beta_1}$ responses at low and high $U$ can consequently be traced to how the reaction-center network accommodates population removed through the same terminal D1 transition. At low $U$, increasing the rate transfers the population imbalance into configurations that enhance the global nonpassivity, so the ergotropy rises while the work falls. At high $U$, the redistribution no longer produces this enhancement, and both quantities decrease. 
	
	\begin{figure}[htbp]
		\centering
		\includegraphics[width=1.0\linewidth]{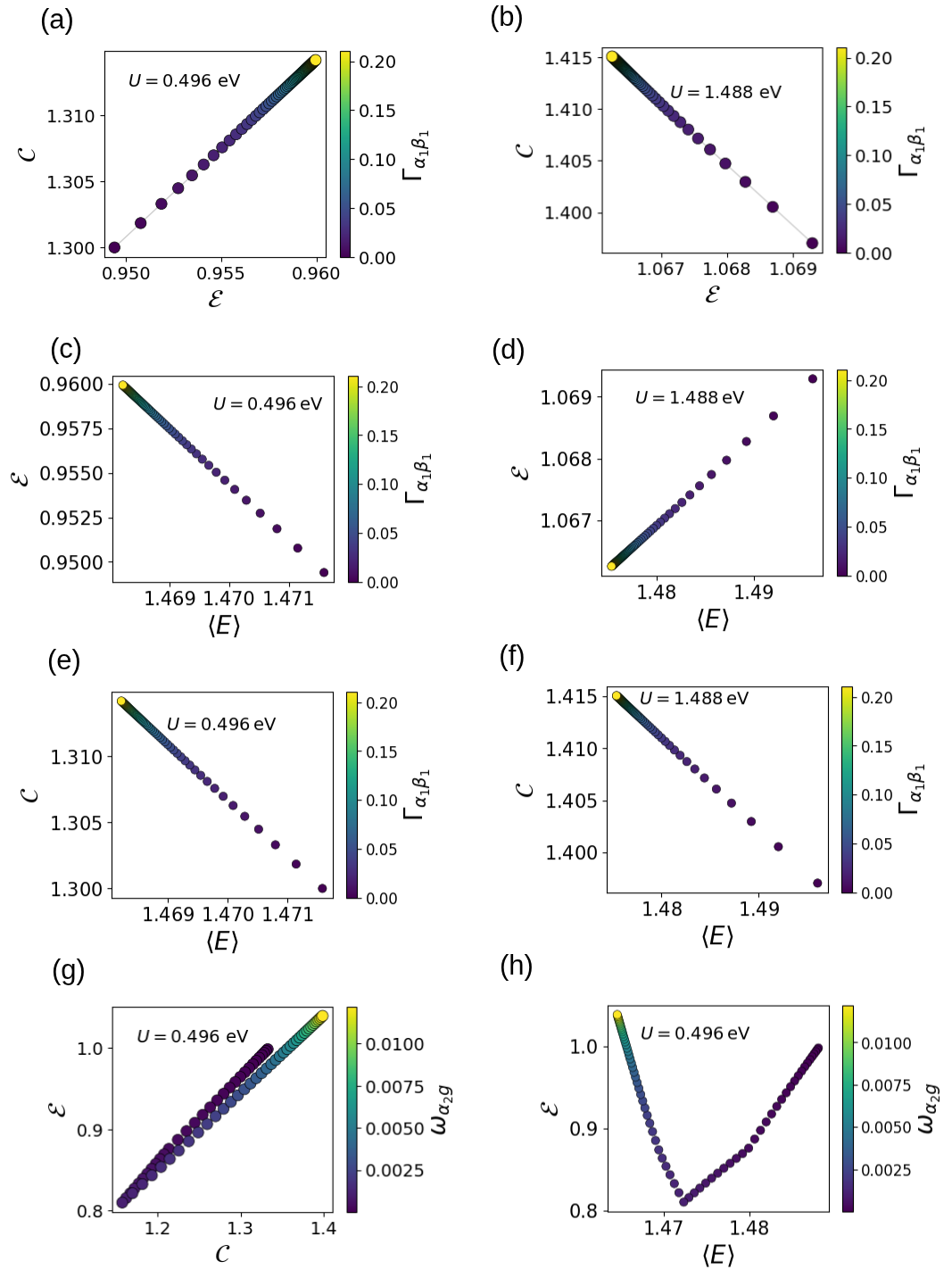}
		\caption{              Piecewise linear relationships among energetic capacitance $\mathcal{C}$, ergotropy $\mathcal{E}$, and mean energy $\langle E\rangle$. (a,b) Energetic capacitance $\mathcal{C}$ as a function of ergotropy $\mathcal{E}$ as the electron-ejection rate $\Gamma_{\alpha_1\beta_1}$ is varied for $U=0.496\,\mathrm{eV}$ and $U=1.488\,\mathrm{eV}$, respectively. (c,d) Ergotropy $\mathcal{E}$ as a function of the mean energy $\langle E\rangle$ for $U=0.496\,\mathrm{eV}$ and $U=1.488\,\mathrm{eV}$, respectively. (e,f) Energetic capacitance $\mathcal{C}$ as a function of the mean energy $\langle E\rangle$ for $U=0.496\,\mathrm{eV}$ and $U=1.488\,\mathrm{eV}$, respectively. The color scale in panels (a)-(f) indicates the corresponding value of $\Gamma_{\alpha_1\beta_1}$. (g) Ergotropy $\mathcal{E}$ as a function of energetic capacitance $\mathcal{C}$ and (h) ergotropy $\mathcal{E}$ as a function of mean energy $\langle E\rangle$ as the cotunneling rate $\omega_{\alpha_2 g}$ is varied at $U=0.496\,\mathrm{eV}$. The color scale in panels (g) and (h) indicates the corresponding value of $\omega_{\alpha_2 g}$.}
		\label{fig:5}
	\end{figure}

	We next examine the energetic capacitance as a function of the  electron-ejection rate $\Gamma_{\alpha_1\beta_1}$ at fixed Coulomb interaction. At low $U$, the capacitance increases with $\Gamma_{\alpha_1\beta_1}$ together with the ergotropy, producing the positive $\mathcal{E}$-$\mathcal{C}$ relation in Fig.~\ref{fig:5}(a). In this regime, removal of population through the D1 radical-pair channel is accompanied by population changes in the correlated acceptor-side and D2-weighted excitonic sectors that increase both quantities.
	
	At high $U$, however, the capacitance continues to increase while the ergotropy decreases, giving the negative $\mathcal{E}$-$\mathcal{C}$ relation in Fig.~\ref{fig:5}(b). This difference reveals information that is not available from the ergotropy alone. The decrease in ergotropy shows that the actual steady state population ordering contains less extractable energy, but by itself does not establish whether the underlying population spectrum has also lost its energetic range. The simultaneous increase in capacitance shows that it has not. Instead, the populations distributed over the $Q_B$-dominated correlated acceptor state and the $Phe_{D2}$- and $Chl_{D2}$-weighted excitonic states support a larger separation between the lowest- and highest-energy arrangements accessible to that population spectrum. The loss of ergotropy at high $U$ therefore arises from a more passive placement of the populations and not from a reduction of their overall energetic differentiation. In physical terms, Coulomb interaction reduces the thermodynamic importance of the population imbalances realized in the steady state while preserving, and even increasing, the energetic range that those populations could span under rearrangement. Energetic capacitance thus distinguishes whether a reduction in ergotropy originates from a narrowing of the population spectrum or from a change in how that spectrum is positioned across the PSIIRC energy levels. The high $U$ results are consistent with the latter mechanism.
	
	\subsection{Linearity among the energetic observables}
	
	We next examine the mutual dependence of the energetic observables evaluated over the same sweep of the electron-ejection rate $\Gamma_{\alpha_1\beta_1}$ at fixed Coulomb interaction. Figures~\ref{fig:5}(a) and \ref{fig:5}(b) show the energetic capacitance as a function of ergotropy, Figs.~\ref{fig:5}(c) and \ref{fig:5}(d) show the ergotropy as a function of the mean energy, and Figs.~\ref{fig:5}(e) and \ref{fig:5}(f) show the energetic capacitance as a function of the mean energy.
	
	At low $U$, the energetic capacitance varies linearly with the ergotropy with a positive slope [Fig.~\ref{fig:5}(a)], whereas the ergotropy varies linearly with the mean energy with a negative slope [Fig.~\ref{fig:5}(c)]. The energetic capacitance consequently exhibits a negative linear dependence on the mean energy [Fig.~\ref{fig:5}(e)]. At high $U$, the energetic-capacitance-ergotropy slope becomes negative [Fig.~\ref{fig:5}(b)], while the ergotropy-mean-energy slope becomes positive [Fig.~\ref{fig:5}(d)]. Their combined dependence again produces a negative energetic-capacitance-mean-energy slope [Fig.~\ref{fig:5}(f)].
	
	The linear relations among the energetic quantities in Fig.~\ref{fig:5} can be understood from their common dependence on the PSIIRC steady state populations. For each value of the  rate $\Gamma_{\alpha_1\beta_1}$ at fixed Coulomb interaction $U$, we define the steady state population vector as $\boldsymbol{\rho}_{\mathrm{ss}}$ and the corresponding energy vector as $\mathbf E$. Since the energy levels do not depend on $\Gamma_{\alpha_1\beta_1}$, varying the ejection rate changes only $\boldsymbol{\rho}_{\mathrm{ss}}$. Within an interval in which the population ordering remains unchanged, the passive and anti-passive rearrangements are represented by fixed $13\times13$ permutation matrices $\mathbf P_{\mathrm p}$ and $\mathbf P_{\mathrm{ap}}$, respectively, so that $\mathbf P_{\mathrm p}\boldsymbol{\rho}_{\mathrm{ss}}$ and $\mathbf P_{\mathrm{ap}}\boldsymbol{\rho}_{\mathrm{ss}}$ give the corresponding passive and anti-passive population vectors. With $\mathbf I$ denoting the $13\times13$ identity matrix, the mean energy, ergotropy, and energetic capacitance are therefore $\langle E\rangle=\mathbf E^T\boldsymbol{\rho}_{\mathrm{ss}}$, $\mathcal E=\mathbf E^T(\mathbf I-\mathbf P_{\mathrm p})\boldsymbol{\rho}_{\mathrm{ss}}$, and $\mathcal C=\mathbf E^T(\mathbf P_{\mathrm{ap}}-\mathbf P_{\mathrm p})\boldsymbol{\rho}_{\mathrm{ss}}$. The nearly linear relations observed in Fig.~\ref{fig:5} indicate that, over the investigated range of $\Gamma_{\alpha_1\beta_1}$, the steady-state population varies predominantly along a single redistribution direction, which we write as $\boldsymbol{\rho}_{\mathrm{ss}}(\Gamma_{\alpha_1\beta_1};U)\simeq\boldsymbol{\rho}_{\mathrm{ref}}(U)+s(\Gamma_{\alpha_1\beta_1};U)\Delta\boldsymbol{\rho}(U)$. Here, $\boldsymbol{\rho}_{\mathrm{ref}}$ is a reference steady state population vector chosen within a fixed ordering interval, $\Delta\boldsymbol{\rho}$ specifies the dominant direction of population redistribution, and the scalar $s$ measures the displacement of the steady state from $\boldsymbol{\rho}_{\mathrm{ref}}$ along this direction, with $\sum_k\Delta\rho_k=0$ imposed by probability conservation. Substitution then gives $\langle E\rangle\simeq\langle E\rangle_0+A_Es$, $\mathcal E\simeq\mathcal E_0+A_{\mathcal E}s$, and $\mathcal C\simeq\mathcal C_0+A_{\mathcal C}s$, where $\langle E\rangle_0=\mathbf E^T\boldsymbol{\rho}_{\mathrm{ref}}$, $\mathcal E_0=\mathbf E^T(\mathbf I-\mathbf P_{\mathrm p})\boldsymbol{\rho}_{\mathrm{ref}}$, and $\mathcal C_0=\mathbf E^T(\mathbf P_{\mathrm{ap}}-\mathbf P_{\mathrm p})\boldsymbol{\rho}_{\mathrm{ref}}$ are the mean energy, ergotropy, and energetic capacitance of the reference steady state, while $A_E=\mathbf E^T\Delta\boldsymbol{\rho}$, $A_{\mathcal E}=\mathbf E^T(\mathbf I-\mathbf P_{\mathrm p})\Delta\boldsymbol{\rho}$, and $A_{\mathcal C}=\mathbf E^T(\mathbf P_{\mathrm{ap}}-\mathbf P_{\mathrm p})\Delta\boldsymbol{\rho}$ quantify the response of the three energetic quantities along the dominant redistribution direction. Eliminating $s$ gives $\mathcal C\simeq\mathcal C_0+(A_{\mathcal C}/A_{\mathcal E})(\mathcal E-\mathcal E_0)$, $\mathcal E\simeq\mathcal E_0+(A_{\mathcal E}/A_E)(\langle E\rangle-\langle E\rangle_0)$, and $\mathcal C\simeq\mathcal C_0+(A_{\mathcal C}/A_E)(\langle E\rangle-\langle E\rangle_0)$, provided the corresponding denominators are nonzero. The signs of these ratios determine the slopes in Fig.~\ref{fig:5}.  At low $U$, $A_{\mathcal C}/A_{\mathcal E}>0$, $A_{\mathcal E}/A_E<0$, and $A_{\mathcal C}/A_E<0$, producing the positive $\mathcal C$-$\mathcal E$ relation in Fig.~\ref{fig:5}(a) and the negative $\mathcal E$-$\langle E\rangle$ and $\mathcal C$-$\langle E\rangle$ relations in Figs.~\ref{fig:5}(c) and \ref{fig:5}(e), whereas at high $U$, $A_{\mathcal C}/A_{\mathcal E}<0$, $A_{\mathcal E}/A_E>0$, and $A_{\mathcal C}/A_E<0$, accounting for the negative $\mathcal C$-$\mathcal E$ relation in Fig.~\ref{fig:5}(b), the positive $\mathcal E$-$\langle E\rangle$ relation in Fig.~\ref{fig:5}(d), and the negative $\mathcal C$-$\langle E\rangle$ relation in Fig.~\ref{fig:5}(f). At a population crossing, the passive or anti-passive permutation changes, so $\mathbf P_{\mathrm p}$, $\mathbf P_{\mathrm{ap}}$, and the corresponding response coefficients must be reevaluated and the construction is therefore applied piecewise within each fixed population-ordering region.

	The reversal of the first two slopes with $U$, together with the persistent
	negative $\mathcal{C}$-$\langle E\rangle$ slope reflects a change
	in how the redistribution of the thirteen PSIIRC
	populations is weighted by the ergotropy and capacitance constructions.
	If a population crossing changes the passive or anti-passive ordering, the
	corresponding permutation matrices change, and the same construction applies
	separately within the new ordering sector.
	The cotunneling-rate sweep makes the role of population ordering particularly explicit. Figures~\ref{fig:5}(g) and \ref{fig:5}(h) show the relations among the ergotropy, energetic capacitance, and mean energy for steady states generated by varying the cotunneling rate $\omega_{\alpha_2 g}$ at low $U$. In contrast to the nearly uniform linear trends obtained when $\Gamma_{\alpha_1\beta_1}$ is varied, the cotunneling sweep separates into distinct linear segments. These segments arise because increasing $\omega_{\alpha_2 g}$ progressively depletes the correlated state $\alpha_2$, causing its population to cross those of other PSIIRC states. Each crossing changes the ordering of the steady state populations and therefore changes the permutation matrices entering the passive and anti-passive constructions. Between successive crossings, the population ordering remains fixed, so the energetic quantities retain a fixed set of population-weighting coefficients and follow an approximately linear relation.
	
	Using the notation introduced above, the steady state  within a given ordering sector $r$ can be written as
	$
	\boldsymbol{\rho}_{\mathrm{ss}}
	\simeq
	\boldsymbol{\rho}_{\mathrm{ref}}^{(r)}
	+
	s_r\Delta\boldsymbol{\rho}^{(r)},
	$
	where $\boldsymbol{\rho}_{\mathrm{ref}}^{(r)}$ is a reference population vector in sector $r$, $\Delta\boldsymbol{\rho}^{(r)}$ is the dominant redistribution direction, and $s_r$ parameterizes the progression along that direction as $\omega_{\alpha_2 g}$ is varied. The corresponding energetic quantities are then
	$
	\mathcal E
	\simeq
	\mathcal E_0^{(r)}
	+
	A_{\mathcal E}^{(r)}s_r,
	$
	$
	\mathcal C
	\simeq
	\mathcal C_0^{(r)}
	+
	A_{\mathcal C}^{(r)}s_r,
	$
	and
	$
	\langle E\rangle
	\simeq
	\langle E\rangle_0^{(r)}
	+
	A_E^{(r)}s_r,
	$
	where the subscript $0$ denotes the value at $\boldsymbol{\rho}_{\mathrm{ref}}^{(r)}$, and the coefficients $A_{\mathcal E}^{(r)}$, $A_{\mathcal C}^{(r)}$, and $A_E^{(r)}$ are from the sector-specific redistribution vector and the corresponding passive and anti-passive permutation matrices. Eliminating $s_r$ gives
	$
	\mathcal C
	\simeq
	\mathcal C_0^{(r)}
	+
	A_{\mathcal C}^{(r)}/A_{\mathcal E}^{(r)}
	\left(
	\mathcal E-\mathcal E_0^{(r)}
	\right),
	$
	and
	$
	\mathcal E
	\simeq
	\mathcal E_0^{(r)}
	+
	A_{\mathcal E}^{(r)}/A_E^{(r)}
	\left(
	\langle E\rangle-\langle E\rangle_0^{(r)}
	\right).
	$
	The slope is therefore fixed only while the population ordering remains unchanged. At a crossing, the relevant permutation matrices change and a new set of coefficients applies, producing the next linear segment in Figs.~\ref{fig:5}(g) and \ref{fig:5}(h). The observed piecewise linearity is consequently a direct energetic signature of successive reorganizations of the PSIIRC population ordering under cotunneling.
	
	The molecular significance of the segmented response follows from the progressive depletion of the correlated $\alpha_2$ state, which is dominated by the doubly reduced $Q_B$ configuration. As $\omega_{\alpha_2 g}$ increases, $\rho_{\alpha_2}$ successively crosses the populations of the D2-weighted excitonic states and, at larger cotunneling rate, the ground-state population. Each crossing changes the rank of $\rho_{\alpha_2}$ in the ordered population list and therefore changes which population is assigned to a given energy level in the passive state. Cotunneling changes not only the magnitude of the $\alpha_2$ population but also its thermodynamic relation to the excitonic and ground-state sectors. Between two crossings, the population ordering remains fixed, so the corresponding energetic quantities follow a linear relation. When $\rho_{\alpha_2}$ crosses an excitonic population, the passive assignment involving the $Q_B$-dominated acceptor sector and that D2-weighted excitonic sector changes, producing a new slope. When $\rho_{\alpha_2}$ subsequently crosses $\rho_g$, the passive placement of the acceptor population relative to the ground state changes again and another linear segment appears. The breaks or kinks in Figs.~\ref{fig:5}(g) and \ref{fig:5}(h) therefore identify the cotunneling rates at which the $Q_B$-dominated $\alpha_2$ population changes its position relative to the D2 excitonic and ground-state populations in the passive ordering. They consequently show which molecular sector becomes the relevant passive state partner of $\alpha_2$ in each cotunneling-rate interval.

	\subsection{Response to the cotunneling rate}

	\begin{figure}[htbp]
		\centering
		\includegraphics[width=1.0\linewidth]{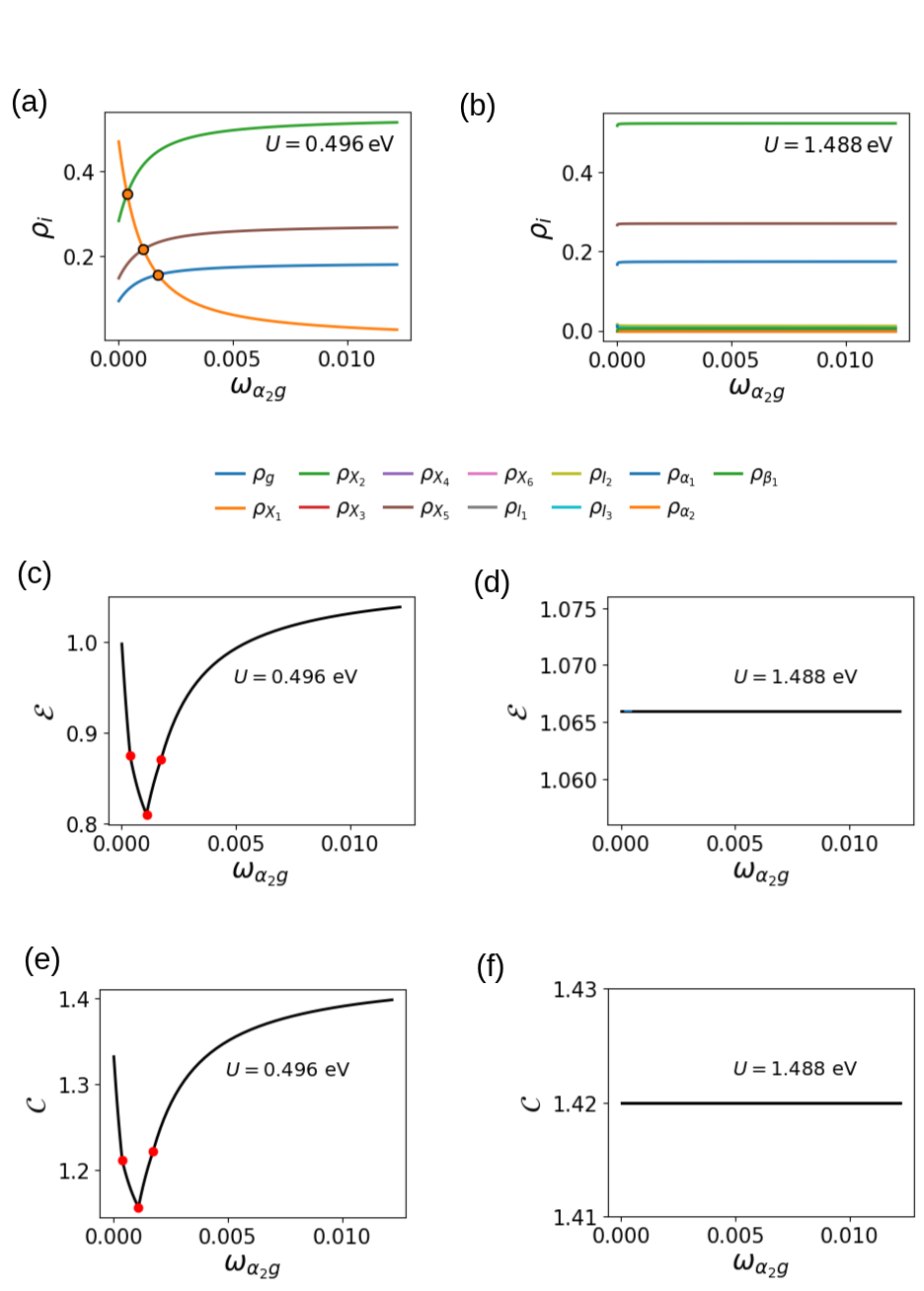}
		\caption{ Panels (a) and (b) show the steady-state populations $\rho_i$ at $U=0.496~\mathrm{eV}$ and $U=1.488~\mathrm{eV}$, respectively as functions of the cotunneling rate $\omega_{\alpha_2 g}$ for two interaction energies.  The marked points in panel (a) indicate the population-crossing positions. Panels (c) and (d) show the corresponding ergotropy $\mathcal{E}$, while panels (e) and (f) show the energetic capacitance $\mathcal{C}$. The red markers in panels (c) and (e) denote the values of $\mathcal{E}$ and $\mathcal{C}$ evaluated at the population-crossing points identified in panel (a).}
		\label{fig:6}
	\end{figure}

	We next isolate the effect of the cotunneling rate
	$\omega_{\alpha_2 g}$ on the PSIIRC steady state. Figure~\ref{fig:6}
	shows the steady state populations, ergotropy, and energetic capacitance
	obtained by varying $\omega_{\alpha_2 g}$ at fixed interaction energy,
	with the low and high $U$ regimes shown separately.
	At low $U$, Fig.~\ref{fig:6}(a), increasing
	$\omega_{\alpha_2 g}$ strongly depletes $\rho_{\alpha_2}$. As a result,
	$\rho_{\alpha_2}$ successively crosses $\rho_{X_2}$,
	$\rho_{X_5}$, and $\rho_g$ at
	$\omega_{\alpha_2 g}=0.0003876~\mathrm{eV}$,
	$0.001097~\mathrm{eV}$, and
	$0.001735~\mathrm{eV}$, respectively. These crossings divide the
	cotunneling-rate sweep into four regions with different passive state
	orderings. For
	$0\leq\omega_{\alpha_2 g}\leq0.0003876~\mathrm{eV}$,
	the passive state ordering is
	$
	\{
	\rho_{\alpha_2},\allowbreak
	\rho_{X_2},\allowbreak
	\rho_{X_5},\allowbreak
	\rho_g,\allowbreak
	\rho_{I_2},\allowbreak
	\rho_{I_1},\allowbreak
	\rho_{\beta_1},\allowbreak
	\rho_{I_3},\allowbreak
	\rho_{X_6},\allowbreak
	\rho_{\alpha_1},\allowbreak
	\rho_{X_1},\allowbreak
	\rho_{X_3},\allowbreak
	\rho_{X_4}
	\}.
	$
	After the first crossing at
	$\omega_{\alpha_2 g}=0.0003876~\mathrm{eV}$, where
	$\rho_{\alpha_2}$ falls below $\rho_{X_2}$, the passive state ordering becomes
	$
	\{
	\rho_{X_2},\allowbreak
	\rho_{\alpha_2},\allowbreak
	\rho_{X_5},\allowbreak
	\rho_g,\allowbreak
	\rho_{I_2},\allowbreak
	\rho_{I_1},\allowbreak
	\rho_{\beta_1},\allowbreak
	\rho_{I_3},\allowbreak
	\rho_{X_6},\allowbreak
	\rho_{\alpha_1},\allowbreak
	\rho_{X_1},\allowbreak
	\rho_{X_3},\allowbreak
	\rho_{X_4}
	\},
	$
	and this configuration persists up to
	$\omega_{\alpha_2 g}=0.001097~\mathrm{eV}$.
	At this second crossing, $\rho_{\alpha_2}$ falls below $\rho_{X_5}$,
	and the passive state ordering changes to
	$
	\{
	\rho_{X_2},\allowbreak
	\rho_{X_5},\allowbreak
	\rho_{\alpha_2},\allowbreak
	\rho_g,\allowbreak
	\rho_{I_2},\allowbreak
	\rho_{I_1},\allowbreak
	\rho_{\beta_1},\allowbreak
	\rho_{I_3},\allowbreak
	\rho_{X_6},\allowbreak
	\rho_{\alpha_1},\allowbreak
	\rho_{X_1},\allowbreak
	\rho_{X_3},\allowbreak
	\rho_{X_4}
	\}.
	$
	This passive configuration is retained up to
	$\omega_{\alpha_2 g}=0.001735~\mathrm{eV}$.
	Following the final crossing at this value, where
	$\rho_{\alpha_2}$ falls below the ground-state population $\rho_g$,
	the ordering becomes
	$
	\{
	\rho_{X_2},\allowbreak
	\rho_{X_5},\allowbreak
	\rho_g,\allowbreak
	\rho_{\alpha_2},\allowbreak
	\rho_{I_2},\allowbreak
	\rho_{I_1},\allowbreak
	\rho_{\beta_1},\allowbreak
	\rho_{I_3},\allowbreak
	\rho_{X_6},\allowbreak
	\rho_{\alpha_1},\allowbreak
	\rho_{X_1},\allowbreak
	\rho_{X_3},\allowbreak
	\rho_{X_4}
	\},
	$
	and remains unchanged over the rest of the investigated sweep up to
	$\omega_{\alpha_2 g}=0.012~\mathrm{eV}$.
	The high $U$ population response is qualitatively different
	[Fig.~\ref{fig:6}(b)]. Over the complete range
	$0\leq\omega_{\alpha_2 g}\leq0.012~\mathrm{eV}$,
	the steady state populations change only weakly and no population
	crossings occur. The passive state ordering therefore remains
	$
	\{
	\rho_{X_2},\allowbreak
	\rho_{X_5},\allowbreak
	\rho_g,\allowbreak
	\rho_{I_2},\allowbreak
	\rho_{I_1},\allowbreak
	\rho_{I_3},\allowbreak
	\rho_{\beta_1},\allowbreak
	\rho_{\alpha_1},\allowbreak
	\rho_{X_6},\allowbreak
	\rho_{X_1},\allowbreak
	\rho_{X_3},\allowbreak
	\rho_{X_4},\allowbreak
	\rho_{\alpha_2}
	\}
	$
	throughout the sweep.
	
	The consequences of the low $U$ population rearrangements for the
	ergotropy are shown in Fig.~\ref{fig:6}(c). For
	$0\leq\omega_{\alpha_2 g}\leq0.0003876~\mathrm{eV}$,
	the dominant contribution is
	$
	\mathcal{E}
	\simeq
	E_{\beta_1}
	\left(
	\rho_{\beta_1}-\rho_{X_2}
	\right)
	+
	E_{\alpha_2}
	\left(
	\rho_{\alpha_2}-\rho_{X_5}
	\right)
	+
	E_{\alpha_1}
	\left(
	\rho_{\alpha_1}-\rho_g
	\right)
	+
	E_{X_2}
	\left(
	\rho_{X_2}-\rho_{X_6}
	\right)
	+
	E_{X_5}
	\left(
	\rho_{X_5}-\rho_{X_3}
	\right).
	$
	After the first crossing, the same five energy weighted population
	differences remain dominant up to
	$\omega_{\alpha_2 g}=0.001097~\mathrm{eV}$. Thus, although the
	passive state ordering changes when $\rho_{\alpha_2}$ falls below
	$\rho_{X_2}$, the dominant ergotropic structure remains unchanged and
	the ergotropy continues to decrease. At the second crossing,
	$\omega_{\alpha_2 g}=0.001097~\mathrm{eV}$, the
	$\beta_1$- and $\alpha_2$-associated terms cease to dominate and the
	leading contribution becomes
	$
	\mathcal{E}
	\simeq
	E_{\alpha_1}
	\left(
	\rho_{\alpha_1}-\rho_g
	\right)
	+
	E_{X_2}
	\left(
	\rho_{X_2}-\rho_{X_6}
	\right)
	+
	E_{X_5}
	\left(
	\rho_{X_5}-\rho_{X_3}
	\right).
	$
	This dominant structure persists up to
	$\omega_{\alpha_2 g}=0.001735~\mathrm{eV}$ and occurs in the region of
	the ergotropy minimum. Following the final crossing at
	$\omega_{\alpha_2 g}=0.001735~\mathrm{eV}$, the dominant structure
	changes again to
	$
	\mathcal{E}
	\simeq
	E_{\alpha_2}
	\left(
	\rho_{\alpha_2}-\rho_g
	\right)
	+
	E_{\alpha_1}
	\left(
	\rho_{\alpha_1}-\rho_{\alpha_2}
	\right)
	+
	E_{X_2}
	\left(
	\rho_{X_2}-\rho_{X_6}
	\right)
	+
	E_{X_5}
	\left(
	\rho_{X_5}-\rho_{X_3}
	\right),
	$
	and this structure is retained up to the largest investigated rate,
	$\omega_{\alpha_2 g}=0.012~\mathrm{eV}$. The $\alpha_2$-associated
	contribution therefore reappears through
	$\rho_{\alpha_2}-\rho_g$, while the $\alpha_1$ contribution changes
	from $\rho_{\alpha_1}-\rho_g$ to
	$\rho_{\alpha_1}-\rho_{\alpha_2}$. The increase of the ergotropy beyond
	the minimum consequently occurs together with a change in the population
	differences entering its dominant terms. At still larger
	$\omega_{\alpha_2 g}$ within this final region, the populations approach
	limiting values and the ergotropy correspondingly saturates. The 
	dominant energy weighted population contributions in the different
	passive state regions are shown in Fig.~6. of the SI.
	
	At high $U$, Fig.~\ref{fig:6}(d), the ergotropy remains essentially
	unchanged throughout
	$0\leq\omega_{\alpha_2 g}\leq0.012~\mathrm{eV}$. This follows directly
	from Fig.~\ref{fig:6}(b), where neither the steady state populations nor
	their passive ordering undergo an appreciable change. 
	
	The energetic capacitance at low $U$ is shown in
	Fig.~\ref{fig:6}(e). It exhibits a corresponding nonmonotonic
	dependence on $\omega_{\alpha_2 g}$, with its changes occurring across
	the same passive state regions identified in Fig.~\ref{fig:6}(a).
	Thus, the cotunneling-induced rearrangements that modify the dominant
	ergotropic population differences also modify the energetic capacitance
	over the same rate ranges.
	
	At high $U$, Fig.~\ref{fig:6}(f), the energetic capacitance remains
	constant throughout
	$0\leq\omega_{\alpha_2 g}\leq0.012~\mathrm{eV}$. As in the ergotropy,
	the absence of a significant response reflects the weak dependence of
	the complete steady state population distribution on
	$\omega_{\alpha_2 g}$ in this interaction regime.

	The physical origin of the low $U$ response is most clearly seen by comparing Figs.~\ref{fig:6}(a), \ref{fig:6}(c), and \ref{fig:6}(e). Figure~\ref{fig:6}(a) shows that increasing $\omega_{\alpha_2 g}$ continuously depletes the correlated $\alpha_2$ state, which is dominated by the doubly reduced $Q_B$ configuration. The ergotropy in Fig.~\ref{fig:6}(c), however, does not simply follow this depletion.  It first decreases, reaches a minimum, and then increases again even though $\rho_{\alpha_2}$ continues to fall. This immediately shows that the thermodynamic importance of $\alpha_2$ is not determined by its population alone.  Around the minimum in Fig.~\ref{fig:6}(c), these contributions lose their dominant role and the ergotropic balance is carried mainly by the $P_{D1}^{+}Phe_{D1}^{-}$ radical pair together with the $Phe_{D2}$- and $Chl_{D2}$-weighted excitonic states. Cotunneling therefore shifts thermodynamic importance away from the correlated $Q_B$ acceptor sector toward the D1 radical-pair and D2 excitonic sectors. At larger $\omega_{\alpha_2 g}$, the ergotropy rises again without any recovery of $\rho_{\alpha_2}$ in Fig.~\ref{fig:6}(a). The renewed importance of the $\alpha_2$-associated term therefore originates from its changed passive state assignment rather than from repopulation of the correlated acceptor state. Figure~\ref{fig:6}(e) shows that the energetic capacitance is reorganized over the same low $U$ range, confirming that cotunneling modifies not only the realized nonpassive energy but also the broader energetic structure supported by the redistributed PSIIRC populations.
	
	The contrast with the high $U$ regime is evident from Figs.~\ref{fig:6}(b), \ref{fig:6}(d), and \ref{fig:6}(f). Figure~\ref{fig:6}(b) shows that varying $\omega_{\alpha_2 g}$ produces only weak changes in the steady state populations and no reordering of their relative magnitudes. Correspondingly, both the ergotropy [Fig.~\ref{fig:6}(d)] and energetic capacitance [Fig.~\ref{fig:6}(f)] remain nearly unchanged.

	\section{Conclusion}

	We have developed an open-quantum-system description of PSIIRC in which the excitonic, charge-transfer, and correlated many-electron sectors are treated within a common kinetic framework. Starting from a Nakajima-Zwanzig projection of the system-environment dynamics onto thirteen relevant many-body states, photoexcitation is allowed into the full excitonic manifold. Population redistribution within this manifold is described using a modified-Redfield treatment, whereas weakly coupled exciton-charge-transfer and charge-transfer transitions are described by Förster-type rates. The vibrational correlation functions retain the experimentally structured PSIIRC spectral profile containing 48 resolved modes rather than replacing the environment by a single phenomenological relaxation constant. Electronic mixing is retained explicitly in constructing the excitonic and charge-transfer eigenstates, while the long-time dynamics are propagated at the population level without treating coherences between the retained many-body states as independent steady state variables. Electron exchange with the electronic reservoirs is treated perturbatively. One-electron processes enter to second order in the molecule-reservoir coupling, whereas correlated two-electron processes are retained to fourth order and generate the regularized cotunneling channels.
	
	The doubly anionic state entering the cotunneling sector is constructed microscopically rather than introduced as an abstract transport level. Its minimal acceptor-side active space contains $\ket{Q_B^{2-}}$, $\ket{Q_A^-Q_B^-}$, and $\ket{Phe_{D1}^-Q_B^-}$, representing two-electron configurations along the $Phe_{D1}$-$Q_A$-$Q_B$ acceptor side. The retained state $\ket{\alpha_2}$ is identified as the eigenstate adiabatically connected to $\ket{Q_B^{2-}}$ and is overwhelmingly dominated by this configuration, with only weak $Q_A^-Q_B^-$ and $Phe_{D1}^-Q_B^-$ admixtures. Its molecular character remains essentially unchanged under substantial variations of the diabatic energies and couplings within the off-resonant weak-hybridization regime. The fourth-order cotunneling dynamics can therefore be associated with a physically specified correlated quinone-acceptor configuration rather than an additional phenomenological state.
	
	The state-resolved passivity analysis identifies where the nonequilibrium energetic resource resides within this molecular network. Important contributions arise from the D1 radical pair $\ket{\alpha_1}\equiv\ket{P_{D1}^{+}Phe_{D1}^{-}}$, the correlated acceptor state $\ket{\alpha_2}$ dominated by $Q_B^{2-}$, and excitonic states with predominantly $Phe_{D2}$ and $Chl_{D2}$ character. The persistence of the D2-weighted excitonic contributions shows that the energetic resource is not confined to the terminal D1 charge-transfer sector. More importantly, population magnitude, transport participation, and thermodynamic importance are found to be distinct properties. The steady-state population of the  $Q_B^{2-}$-dominated $\alpha_2$, can be relatively small and decrease further while its energy-weighted departure from the passive assignment remains a dominant contribution to the ergotropy. Conversely, a strongly populated state need not be thermodynamically important if its population is already close to that required by passive ordering. Ergotropy is therefore controlled not simply by how much population a molecular state carries, but by where that population occurs relative to the energy ordering of the complete PSIIRC state space.
	
	An important consequence is the masking of microscopic population reorganizations in conventional energetic observables. Population crossings can change the passive configuration without producing a sharp feature in the ergotropy or energetic capacitance when the leading energy weighted population differences remain largely unchanged. A smooth energetic response therefore does not imply an unchanged microscopic population ordering. Only when the reordering changes the molecular sectors that dominate the nonpassive energy does a pronounced change appear in the energetic response. Passivity thus exposes reorganizations of the PSIIRC steady state that can remain hidden in current, mean energy, ergotropy, capacitance, or an individual population considered in isolation.
	
	This behavior is also reflected analytically in the correlations among mean energy, ergotropy, and energetic capacitance. Within a region of fixed population ordering, these quantities are different linear projections of the same thirteen-dimensional steady state population vector. Because the calculated population  is approximately dominated by a single redistribution direction, the corresponding energetic quantities exhibit nearly linear pairwise relations. Their slopes specify how that molecular redistribution is projected onto mean energy, nonpassivity, and energetic capacitance. When the population ordering changes, the passive or anti-passive permutation changes with it, modifying these projections and producing a new linear branch. The resulting piecewise-linear structure under variation of the cotunneling rate therefore provides a direct thermodynamic signature of microscopic population reorganizations that can be masked in the individual response curves.
	
	The cotunneling-rate dependence makes these ideas particularly transparent. Increasing $\omega_{\alpha_2 g}$ continuously depletes the correlated state dominated by $Q_B^{2-}$, yet the ergotropy does not follow $\rho_{\alpha_2}$ monotonically. At low $U$, the decreasing quinone-acceptor population changes its ordering relative to the predominantly $Phe_{D2}$- and $Chl_{D2}$-weighted excitonic populations and eventually relative to the ground-state population. Some of these reorderings alter the passive configuration while leaving the principal molecular contributions to ergotropy essentially unchanged, whereas others redistribute thermodynamic importance among the correlated $Q_B$ sector, the $P_{D1}^{+}Phe_{D1}^{-}$ radical pair, and the D2 excitonic manifold. Most notably, the recovery of ergotropy at larger $\omega_{\alpha_2 g}$ occurs while the population of the $Q_B^{2-}$-dominated state continues to decrease. Its renewed contribution therefore results from its changed position in the passive ordering rather than from repopulation. This provides a direct molecular demonstration that a weakly populated, and even progressively depleted, state can carry substantial thermodynamic importance.
	
	Energetic capacitance provides complementary information about this population reorganization. In regimes where $\mathcal{C}$ increases while $\mathcal{E}$ decreases, the loss of ergotropy cannot be attributed to a collapse of the energetic differentiation of the PSIIRC population spectrum. The population spectrum can support a larger separation between its passive and anti-passive limits while the realized steady state moves closer to passive ordering. The combined behavior of $\mathcal{E}$ and $\mathcal{C}$ therefore distinguishes a reduction of realized nonpassivity from a reduction of the energetic range available to the molecular population distribution, a distinction that cannot be obtained from ergotropy alone.
	
	The central physical picture is consequently more specific than a change in transport efficiency or population magnitude. Cotunneling through the correlated $Q_B$ sector reorganizes the thermodynamic importance of molecular populations distributed across the D1 charge-separated and D2 excitonic parts of PSIIRC. These reorganizations may be invisible in individual energetic curves, yet become identifiable through passive state changes and the linear or piecewise-linear relations among the energetic observables. The most populated state, the state most directly involved in charge transport, the state most strongly depleted by cotunneling, and the state carrying the largest contribution to the extractable nonequilibrium energy need not coincide. Passivity therefore provides a molecular criterion for identifying where thermodynamic importance resides in PSIIRC that cannot be inferred from population magnitude or transport observables alone.
	
	\FloatBarrier
	\clearpage
	\appendix
	\onecolumngrid
	\begin{center}
		{\large\bfseries Supplementary Information}
	\end{center}
	
	\section{Microscopic construction of the finite PSIIRC system}
	\label{S-sec:system}

	The reduced PSIIRC model is formulated as a finite many-body system coupled
	to vibrational, optical, and electronic environments.~\cite{breuer2002theory,Nitzan2006}  The cofactors that
	participate explicitly in excitation, primary charge separation, charge
	accumulation, and electron ejection are a part of the finite
	system. Quinone configurations used to construct the
	doubly reduced state $\ket{\alpha_2}$ and are included in the system. Downstream electronic
	degrees of freedom are represented by an effective drain reservoir.  The construction proceeds in two logically distinct steps.  First, the
	strong electronic couplings within the pigment manifold are diagonalized
	to obtain the six delocalized excitonic eigenstates.  Second, the weaker
	electronic couplings connecting these dressed excitons to charge-transfer
	configurations. The weak couplings within the charge-transfer network
	are retained as perturbations that generate population-transfer rates.
	The diagonal form of the system Hamiltonian is used later in the
	Nakajima-Zwanzig construction is a consequence of the chosen
	eigenstate basis and does not imply that the microscopic molecular
	Hamiltonian contained no coherent couplings.~\cite{Nakajima1958,Zwanzig1960,Novoderezhkin2007,Kalita2026,stones2017performance,Sharma2025}
	
	\subsection{Localized pigment Hamiltonian}
	\label{pigmenthamiltonian}
	
	The one-excitation pigment basis is
	$
	\mathcal B_{\rm pig}
	=
	\left\{
	\ket{P_{D1}},
	\ket{P_{D2}},
	\ket{\mathrm{Chl}_{D1}},
	\ket{\mathrm{Chl}_{D2}},
	\ket{\mathrm{Pheo}_{D1}},
	\ket{\mathrm{Pheo}_{D2}}
	\right\}.
	$
	In this basis the electronic pigment Hamiltonian is
	\begin{align}
		H_{\rm pig}
		&=
		\sum_{i=1}^{6}
		\epsilon_i\ket i\bra i
		+
		\sum_{i<j}
		\left(
		J_{ij}\ket i\bra j
		+
		J_{ij}^{*}\ket j\bra i
		\right),
		\label{S-eq:H_pig_general}
	\end{align}
	where $\epsilon_i$ are the localized excitation energies and $J_{ij}$ are
	the electronic couplings between pigments.  In the ordered basis
	$
	\mathcal O_{\rm pig}
	=
	\left(
	P_{D1},
	P_{D2},
	\mathrm{Chl}_{D1},
	\mathrm{Chl}_{D2},
	\mathrm{Pheo}_{D1},
	\mathrm{Pheo}_{D2}
	\right),
	$
	the off-diagonal coupling matrix used in the PSIIRC parametrization is
	\begin{align}
		J_{\rm pig}
		&=
		\begin{pmatrix}
			0    & 150  & -42 & -55 & -6  & 17 \\
			150  & 0    & -56 & -36 & 20  & -2 \\
			-42  & -56  & 0   & 7   & 46  & -4 \\
			-55  & -36  & 7   & 0   & -5  & 37 \\
			-6   & 20   & 46  & -5  & 0   & -3 \\
			17   & -2   & -4  & 37  & -3  & 0
		\end{pmatrix}
		\ {\rm cm}^{-1}.
		\label{S-eq:J_pig_numeric}
	\end{align}
	The coherent pigment Hamiltonian is diagonalized according to
	\begin{align}
		H_{\rm pig}\ket{X_n}
		&=
		E_{X_n}\ket{X_n},
		\qquad
		n=1,\ldots,6,
		\\
		\ket{X_n}
		&=
		\sum_{i=1}^{6}
		c_i^{(n)}\ket i,
		\\
		c_i^{(n)}
		&=
		\braket{i|X_n},
		\\
		\sum_i
		c_i^{(n)*}c_i^{(m)}
		&=
		\delta_{nm}.
		\label{S-eq:exciton_diagonalization}
	\end{align}
	The coefficients $c_i^{(n)}$ retain the coherent electronic
	mixing generated by $J_{ij}$.  They subsequently determine both the
	state-resolved exciton-charge-transfer couplings and the vibrational
	dressing of each exciton.
	The numerical excitonic parametrization used in the calculation is
	summarized in Table~\ref{S-tab:exciton_parameters}.  The state-specific
	reorganization energies are obtained from the same excitonic coefficients $
	\lambda_{X_n}
	=
	\lambda_{\rm site}
	\sum_i
	\left|
	c_i^{(n)}
	\right|^4,
	\qquad
	\lambda_{\rm site}
	=
	540~{\rm cm}^{-1}.
	$
	
	\begin{table}[H]
		\centering
		\caption{Excitonic compositions, energies, reorganization energies, and
			projected couplings to the two primary charge-transfer states.  Energies
			and couplings are in cm$^{-1}$.}
		\label{S-tab:exciton_parameters}
		\begin{ruledtabular}
			\begin{tabular}{lrrrrrrrrrr}
				State &
				$P_{D1}$ &
				$P_{D2}$ &
				$\mathrm{Chl}_{D1}$ &
				$\mathrm{Chl}_{D2}$ &
				$\mathrm{Pheo}_{D1}$ &
				$\mathrm{Pheo}_{D2}$ &
				$E$ &
				$\lambda$ &
				$V_{XI_1}$ &
				$V_{XI_2}$
				\\
				\hline
				$X_1$ & $-0.029$ & $0.251$  & $0.800$  & $-0.018$ & $-0.541$ & $0.048$
				& $14412.5$ & $269.6$ & $18.13$  & $9.99$
				\\
				$X_2$ & $-0.185$ & $0.057$  & $-0.061$ & $-0.404$ & $0.040$  & $0.891$
				& $14459.7$ & $355.4$ & $-1.47$  & $-5.76$
				\\
				$X_3$ & $0.265$  & $-0.145$ & $0.560$  & $0.112$  & $0.754$  & $0.120$
				& $14517.0$ & $230.7$ & $91.98$  & $5.40$
				\\
				$X_4$ & $-0.558$ & $0.735$  & $-0.007$ & $0.116$  & $0.345$  & $-0.126$
				& $14540.8$ & $217.8$ & $23.66$  & $7.97$
				\\
				$X_5$ & $0.134$  & $0.096$  & $-0.121$ & $0.876$  & $-0.134$ & $0.416$
				& $14565.0$ & $334.7$ & $-17.85$ & $10.35$
				\\
				$X_6$ & $-0.752$ & $-0.603$ & $0.165$  & $0.209$  & $-0.003$ & $-0.012$
				& $14864.0$ & $245.5$ & $-11.76$ & $-60.98$
			\end{tabular}
		\end{ruledtabular}
	\end{table}
	The pigment-site parameters and excitonic coefficients used in this subsection were taken from the PSIIRC parametrization reported by Stones \textit{et al.}.~\cite{stones2017performance} These quantities were reproduced here for completeness and subsequently used as input parameters in our numerical simulations
	
	\subsection{Weak pigment-charge-transfer and charge-transfer couplings}
	\label{pigmentchargetransfer}
	
	The microscopic electronic parametrization contains the weak couplings that initiate charge separation. The nonzero pigment-CT matrix elements retained in the minimal model are $V_{P_{D1},I_2}=45~{\rm cm}^{-1}$, $V_{P_{D2},I_2}=45~{\rm cm}^{-1}$, $V_{\mathrm{Chl}_{D1},I_1}=70~{\rm cm}^{-1}$, and $V_{\mathrm{Pheo}_{D1},I_1}=70~{\rm cm}^{-1}$.
	Projection into the excitonic basis gives
	$
	V_{X_n I_j}
	=
	\bra{X_n}
	V_{\rm el}
	\ket{I_j}
	=
	\sum_i
	c_i^{(n)*}
	V_{iI_j},
	\qquad
	j=1,2.
	\label{S-eq:projected_XI}
	$
	For the above coupling pattern,
	\begin{align}
		V_{X_nI_1}
		&=
		70
		\left[
		c_{\mathrm{Chl}_{D1}}^{(n)*}
		+
		c_{\mathrm{Pheo}_{D1}}^{(n)*}
		\right],
		\\
		V_{X_nI_2}
		&=
		45
		\left[
		c_{P_{D1}}^{(n)*}
		+
		c_{P_{D2}}^{(n)*}
		\right].
		\label{S-eq:projected_XI_explicit}
	\end{align}
	The weak couplings retained inside the charge-separated network are $V_{I_2I_3}=70~{\rm cm}^{-1}$, $V_{I_1\alpha_1}=40~{\rm cm}^{-1}$, and $V_{I_3\alpha_1}=40~{\rm cm}^{-1}$. These couplings are not diagonalized again. Instead, they are retained as weak transfer interactions and therefore constitute the electronic matrix elements that appear explicitly in the Förster rates derived below. The pigment-site parameters and excitonic coefficients used in this subsection were taken from the PSIIRC parametrization reported by Stones \textit{et al.}.~\cite{stones2017performance} These quantities are reproduced here for completeness and are used as input parameters in our numerical simulations.
	
	\subsection{ Charge-transfer states and energies}
	\label{retainedchargetransfer}
	
	The primary charge-transfer configurations are $\ket{I_1}\equiv\ket{\mathrm{Chl}_{D1}^{+}\mathrm{Phe}_{D1}^{-}}$, $\ket{I_2}\equiv\ket{P_{D2}^{+}P_{D1}^{-}}$, and $\ket{I_3}\equiv\ket{P_{D1}^{+}\mathrm{Chl}_{D1}^{-}}$, while the terminal one-electron configurations are $\ket{\alpha_1}\equiv\ket{P_{D1}^{+}\mathrm{Phe}_{D1}^{-}}$ and $\ket{\beta_1}\equiv\ket{P_{D1}^{+}\mathrm{Phe}_{D1}}$. The corresponding retained energies are $E_g=0$, $E_{I_1}=E_{I_2}=14372~{\rm cm}^{-1}$, $E_{I_3}=14222~{\rm cm}^{-1}$, $E_{\alpha_1}=13972~{\rm cm}^{-1}$, and $E_{\beta_1}=2681~{\rm cm}^{-1}$. The charge-separated configurations are assigned state-dependent reorganization energies through the scaling factors $\nu_a$, such that $\lambda_a=\nu_a\lambda_{\rm site}$ and $g_a(t)=\nu_a g(t)$. The adopted values are $\nu_{I_1}=3$ with $\lambda_{I_1}=1620~{\rm cm}^{-1}$, $\nu_{I_2}=1.5$ with $\lambda_{I_2}=810~{\rm cm}^{-1}$, $\nu_{I_3}=3$ with $\lambda_{I_3}=1620~{\rm cm}^{-1}$, and $\nu_{\alpha_1}=4$ with $\lambda_{\alpha_1}=2160~{\rm cm}^{-1}$. These state-dependent line-broadening functions enter the fluorescence and absorption profiles used in the internal electron-transfer rates.~\cite{stones2017performance}
	
	\subsection{Acceptor-side one-electron and two-electron sectors}
	\label{S-sec:alpha2}
	
	The electronic manifold entering the cotunneling description is separated into one-electron $(N+1)$ and correlated two-electron $(N+2)$ charge sectors. These sectors have distinct roles in the reduced model. The $(N+1)$ sector provides the intermediate one-electron configurations entering the fourth-order cotunneling propagators, whereas the $(N+2)$ sector is used to construct the correlated doubly reduced state $\ket{\alpha_2}$ retained in the kinetic description. Because a complete \textit{ab initio} construction of all charged PSIIRC configurations is outside the scope of the present model, the reduced Hamiltonians are parameterized using representative energetic and coupling scales guided by the known acceptor-side electron-transfer sequence and its redox energetics. The numerical values should therefore be interpreted as an effective parametrization of the relevant charge sectors rather than as direct measurements of isolated molecular orbital energies or microscopic electron-transfer matrix elements.
	
	\subsubsection{Virtual one-electron $(N+1)$ sector}
	
	The one-electron active space follows the established acceptor-side sequence
	$\mathrm{Pheo}_{D1}\rightarrow Q_A\rightarrow Q_B$.
	Following primary charge separation, $\mathrm{Pheo}_{D1}^{-}$ transfers an electron to the tightly bound quinone $Q_A$, which subsequently transfers it to the terminal quinone $Q_B$.~
	\cite{Diner2002,DeCausmaecker2019,Brinkert2016,Kato2009Pheo,Kato2010QA}
	We therefore retain the three localized reduced configurations
	$
	\mathcal B_{\rm acc}^{(N+1)}
	=
	\left\{
	\ket{\mathrm{Pheo}_{D1}^{-}},
	\ket{Q_A^{-}},
	\ket{Q_B^{-}}
	\right\}.
	$
	This is the smallest localized subspace that contains the complete physiological acceptor sequence from the primary electron acceptor to the terminal quinone. The truncation is therefore based first on connectivity. Configurations outside this sequence are not retained in the acceptor-side virtual subspace considered here.
	
	The relative placement of the diagonal energies is guided by the known redox energetics of this branch. Reported midpoint potentials for $\mathrm{Pheo}_{D1}/\mathrm{Pheo}_{D1}^{-}$ depend appreciably on preparation and experimental conditions, with values in approximately the $-0.50$ to $-0.60~\mathrm{V}$ range, whereas spectro-electrochemical measurements place the $Q_A/Q_A^{-}$ potential near $-0.14~\mathrm{V}$ in oxygen-evolving PSIIRC. \cite{Kato2009Pheo,Kato2010QA,Brinkert2016} Corresponding estimates of the free-energy decrease for $\mathrm{Pheo}_{D1}^{-}\rightarrow Q_A$ are typically on the scale of several tenths of an electron volt. The subsequent $Q_A^{-}\rightarrow Q_B$ step has conventionally been assigned a substantially smaller energetic separation of approximately $0.07~\mathrm{eV}$, although more recent measurements and analyses suggest values approaching $0.18$-$0.23~\mathrm{eV}$. \cite{DeCausmaecker2019,kato2016redox} These values establish the direction and characteristic hierarchy of the acceptor-side energetics, but they should not be identified directly with the diagonal matrix elements of a diabatic electronic Hamiltonian, since redox potentials are thermodynamic free-energy quantities that also contain environmental and structural contributions.
	
	The diagonal entries used here are therefore chosen to reproduce the same energetic ordering,
	$
	E_{\mathrm{Pheo}_{D1}^{-}}
	>
	E_{Q_A^{-}}
	>
	E_{Q_B^{-}},
	$
	with a large separation between $\mathrm{Pheo}_{D1}^{-}$ and $Q_A^{-}$ and a smaller separation between $Q_A^{-}$ and $Q_B^{-}$. The Hamiltonian is referenced to the energy convention used in the transport model, so that a uniform change of energy zero would require the corresponding reservoir energies to be shifted consistently. We use
	\begin{equation}
		H^{(\alpha')}=
		\begin{pmatrix}
			0.63 & 0.020 & 0.002\\
			0.020 & 0.12 & 0.003\\
			0.002 & 0.003 & 0.03
		\end{pmatrix}
		\ {\rm eV},
		\label{S-eq:Halphaprime}
	\end{equation}
	which gives diabatic separations
	\begin{equation}
		E_{\mathrm{Pheo}_{D1}^{-}}-E_{Q_A^{-}}
		=
		0.51~{\rm eV},
		\qquad
		E_{Q_A^{-}}-E_{Q_B^{-}}
		=
		0.09~{\rm eV}.
	\end{equation}
	The $0.09~\mathrm{eV}$ separation is close to the conventional energetic scale used for the $Q_A^{-}\rightarrow Q_B$ step, whereas the $0.51~\mathrm{eV}$ separation is larger than modern estimates of the corresponding $\mathrm{Pheo}_{D1}^{-}\rightarrow Q_A$ free-energy decrease. We therefore treat the latter as an effective diabatic detuning rather than as a direct representation of a measured redox free-energy difference. The purpose of these diagonal values is to preserve the acceptor-side energetic hierarchy and an off-resonant separation of the localized configurations, rather than to provide a unique fit to a particular experimental preparation.
	
	The off-diagonal matrix elements require a different interpretation. They are not identified with microscopic transfer couplings calculated for the individual physiological electron-transfer steps. Recent superexchange calculations yield effective couplings on the sub-meV scale for both $\mathrm{Pheo}_{D1}^{-}\rightarrow Q_A$ and $Q_A^{-}\rightarrow Q_B$ transfer in PSIIRC. \cite{Saito2024,Sugo2022} The larger off-diagonal elements used in Eq.~\eqref{S-eq:Halphaprime} are instead representative effective hybridizations introduced to construct the reduced diabatic Hamiltonian. They deliberately provide a more strongly mixed test than the reported physiological transfer couplings while remaining small compared with the corresponding diabatic energy separations. Consequently, their role is to test whether the molecular identity of the virtual states remains well defined within a weak-hybridization regime, rather than to reproduce measured electron-transfer matrix elements.
	
	This weak-mixing condition can be quantified directly. The three coupling-to-detuning ratios are
	\begin{equation}
		\frac{0.020}{0.51}
		\simeq
		3.9\times10^{-2},
		\qquad
		\frac{0.003}{0.09}
		\simeq
		3.3\times10^{-2},
		\qquad
		\frac{0.002}{0.60}
		\simeq
		3.3\times10^{-3}.
		\label{S-eq:Nplus1_mixing_ratios}
	\end{equation}
	Thus, even with the effective hybridizations used here, the largest ratio remains below $0.04$. The corresponding second-order shifts of the $Q_B^{-}$-, $Q_A^{-}$-, and $\mathrm{Pheo}_{D1}^{-}$-like levels are approximately $-0.10$, $-0.69$, and $+0.79~\mathrm{meV}$, respectively, which are small compared with the $90$-$600~\mathrm{meV}$ diabatic separations. The molecular identity of the resulting eigenstates is therefore controlled primarily by the diagonal energy hierarchy rather than by fine tuning of the off-diagonal matrix elements.
	
	Diagonalization of Eq.~\eqref{S-eq:Halphaprime} gives eigenvalues of approximately
	$0.0299$, $0.1193$, and $0.6308~{\rm eV}$.
	The two low-energy eigenstates are
	\begin{align}
		\ket{\alpha_2'}
		&\simeq
		-0.002\,\ket{\mathrm{Pheo}_{D1}^{-}}
		-0.033\,\ket{Q_A^{-}}
		+0.999\,\ket{Q_B^{-}},
		\label{S-eq:alpha2prime_state}
		\\
		\ket{\alpha_1'}
		&\simeq
		-0.039\,\ket{\mathrm{Pheo}_{D1}^{-}}
		+0.999\,\ket{Q_A^{-}}
		+0.033\,\ket{Q_B^{-}}.
		\label{S-eq:alpha1prime_state}
	\end{align}
	The corresponding squared amplitudes show that $\ket{\alpha_2'}$ contains approximately $99.89\%$ $Q_B^{-}$ character and $\ket{\alpha_1'}$ approximately $99.74\%$ $Q_A^{-}$ character. The remaining eigenstate lies near $0.631~\mathrm{eV}$ and is approximately $99.85\%$ $\mathrm{Pheo}_{D1}^{-}$-like. Thus, within this reduced Hamiltonian, the two low-energy branches retain essentially pure $Q_B^{-}$ and $Q_A^{-}$ molecular character, while the higher branch remains predominantly $\mathrm{Pheo}_{D1}^{-}$-like.
	
	The fourth-order cotunneling propagators are consequently projected onto the two quinone-dominated virtual branches $\ket{\alpha_1'}$ and $\ket{\alpha_2'}$, consistent with the $Q_B$-centered correlated-transfer sector constructed below. The $\mathrm{Pheo}_{D1}^{-}$-like eigenstate is not retained as an additional virtual channel in this reduced description. Its exclusion should be understood as part of the quinone-centered projection rather than as a statement that its contribution to every possible fourth-order process is identically negligible. The large energetic separation and weak quinone admixture are consistent with this truncation, provided that the omitted branch does not acquire a substantially larger coupling through the specific system-reservoir operators entering the cotunneling kernel.
	
	The numerical rate calculation uses the corresponding unrounded virtual energies,
	$E_{\alpha_2}'\simeq0.028~{\rm eV}$ and
	$E_{\alpha_1}'\simeq0.116~{\rm eV}$.
	The small differences from the eigenvalues obtained from Eq.~\eqref{S-eq:Halphaprime} arise because the matrix displayed above contains rounded parameters, whereas the rates are evaluated using the underlying numerical values.
	
	The primed states are not additional stationary populations of the thirteen-state kinetic model. They are virtual $(N+1)$ intermediate channels used only to resolve the fourth-order cotunneling amplitude. Their role is therefore distinct from that of the physical correlated $(N+2)$ state $\ket{\alpha_2}$ constructed below.
	
	\subsubsection{$Q_B$-centered correlated $(N+2)$ sector}
	
	The $(N+2)$ sector is constructed independently as a reduced
	acceptor-side active space. Its definition precedes diagonalization
	and is based on the known topology and redox functionality of the PSII
	acceptor branch. Although $Q_A$ and $Q_B$ are both plastoquinones,
	their physiological roles are different. $Q_A$ functions primarily as
	a tightly bound one-electron relay, whereas $Q_B$ is the exchangeable
	terminal quinone that receives two reducing equivalents in successive
	photochemical turnovers. Following the first
	$Q_A^{-}\rightarrow Q_B$ transfer, the semiquinone $Q_B^{-}$ is
	formed. During the subsequent turnover, a newly generated $Q_A^{-}$
	supplies the second reducing equivalent. Proton-coupled reduction
	ultimately produces plastoquinol, $Q_BH_2$, which leaves the
	$Q_B$-binding pocket and enters the membrane plastoquinone pool.
	\cite{Diner2002,DeCausmaecker2019,kato2016redox}
	
	The cotunneling construction does not attempt to reproduce this
	complete proton-coupled physiological $Q_B$ cycle. Instead, it isolates
	the electronic subspace relevant to a proposed $Q_B$-centered
	doubly reduced state. We therefore define the $(N+2)$ active space by
	two restrictions. First, at least one of the two excess electrons is
	required to occupy $Q_B$, so that every retained configuration belongs
	to the $Q_B$-centered reduction branch. Second, the remaining excess
	electron is restricted to the redox-active sites of the same
	sequential acceptor pathway, namely $Q_B$, $Q_A$, or
	$\mathrm{Pheo}_{D1}$. Symbolically, this projection may be written as
	\begin{equation}
		\mathcal H_{\rm acc}^{(N+2)}
		\longrightarrow
		\mathcal H_{Q_B}^{(N+2)}
		=
		\operatorname{span}
		\left\{
		(Q_B,Q_B),\,
		(Q_A,Q_B),\,
		(\mathrm{Pheo}_{D1},Q_B)
		\right\}.
		\label{S-eq:Nplus2_projection}
	\end{equation}
	The corresponding localized diabatic basis is therefore
	$
	\mathcal B_{Q_B}^{(N+2)}
	=
	\left\{
	\ket{Q_B^{2-}},
	\ket{Q_A^{-}Q_B^{-}},
	\ket{\mathrm{Pheo}_{D1}^{-}Q_B^{-}}
	\right\}.
	$
	This restriction is an active-space approximation rather than a claim
	that the complete $(N+2)$ Hilbert space of PSII contains only these
	three configurations. Configurations without $Q_B$ occupation, states
	involving other reaction-center pigments, and explicit protonated
	quinone configurations are outside the $Q_B$-centered electronic
	sector considered here and are therefore projected out.
	
	The three retained diabatic configurations have specific roles in this
	construction. The configuration $\ket{Q_B^{2-}}$ represents the
	electronic double-occupation limit of the terminal quinone and defines
	the molecular character sought for $\ket{\alpha_2}$. It should not be
	identified with an isolated long-lived physiological intermediate.
	Under physiological conditions, acquisition of the second reducing
	equivalent at $Q_B$ is coupled to proton uptake and ultimately produces
	$Q_BH_2$. The configuration $\ket{Q_A^{-}Q_B^{-}}$ describes charge
	distributed over the two quinone acceptors and represents the electronic
	configuration preceding completion of the second reduction of $Q_B$.
	Finally, $\ket{\mathrm{Pheo}_{D1}^{-}Q_B^{-}}$ retains the upstream
	primary acceptor within the same Pheo$_{D1}$-$Q_A$-$Q_B$ topology and
	provides an additional charge-separated electronic configuration that
	can mix with the $Q_B$-localized branch. The latter configuration is
	therefore a model-defined diabatic basis state and its inclusion does not
	imply that $\mathrm{Pheo}_{D1}^{-}Q_B^{-}$ is an experimentally
	isolated intermediate of the physiological $Q_B$ cycle.
	
	After this active space has been specified, electronic mixing is
	introduced through the effective Hamiltonian
	\begin{equation}
		H^{(\alpha_2)}=
		\begin{pmatrix}
			1.50 & 0.003 & 0.002\\
			0.003 & 0.55 & 0.008\\
			0.002 & 0.008 & 0.85
		\end{pmatrix}
		\ {\rm eV},
		\label{S-eq:Halpha2}
	\end{equation}
	written in the ordered basis
	$\{
	\ket{Q_B^{2-}},
	\ket{Q_A^{-}Q_B^{-}},
	\ket{\mathrm{Pheo}_{D1}^{-}Q_B^{-}}
	\}$.
	The diagonal elements are effective diabatic energies of the three localized electronic configurations rather than experimentally measured energies of physiological intermediates. To assess whether identification of the $Q_B^{2-}$-like branch depends on fine tuning of this parametrization, we tested its molecular character against independent variations of the diabatic energies and electronic couplings. At the nominal parametrization, the selected eigenstate contains approximately $99.998\%$ $Q_B^{2-}$ weight. Allowing each diagonal energy to vary independently by $\pm0.20~\mathrm{eV}$ while simultaneously varying the off-diagonal couplings between $0.5$ and $2$ times their nominal values leaves the selected branch greater than approximately $99.95\%$ $Q_B^{2-}$-like over the tested parameter grid. This test establishes robustness within the off-resonant weak-hybridization regime sampled by the effective model and it is not intended to assign an experimental uncertainty to the matrix elements. The reduced single-branch description would cease to be appropriate if an omitted or retained $(N+2)$ configuration became nearly resonant with $Q_B^{2-}$ or coupled sufficiently strongly to invalidate this separation of energy scales.
	
	For the parametrization, the $Q_B^{2-}$ diabatic configuration
	is separated from the $Q_A^{-}Q_B^{-}$ and
	$\mathrm{Pheo}_{D1}^{-}Q_B^{-}$ configurations by approximately
	$0.95~{\rm eV}$ and $0.65~{\rm eV}$, respectively, whereas the
	corresponding direct couplings are only $3~{\rm meV}$ and
	$2~{\rm meV}$. The associated ratios,
	\begin{align}
		\frac{|V_{Q_B^{2-},\,Q_A^-Q_B^-}|}
		{|E_{Q_B^{2-}}-E_{Q_A^-Q_B^-}|}
		&\simeq
		3.2\times10^{-3},
		\\
		\frac{|V_{Q_B^{2-},\,\mathrm{Pheo}_{D1}^-Q_B^-}|}
		{|E_{Q_B^{2-}}-E_{\mathrm{Pheo}_{D1}^-Q_B^-}|}
		&\simeq
		3.1\times10^{-3},
		\label{S-eq:Nplus2_mixing_ratios}
	\end{align}
	show that the selected branch lies deeply in the weak-mixing regime for
	this effective Hamiltonian. This estimate is a consequence of the
	adopted parametrization and is not used to select the basis itself.
	Diagonalization gives three correlated eigenstates. The eigenstate
	adiabatically connected to the localized $\ket{Q_B^{2-}}$
	configuration is
	\begin{align}
		\ket{\alpha_2}
		&\simeq
		0.99999\,\ket{Q_B^{2-}}
		+
		0.00318\,\ket{Q_A^{-}Q_B^{-}}
		+
		0.00312\,\ket{\mathrm{Pheo}_{D1}^{-}Q_B^{-}},
		\label{S-eq:alpha2_state}
		\\
		E_{\alpha_2}
		&\simeq
		1.50002~{\rm eV}
		\simeq
		1.21\times10^{4}~{\rm cm}^{-1}.
		\label{S-eq:E_alpha2}
	\end{align}
	The precise overall sign of the eigenvector is arbitrary. The squared
	amplitude of the $Q_B^{2-}$ component is approximately $0.99998$,
	confirming that the selected eigenstate retains overwhelmingly the
	molecular character used to define the target branch. The small
	$Q_A^{-}Q_B^{-}$ and
	$\mathrm{Pheo}_{D1}^{-}Q_B^{-}$ components quantify the weak
	hybridization with the other configurations retained in the active
	space.
	
	It is important that the identification
	$\ket{\alpha_2}\equiv\ket{\Psi_{Q_B^{2-}}}$ is made by continuity and
	molecular character rather than by eigenvalue ordering. In the
	zero-coupling limit, this eigenstate continuously approaches
	$\ket{Q_B^{2-}}$, switching on the off-diagonal matrix elements
	introduces only the small admixtures shown in
	Eq.~\eqref{S-eq:alpha2_state}. Consequently, the fact that this branch
	is not the lowest eigenvalue of Eq.~\eqref{S-eq:Halpha2} does not alter
	its identification with the proposed $Q_B^{2-}$-like
	$\ket{\alpha_2}$ state. The diagonalization therefore provides an
	a posteriori consistency check of the physically specified active
	space rather than the rationale for choosing that space.
	
	\subsubsection{Separation of retained-state and virtual-state energies}
	
	The $(N+2)$ energy, the virtual $(N+1)$ energies, and the Coulomb parameter appearing in the cotunneling kernel represent different quantities in the reduced description, such that $E_{\alpha_2}\neq E_{\alpha_1}'$, $E_{\alpha_2}\neq E_{\alpha_2}'$, and $E_{\alpha_2}\neq U$. Here, $E_{\alpha_2}$ is the assigned energy of the retained $Q_B^{2-}$-like correlated state, whereas $E_{\alpha_1}'$ and $E_{\alpha_2}'$ characterize the virtual one-electron channels entering the fourth-order cotunneling propagators. The parameter $U$ accounts for the interaction associated with the correlated two-electron transfer sector and modifies the corresponding intermediate-state energy denominators and it is neither an electronic hopping matrix element nor the energy $E_{\alpha_2}$ itself. Thus, the primed $(N+1)$ channels provide the virtual one-electron structure required by the cotunneling amplitude, whereas $\ket{\alpha_2}$ represents the retained correlated $(N+2)$ state reached by that process.
	
	The numerical Hamiltonians employed here should therefore be understood as literature-informed effective Hamiltonians rather than as matrices whose individual elements are obtained directly from a single experiment or electronic-structure calculation. The literature constrains the one-electron sector most directly by identifying the acceptor sequence, establishing that electron transfer from $\mathrm{Pheo}_{D1}^{-}$ to $Q_A$ is strongly downhill on the scale of several tenths of an electron volt, and indicating a smaller but preparation-dependent driving force for $Q_A^{-}\rightarrow Q_B$. The adopted one-electron parametrization preserves these features through $E_{\mathrm{Pheo}^-}>E_{Q_A^-}>E_{Q_B^-}$. The numerical differences are effective diabatic detunings and are not identified one-to-one with measured midpoint-potential differences.
	
	Second, the off-diagonal elements reflect the spatial and kinetic
	connectivity of the acceptor branch rather than independently measured
	physiological transfer matrix elements. Since electron transfer proceeds
	through $\mathrm{Pheo}_{D1}\rightarrow Q_A\rightarrow Q_B$, coupling
	between these electronically connected configurations is retained,
	while the direct Pheo$_{D1}$-$Q_B$ term is chosen smaller. The numerical
	values are intentionally kept within a regime in which all
	coupling-to-detuning ratios are small. Their purpose is therefore to
	generate a stable weakly hybridized virtual-state basis, not to reproduce
	the sub-meV effective superexchange couplings calculated for the
	physiological transfer steps.
	
	The same distinction is more important for the $(N+2)$ sector. Its
	basis is fixed from the $Q_B$-centered active-space projection described
	above, but direct experimental energies are not available for the bare
	unprotonated configurations $\ket{Q_B^{2-}}$,
	$\ket{Q_A^-Q_B^-}$, and
	$\ket{\mathrm{Pheo}_{D1}^-Q_B^-}$ as isolated physiological states.
	In particular, physiological acquisition of the second reducing
	equivalent at $Q_B$ is proton coupled and ultimately produces $Q_BH_2$.
	The diagonal elements of Eq.~\eqref{S-eq:Halpha2} are therefore model
	parameters chosen to place the targeted $Q_B^{2-}$-like electronic
	branch in the required energy range and to keep it off resonant from
	the other retained two-electron configurations. The small off-diagonal
	elements then test whether this molecular character survives weak
	electronic hybridization rather than attempting to reproduce a measured
	two-electron Hamiltonian.
	
	Thus, the literature determines the identity and connectivity of the
	acceptor configurations and provides the relevant thermodynamic scales
	for the established one-electron transfer sequence, whereas the precise
	matrix elements define the reduced-model parametrization within those
	constraints. The one-electron Hamiltonian is therefore more directly
	anchored to known redox energetics than the two-electron Hamiltonian,
	for which the principal physical requirements are the $Q_B$-centered
	active-space construction, weak hybridization, and robustness of the
	retained molecular character. The validity of the reduced description
	is consequently assessed by these scale separations and by the
	stability of the resulting eigenstates, rather than by assigning each
	matrix element to a uniquely measured microscopic quantity.
	
	\subsection{Thirteen-state reference Hamiltonian}
	\label{thirteenstate}
	
	The final finite-system basis is $
	\mathcal B_S
	=
	\left\{
	\ket g,
	\ket{X_1},\ldots,\ket{X_6},
	\ket{I_1},
	\ket{I_2},
	\ket{I_3},
	\ket{\alpha_1},
	\ket{\alpha_2},
	\ket{\beta_1}
	\right\}.
	$
	The reference system Hamiltonian used in the reduced dynamics is
	\begin{align}
		H_S
		&=
		E_g\ket g\bra g
		+
		\sum_{n=1}^{6}
		E_{X_n}
		\ket{X_n}\bra{X_n}
		+
		\sum_{j=1}^{3}
		E_{I_j}
		\ket{I_j}\bra{I_j}
		\nonumber\\
		&\quad+
		E_{\alpha_1}
		\ket{\alpha_1}\bra{\alpha_1}
		+
		E_{\alpha_2}
		\ket{\alpha_2}\bra{\alpha_2}
		+
		E_{\beta_1}
		\ket{\beta_1}\bra{\beta_1}.
		\label{S-eq:H13}
	\end{align}
	
	\begin{table}[ht]
		\centering
		\caption{Energies of the non-excitonic states retained in the thirteen-state reference Hamiltonian. The excitonic energies $E_{X_n}$ are given in the preceding excitonic-state construction.}
		\label{tab:retained_state_energies}
		\begin{tabular}{cc}
			\hline\hline
			State & Energy (${\rm cm}^{-1}$) \\
			\hline
			$\ket{g}$          & $0$ \\
			$\ket{I_1}$        & $14372$ \\
			$\ket{I_2}$        & $14372$ \\
			$\ket{I_3}$        & $14222$ \\
			$\ket{\alpha_1}$   & $13972$ \\
			$\ket{\alpha_2}$   & $12098$ \\
			$\ket{\beta_1}$    & $2681$ \\
			\hline\hline
		\end{tabular}
	\end{table}

	\section{Environmental Hamiltonians and interaction terms}
	\label{S-sec:environment}

	The total Hamiltonian is partitioned as
	
	\begin{align}
		H
		&=
		H_0+V,
		\\
		H_0
		&=
		H_S
		+
		H_{\rm vib}
		+
		H_{\rm ant}
		+
		H_L
		+
		H_R,
		\\
		V
		&=
		V_{\rm vib}
		+
		V_{\rm F}
		+
		V_{\rm ant}
		+
		V_{\rm tun}.
		\label{S-eq:total_H_partition}
	\end{align}
	
	Here, $H_S$ is the thirteen-state reference Hamiltonian of the finite PSIIRC introduced above, $H_{\rm vib}$ describes the vibrational environment coupled to the reaction-center states, and $H_{\rm ant}$ represents the light-harvesting antenna degrees of freedom that supply excitation to the reaction center. The Hamiltonians $H_L$ and $H_R$ describe the electronic source and drain reservoirs, respectively, with the source associated with electron injection from the OEC side and the drain representing electron transfer toward the downstream PQ pool. The corresponding interaction terms describe distinct physical processes. $V_{\rm vib}$ couples the excitonic states to the vibrational environment, $V_{\rm F}$ contains the weak internal electronic couplings between retained many-body states, $V_{\rm ant}$ transfers excitation from the antenna to the reaction center, and $V_{\rm tun}$ couples the finite system to the electronic source and drain reservoirs.
	
	\subsection{Vibrational bath, 48-mode spectral profile, and Matsubara terms}

	The vibrational environment is represented by harmonic modes,
	
	\begin{align}
		H_{\rm vib}
		&=
		\sum_{i,\xi}
		\hbar\omega_{i\xi}
		b_{i\xi}^{\dagger}b_{i\xi}.
		\label{S-eq:H_vib}
	\end{align}
	
	The local energy-gap fluctuation operator is
	
	\begin{align}
		B_i
		&=
		\sum_\xi
		g_{i\xi}
		\left(
		b_{i\xi}
		+
		b_{i\xi}^{\dagger}
		\right),
	\end{align}
	
	so that
	$V_{\rm vib}
	=
	\sum_i
	\ket i\bra i
	\otimes B_i.
	\label{S-eq:V_vib_site} 
	$In the exciton basis,
	\begin{align}
		V_{\rm vib}
		&=
		\sum_{p,q=1}^{6}
		\ket{X_p}\bra{X_q}
		\otimes B_{pq},
		\\
		B_{pq}
		&=
		\sum_i
		c_i^{(p)*}c_i^{(q)}
		B_i.
		\label{S-eq:Bpq}
	\end{align}
	The site correlation function is $C_i(t) = 
	\left\langle
	B_i(t)B_i(0)
	\right\rangle_{\rm vib}.
	\label{S-eq:C_site}
	$
	The detailed spectral profile used in the PSIIRC rate calculation is represented through line-broadening function~\cite{stones2017performance}
	\begin{align}
		g(t)
		&=
		g_D(t)
		+
		\sum_{j=1}^{48}
		g_j(t),
		\label{S-eq:g_total}
	\end{align}
	where $g_D(t)$ is the Drude contribution and $g_j(t)$ is the contribution
	from the $j^{th}$ underdamped experimentally observed mode.  In the
	spectroscopic convention used for the line-shape construction,
	\begin{align}
		g_D(t)
		&=
		\frac{c_0^D}{\Omega_D^2}
		\left(
		e^{-\Omega_Dt}
		+
		\Omega_Dt
		-
		1
		\right)
		+
		\sum_{k=1}^{\infty}
		\frac{c_k^D}{\nu_k^2}
		\left(
		e^{-\nu_kt}
		+
		\nu_kt
		-
		1
		\right),
		\label{S-eq:g_D}
	\end{align}
	and
	\begin{align}
		g_j(t)
		&=
		\sum_{\sigma=\pm}
		\frac{c_{0j}^{\sigma}}{\nu_{j\sigma}^2}
		\left(
		e^{-\nu_{j\sigma}t}
		+
		\nu_{j\sigma}t
		-
		1
		\right)
		+
		\sum_{k=1}^{\infty}
		\frac{c_{kj}}{\nu_k^2}
		\left(
		e^{-\nu_kt}
		+
		\nu_kt
		-
		1
		\right).
		\label{S-eq:g_j}
	\end{align}
	
	The coefficients are
	
	\begin{align}
		c_0^D
		&=
		\lambda_D\Omega_D
		\left[
		\cot
		\left(
		\frac{\beta\Omega_D}{2}
		\right)
		-
		i
		\right],
		\\
		c_k^D
		&=
		\frac{4\lambda_D\Omega_D}{\beta}
		\frac{\nu_k}{\nu_k^2-\Omega_D^2},
		\label{S-eq:Drude_coefficients}
	\end{align}
	
	and
	
	\begin{align}
		c_{0j}^{\pm}
		&=
		\pm i
		\frac{\lambda_j\omega_j^2}{2\zeta_j}
		\left[
		\cot
		\left(
		\frac{\beta\nu_{j\pm}}{2}
		\right)
		-
		i
		\right],
		\\
		c_{kj}
		&=
		-\frac{4\lambda_j\gamma_j\omega_j^2}{\beta}
		\frac{\nu_k}{
			\left(
			\omega_j^2+\nu_k^2
			\right)^2
			-
			\gamma_j^2\nu_k^2
		},
		\\
		\nu_{j\pm}
		&=
		\frac{\gamma_j}{2}
		\pm
		i\zeta_j,
		\\
		\zeta_j
		&=
		\sqrt{
			\omega_j^2-\frac{\gamma_j^2}{4}
		}.
		\label{S-eq:underdamped_coefficients}
	\end{align}
	The Matsubara frequencies are given by $\nu_k=2\pi k/\beta$ with $k=1,2,\ldots$. These line-broadening expressions and their Matsubara decomposition follow the PSIIRC spectral-density treatment used by Stones \textit{et al.}.~\cite{stones2017performance}
	The Matsubara summations are truncated only after the retained terms yield numerically converged line-broadening functions at the working temperature. Their role is to represent the finite-temperature correction to the bath correlation function, and they are not additional physical pigment modes. The Drude parameters used here are those employed in the PSIIRC spectral density of Stones \textit{et al.}~\cite{stones2017performance}, with $\lambda_D=35~{\rm cm}^{-1}$ and $\Omega_D=40~{\rm cm}^{-1}$. The 48 vibrational-mode frequencies and Huang-Rhys factors used in the structured PSIIRC spectral density are taken from the parametrization employed by Stones \textit{et al.}~\cite{stones2017performance}, which is based on the experimentally constrained PSIIRC spectral density reported by Novoderezhkin \textit{et al.}.~\cite{Novoderezhkin2005} For an underdamped mode, the corresponding reorganization contribution is $\lambda_j=s_j\omega_j$.
	\begin{table}[H]
		\centering
		\caption{Frequencies $\omega_j$ in cm$^{-1}$ and Huang-Rhys factors
			$s_j$ for the 48 experimentally resolved PSIIRC modes.}
		\label{S-tab:48modes}
		\begin{ruledtabular}
			\begin{tabular}{rrrrrrrrrr}
				$\omega_j$ & $s_j$ &
				$\omega_j$ & $s_j$ &
				$\omega_j$ & $s_j$ &
				$\omega_j$ & $s_j$ &
				$\omega_j$ & $s_j$
				\\
				\hline
				97  & 0.0371 & 573 & 0.0094 & 995  & 0.0293 & 1252 & 0.0051 & 1524 & 0.0067 \\
				138 & 0.0455 & 585 & 0.0034 & 1052 & 0.0131 & 1260 & 0.0064 & 1537 & 0.0222 \\
				213 & 0.0606 & 604 & 0.0034 & 1069 & 0.0064 & 1286 & 0.0047 & 1553 & 0.0091 \\
				260 & 0.0539 & 700 & 0.0050 & 1110 & 0.0192 & 1304 & 0.0057 & 1573 & 0.0044 \\
				298 & 0.0488 & 722 & 0.0074 & 1143 & 0.0303 & 1322 & 0.0202 & 1580 & 0.0044 \\
				342 & 0.0380 & 742 & 0.0269 & 1181 & 0.0179 & 1338 & 0.0037 & 1612 & 0.0044 \\
				388 & 0.0202 & 752 & 0.0219 & 1190 & 0.0084 & 1354 & 0.0057 & 1645 & 0.0034 \\
				425 & 0.0168 & 795 & 0.0077 & 1208 & 0.0121 & 1382 & 0.0067 & 1673 & 0.0010 \\
				518 & 0.0303 & 916 & 0.0286 & 1216 & 0.0111 & 1439 & 0.0067 &      &        \\
				546 & 0.0030 & 986 & 0.0162 & 1235 & 0.0034 & 1487 & 0.0074 &      &
			\end{tabular}
		\end{ruledtabular}
	\end{table}
	For an excitonic eigenstate, the corresponding line-broadening function
	is
	\begin{align}
		g_{X_n}(t)
		&=
		\sum_i
		\left|
		c_i^{(n)}
		\right|^4
		g(t).
		\label{S-eq:g_exciton}
	\end{align}
	The 48-mode information does not modify the electronic eigenstates
	themselves.  It enters after the electronic diagonalization by dressing the energy gaps and transition lineshapes used in the rate kernels.
	The structured vibrational spectral profile, including the Drude
	background and the 48 underdamped PSIIRC modes, follows the parametrization
	used earlier ~\cite{stones2017performance, Sharma2025,Kalita2026}, based on the
	experimentally constrained spectral density of Novoderezhkin
	\textit{et al.}.~\cite{Novoderezhkin2005} The resulting line-broadening functions
	are used here in the calculation of the vibrationally assisted transition
	rates.
	
	\subsection{Weak internal electronic interaction}
	
	The weak internal electronic links retained in the PSIIRC model follow the
	charge-transfer connectivity used in previous PSIIRC descriptions.
	~\cite{stones2017performance,Kalita2026}
	\begin{align}
		V_{\rm F}
		&=
		\sum_{n=1}^{6}
		\left[
		V_{I_1X_n}
		\ket{I_1}\bra{X_n}
		+
		V_{I_2X_n}
		\ket{I_2}\bra{X_n}
		+
		{\rm H.c.}
		\right]
		\nonumber\\
		&\quad+
		V_{I_3I_2}
		\ket{I_3}\bra{I_2}
		+
		V_{\alpha_1I_1}
		\ket{\alpha_1}\bra{I_1}
		+
		V_{\alpha_1I_3}
		\ket{\alpha_1}\bra{I_3}
		+
		{\rm H.c.}
		\label{S-eq:V_F}
	\end{align}
	No direct $X_n\leftrightarrow I_3$ coupling is retained, consistent with the
	negligible electronic coupling and spectral overlap assigned to this pathway
	in the adopted PSIIRC parametrization.~\cite{stones2017performance,Kalita2026}
	
	\subsection{Nonequilibrium antenna reservoir and matter-antenna coupling}
	\label{S-sec:antenna}

	In the present description, photoexcitation of the PSIIRC is mediated
	by the surrounding light-harvesting antenna rather than by direct
	coupling of the reaction centre to an equilibrium free-space radiation
	bath.~\cite{Dorfman2013}  The antenna is represented as an effective nonequilibrium
	bosonic environment whose collective excitation modes are described by.~\cite{breuer2002theory,Nitzan2006}
	\begin{align}
		H_{\rm ant}
		&=
		\sum_{\alpha}
		\hbar\Omega_{\alpha}
		A_{\alpha}^{\dagger}
		A_{\alpha},
		\label{S-eq:H_ant}
	\end{align}
	where $A_{\alpha}^{\dagger}$ and $A_{\alpha}$ create and annihilate,
	respectively, an excitation in the effective antenna mode $\alpha$ of
	frequency $\Omega_{\alpha}$.  These operators should be interpreted as
	collective excitation-transfer modes of the light-harvesting
	environment and not necessarily as bare free-space photon operators.
	Under continuous illumination, the occupation of an antenna mode may
	be described phenomenologically by
	\begin{align}
		\frac{dN_{\alpha}}{dt}
		&=
		G_{\alpha}
		-
		\kappa_{{\rm tr},\alpha}
		N_{\alpha}
		-
		\kappa_{{\rm loss},\alpha}
		N_{\alpha},
		\label{S-eq:antenna_kinetics}
	\end{align}
	
	where $G_{\alpha}$ is the generation rate of antenna excitation,
	$\kappa_{{\rm tr},\alpha}$ describes transfer out of the antenna
	toward the reaction centre, and $\kappa_{{\rm loss},\alpha}$ contains
	the remaining radiative and nonradiative relaxation channels.  The
	stationary occupation is therefore
	\begin{align}
		N_{\alpha}^{\rm eff}
		&=
		\frac{
			G_{\alpha}
		}{
			\kappa_{{\rm tr},\alpha}
			+
			\kappa_{{\rm loss},\alpha}
		}
		\nonumber\\
		&=
		{\rm Tr}_{\rm ant}
		\left[
		\rho_{\rm ant}^{\rm ss}
		A_{\alpha}^{\dagger}
		A_{\alpha}
		\right].
		\label{S-eq:Nalpha_eff}
	\end{align}
	When the individual mode dependence of the antenna occupation is not resolved, we employ the reduced description $N_{\alpha}^{\rm eff}\longrightarrow N_{\rm eff}$, where $N_{\rm eff}$ denotes the effective nonequilibrium occupation sampled by the relevant PSIIRC optical transition.
	For each excitonic transition we define the system raising and lowering
	operators
	
	\begin{align}
		S_n^{+}
		&=
		\ket{X_n}\bra g,
		\\
		S_n^{-}
		&=
		\ket g\bra{X_n},
		\label{S-eq:optical_system_operators}
	\end{align}
	and the antenna operator sampled by that transition,
	\begin{align}
		B_n
		&=
		\sum_{\alpha}
		g_{n\alpha}
		A_{\alpha},
		\\
		B_n^{\dagger}
		&=
		\sum_{\alpha}
		g_{n\alpha}^{*}
		A_{\alpha}^{\dagger}.
		\label{S-eq:Bn_antenna}
	\end{align}
	
	The interaction Hamiltonian can consequently be written in the
	rotating-wave form
	\begin{align}
		V_{\rm ant}
		&=
		\sum_n
		\left[
		S_n^{+}\otimes B_n
		+
		S_n^{-}\otimes B_n^{\dagger}
		\right]
		\nonumber\\
		&=
		\sum_{n,\alpha}
		\left[
		g_{n\alpha}
		\ket{X_n}\bra g
		A_{\alpha}
		+
		g_{n\alpha}^{*}
		\ket g\bra{X_n}
		A_{\alpha}^{\dagger}
		\right].
		\label{S-eq:V_ant}
	\end{align}
	Thus the reaction-centre transition operator and the antenna operator
	remain distinct objects.  The coefficient $g_{n\alpha}$ specifies the
	coupling between excitonic transition $n$ and collective antenna mode
	$\alpha$ and is, in general, dependent on both indices.
	The origin of $g_{n\alpha}$ follows from the electric-dipole
	interaction.  Writing the positive-frequency component of the local
	antenna field at the reaction centre as
	\begin{align}
		\mathbf E_{\rm ant}^{(+)}
		\left(
		\mathbf r_{\rm RC}
		\right)
		&=
		\sum_{\alpha}
		\boldsymbol{\mathcal E}_{\alpha}^{\rm loc}
		\left(
		\mathbf r_{\rm RC}
		\right)
		A_{\alpha},
		\label{S-eq:E_ant_positive}
	\end{align}
	with
	
	\begin{align}
		\mathbf E_{\rm ant}^{(-)}
		\left(
		\mathbf r_{\rm RC}
		\right)
		&=
		\left[
		\mathbf E_{\rm ant}^{(+)}
		\left(
		\mathbf r_{\rm RC}
		\right)
		\right]^{\dagger},
	\end{align}
	the microscopic matter-field interaction is
	\begin{align}
		V_{\rm dip}
		&=
		-
		\boldsymbol{\mu}
		\cdot
		\left[
		\mathbf E_{\rm ant}^{(+)}
		+
		\mathbf E_{\rm ant}^{(-)}
		\right].
		\label{S-eq:dipole_ant}
	\end{align}
	Projection onto the transition
	$\ket g\leftrightarrow\ket{X_n}$ and application of the
	rotating-wave approximation recover Eq.~\eqref{S-eq:V_ant}~\cite{Nitzan2006,breuer2002theory}, with
	\begin{align}
		g_{n\alpha}
		&=
		-
		\boldsymbol{\mu}_{gX_n}
		\cdot
		\boldsymbol{\mathcal E}_{\alpha}^{\rm loc}
		\left(
		\mathbf r_{\rm RC}
		\right),
		\label{S-eq:g_ant_general}
	\end{align}
	where
	$
	\boldsymbol{\mu}_{gX_n}
	=
	\bra g
	\boldsymbol{\mu}
	\ket{X_n}.
	$
	Since the excitonic states are obtained from the localized pigment
	basis,
	\begin{align}
		\ket{X_n}
		&=
		\sum_m
		c_m^{(n)}
		\ket m,
	\end{align}
	The excitonic transition dipoles are constructed from the pigment-site
	transition dipoles and the excitonic expansion coefficients.
	~\cite{Gelzinis2013,stones2017performance}
	the corresponding transition dipole is
	$
	\boldsymbol{\mu}_{gX_n}
	=
	\sum_m
	c_m^{(n)}
	\boldsymbol{\mu}_{gm},
	\label{S-eq:exciton_dipole}
	$
	where $\boldsymbol{\mu}_{gm}$ is the transition dipole associated with
	localized pigment state $\ket m$.  The same coefficients
	$c_m^{(n)}$ that determine the electronic composition of the excitonic
	states therefore also determine their optical coupling strengths.
	To separate the local field magnitude from its polarization, we write
	\begin{align}
		\boldsymbol{\mathcal E}_{\alpha}^{\rm loc}
		\left(
		\mathbf r_{\rm RC}
		\right)
		&=
		\mathcal E_{\alpha}^{\rm eff}
		\mathbf e_{\alpha}^{\rm loc},
		\label{S-eq:E_local_factorization}
	\end{align}
	where $\mathbf e_{\alpha}^{\rm loc}$ is the local polarization unit
	vector.  The microscopic coupling then becomes
	\begin{align}
		g_{n\alpha}
		&=
		-
		\mathcal E_{\alpha}^{\rm eff}
		\,
		\boldsymbol{\mu}_{gX_n}
		\cdot
		\mathbf e_{\alpha}^{\rm loc}
		\nonumber\\
		&=
		-
		\mathcal E_{\alpha}^{\rm eff}
		\,
		\mu_{gX_n}
		\cos\theta_{n\alpha},
		\label{S-eq:g_ant_orientation}
	\end{align}
	where $\mu_{gX_n}=\left|\boldsymbol{\mu}_{gX_n}\right|$ and $\theta_{n\alpha}$ is the angle between the excitonic transition dipole and the local polarization of antenna mode $\alpha$. Equation~\eqref{S-eq:g_ant_orientation} gives the more general microscopic form of the coupling. Since the detailed spatial field distribution and polarization of every collective antenna mode are not resolved explicitly in the present reduced model, their effect is absorbed into a frequency-dependent effective coefficient.
	\begin{align}
		g_0
		\left(
		\Omega_{\alpha}
		\right)
		&\equiv
		-
		\mathcal E_{\alpha}^{\rm eff}
		\,
		\mathcal O_{\alpha},
		\label{S-eq:g0_definition}
	\end{align}
	where $\mathcal O_{\alpha}$ denotes the effective orientational and
	local-field projection appropriate to mode $\alpha$.  The working
	factorized approximation is then
	\begin{align}
		g_{n\alpha}
		&\simeq
		g_0
		\left(
		\Omega_{\alpha}
		\right)
		\mu_{gX_n}.
		\label{S-eq:g_ant_factorized}
	\end{align}
	The approximation in Eq.~\eqref{S-eq:g_ant_factorized} therefore does
	not assume identical microscopic couplings for all excitonic states.
	The exciton dependence remains explicitly contained in
	$\mu_{gX_n}$, whereas the unresolved spatial, polarization, and
	frequency dependence of the antenna field is contained in
	$g_0(\Omega_{\alpha})$.  If the complete spatial and polarization
	structure of the antenna were retained, the factorized form would be
	replaced by the full $n$- and $\alpha$-dependent coupling of
	Eq.~\eqref{S-eq:g_ant_general}.
	
	For comparison with a conventionally normalized electromagnetic mode,
	the characteristic field-per-excitation scale may be estimated as
	\begin{align}
		\mathcal E_{\alpha}^{\rm eff}
		&\sim
		L_{\alpha}
		\sqrt{
			\frac{
				\hbar\Omega_{\alpha}
			}{
				2
				\epsilon_0
				\epsilon_r
				V_{\alpha}^{\rm eff}
			}
		},
		\label{S-eq:Eeff_estimate}
	\end{align}
	where $V_{\alpha}^{\rm eff}$ is an effective mode volume,
	$\epsilon_r$ is the dielectric permittivity of the local environment,
	and $L_{\alpha}$ represents a local-field renormalization factor.  This
	gives the order-of-magnitude estimate
	\begin{align}
		\left|
		g_{n\alpha}
		\right|
		&\sim
		\mu_{gX_n}
		L_{\alpha}
		\sqrt{
			\frac{
				\hbar\Omega_{\alpha}
			}{
				2
				\epsilon_0
				\epsilon_r
				V_{\alpha}^{\rm eff}
			}
		}
		\left|
		\cos\theta_{n\alpha}
		\right|.
		\label{S-eq:g_ant_estimate}
	\end{align}
	Equation~\eqref{S-eq:Eeff_estimate} is used only as a physical scaling
	estimate.  The operators $A_{\alpha}$ in the present model describe
	effective collective antenna excitations rather than rigorously
	quantized free-space electromagnetic modes, so that neither a unique
	mode volume nor a unique local polarization field is generally
	available.
	A more direct characterization of the effective coupling is obtained
	through the antenna spectral density,
	\begin{align}
		J_{{\rm ant},n}(\omega)
		&=
		\sum_{\alpha}
		\left|
		g_{n\alpha}
		\right|^2
		\delta
		\left(
		\omega-\Omega_{\alpha}
		\right).
		\label{S-eq:J_ant_general}
	\end{align}
	Under the factorization of
	Eq.~\eqref{S-eq:g_ant_factorized},
	\begin{align}
		J_{{\rm ant},n}(\omega)
		&\simeq
		D_{\rm ant}(\omega)
		\left|
		g_0(\omega)
		\right|^2
		\left|
		\mu_{gX_n}
		\right|^2,
		\label{S-eq:J_ant_factorized}
	\end{align}
	where $D_{\rm ant}(\omega)$ is the effective density of antenna
	excitations, with the unresolved orientational factor understood to be
	included in $g_0(\omega)$.
	Consequently, $g_0(\omega)$ need not be assigned independently at the
	Hamiltonian level.  It may instead be fixed by the effective optical
	pumping strength evaluated from the antenna correlation function.  For
	a transition resonant at
	\begin{align}
		\Omega_{X_ng}
		&=
		\frac{
			E_{X_n}-E_g
		}{
			\hbar
		},
	\end{align}
	the excitation rate obtained from the second-order reservoir kernel is~\cite{breuer2002theory,Nitzan2006}
	\begin{align}
		\gamma_{g\rightarrow X_n}
		&=
		\frac{
			2\pi
		}{
			\hbar^2
		}
		J_{{\rm ant},n}
		\left(
		\Omega_{X_ng}
		\right)
		N_n^{\rm eff},
		\label{S-eq:gamma_from_Jant}
	\end{align}
	where $N_n^{\rm eff}$ denotes the effective antenna occupation sampled
	at the transition frequency.  Therefore the kinetic model constrains
	the combination
	\begin{align}
		J_{{\rm ant},n}
		\left(
		\Omega_{X_ng}
		\right)
		N_n^{\rm eff},
	\end{align}
	rather than determining $g_0$ and the nonequilibrium antenna occupation
	independently.
	
	The interaction Hamiltonian in Eq.~\eqref{S-eq:V_ant} retains optical coupling between the ground state and the complete six-state excitonic manifold, $\ket g\leftrightarrow\ket{X_n}$ with $n=1,\ldots,6$. No higher excitonic transition is removed at the Hamiltonian level. Instead, the relative strengths of the six pumping channels are determined by their state-dependent transition dipole moments. Under the factorized coupling approximation
	\begin{align}
		g_{n\alpha}
		&\simeq
		g_0(\Omega_\alpha)
		\mu_{gX_n},
	\end{align}
	
	the mode-resolved coupling strength satisfies
	
	\begin{align}
		\left|
		g_{n\alpha}
		\right|^2
		&\simeq
		\left|
		g_0(\Omega_\alpha)
		\right|^2
		\left|
		\mu_{gX_n}
		\right|^2.
		\label{S-eq:g_ant_squared}
	\end{align}
	The common optical coupling is calibrated by taking the
	$\ket g\leftrightarrow\ket{X_1}$ excitation rate as
	$\gamma_{g\rightarrow X_1}=75~{\rm cm}^{-1}$, consistent with the
	photo-excitation rate employed in previous PSIIRC junction models.
	~\cite{stones2017performance,Sharma2025} The remaining five optical
	rates are then obtained in the present model from their corresponding
	exciton transition-dipole strengths.
	The physical description of antenna-mediated excitation of the reaction
	centre follows established photosynthetic light-harvesting and
	nonequilibrium photocell descriptions.~\cite{Dorfman2013,creatore2013efficient}
	The pigment transition-dipole vectors and excitonic coefficients used to
	construct the state-resolved optical couplings are taken from the PSIIRC
	parametrizations of Gelzinis \textit{et al.} and Stones \textit{et al.}.
	~\cite{Gelzinis2013,stones2017performance} The effective antenna
	occupation and the factorized coupling $g_{n\alpha}\simeq
	g_0(\Omega_\alpha)\mu_{gX_n}$ constitute the reduced description adopted
	in the present work.

	\subsection{Electronic Source and Drain Reservoirs}
	
	The electronic source and drain are modeled as non-interacting fermionic reservoirs following the standard open-junction description of electronic transport.~\cite{harbola2006quantum,Timm2008,Harbola2008} In the present PSIIRC mapping, the source represents the donor-side oxygen-evolving complex, which supplies reducing equivalents required to restore the oxidized reaction-center sector, whereas the drain represents the downstream plastoquinone pool that accepts electronic charge transferred through the acceptor side.~\cite{stones2017performance,Sharma2025} The explicitly retained $Q_B$-centered configurations, including the one- and two-electron states introduced above, are treated as part of the finite PSIIRC system rather than as part of the drain. Only the electronic degrees of freedom downstream of this retained acceptor sector are coarse-grained into the effective drain reservoir. Throughout the transport formulation, the labels left and source are used synonymously, $L\equiv{\rm source}$, and the labels right and drain are used synonymously, $R\equiv{\rm drain}$. The notation $L$ and $R$ is therefore a transport convention and does not introduce an additional spatial partition beyond the physical source-system-drain structure.
	
	The two electronic reservoirs are represented by
	
	\begin{align}
		H_\ell
		&=
		\sum_k
		\epsilon_{k\ell}
		c_{k\ell}^{\dagger}c_{k\ell},
		\qquad
		\ell=L,R,
		\label{S-eq:H_leads}
	\end{align}
	where $c_{k\ell}^{\dagger}$ and $c_{k\ell}$ create and annihilate an electron in reservoir mode $k$ of reservoir $\ell$, respectively. Thus, $H_L$ describes the effective electronic degrees of freedom associated with the OEC-side source, whereas $H_R$ describes the downstream PQ-pool degrees of freedom lying outside the explicitly retained PSIIRC acceptor manifold.
	
	The reservoirs are taken to remain in grand-canonical reference states,
	\begin{align}
		\rho_\ell
		&=
		\frac{
			\exp
			\left[
			-\beta_\ell
			\left(
			H_\ell-\mu_\ell N_\ell
			\right)
			\right]
		}{
			{\rm Tr}_\ell
			\exp
			\left[
			-\beta_\ell
			\left(
			H_\ell-\mu_\ell N_\ell
			\right)
			\right]
		},
		\\
		f_\ell(\epsilon)
		&=
		\frac{1}{
			e^{\beta_\ell(\epsilon-\mu_\ell)}+1
		},
		\label{S-eq:Fermi}
	\end{align}
	where $\beta_\ell$ and $\mu_\ell$ are the inverse temperature and chemical potential of reservoir $\ell$, and $N_\ell$ is its electron-number operator. The corresponding Fermi function $f_\ell(\epsilon)$ determines the occupation available for electron exchange at the source and drain interfaces.
	Electron exchange between the finite PSIIRC system and these reservoirs is described by the standard bilinear tunneling interaction~\cite{harbola2006quantum,Timm2008},
	\begin{align}
		V_{\rm tun}
		&=
		V_L+V_R,
		\\
		V_\ell
		&=
		\sum_{k,\chi}
		\left[
		t_{k\chi}^{(\ell)}
		c_{k\ell}^{\dagger}d_\chi
		+
		t_{k\chi}^{(\ell)*}
		d_\chi^{\dagger}c_{k\ell}
		\right],
		\label{S-eq:V_tun}
	\end{align}
	where $d_\chi$ acts on the relevant electronic degree of freedom of the finite PSIIRC system and $t_{k\chi}^{(\ell)}$ is the corresponding system-reservoir tunneling amplitude. On the source side, these terms represent electronic replenishment from the OEC-side reservoir into the oxidized donor sector of the reaction center. On the drain side, they represent electronic transfer from the retained reaction-center and acceptor-side configurations toward the downstream PQ pool. In the reduced kinetic network, this partition underlies the source-mediated restoration of the oxidized reaction center and the drain-mediated electronic extraction processes, while the internal excitonic, charge-transfer, and $Q_B$-centered configurations remain part of the finite system.
	
	\section{General Nakajima-Zwanzig Construction}
	\label{S-sec:NZ}
	
	The total density operator obeys
	
	\begin{align}
		\frac{d}{dt}
		\rho_{\rm tot}(t)
		&=
		-\frac{i}{\hbar}
		\left[
		H,\rho_{\rm tot}(t)
		\right]
		\equiv
		-i\mathcal L
		\rho_{\rm tot}(t).
		\label{S-eq:von_Neumann}
	\end{align}
	
	The projection operators are
	
	\begin{align}
		\mathcal P A
		&=
		\rho_B
		\otimes
		{\rm Tr}_B A,
		\\
		\mathcal Q
		&=
		1-\mathcal P,
		\label{S-eq:PQ}
	\end{align}
	with
	$
	\rho_B
	=
	\rho_{\rm vib}
	\otimes
	\rho_{\rm ant}^{\rm ss}
	\otimes
	\rho_L
	\otimes
	\rho_R.
	\label{S-eq:rho_B}
	$
	For the initially factorized condition
	$
	\rho_{\rm tot}(0)
	=
	\rho_S(0)\otimes\rho_B,
	\\
	\mathcal Q\rho_{\rm tot}(0)
	=
	0,
	$
	the exact Nakajima-Zwanzig equation is~\cite{Nakajima1958,Zwanzig1960,breuer2002theory},
	
	\begin{align}
		\frac{d}{dt}
		\mathcal P\rho_{\rm tot}(t)
		&=
		-i\mathcal P\mathcal L\mathcal P
		\rho_{\rm tot}(t)
		+
		\int_0^t
		d\tau\,
		\mathcal K(\tau)
		\mathcal P\rho_{\rm tot}(t-\tau),
		\\
		\mathcal K(\tau)
		&=
		-\mathcal P\mathcal L
		e^{-i\mathcal Q\mathcal L\tau}
		\mathcal Q\mathcal L\mathcal P.
		\label{S-eq:NZ_exact}
	\end{align}
	
	For a generic interaction channel $\chi$,
	
	\begin{align}
		V_\chi
		&=
		\sum_{a,b}
		\ket a\bra b
		\otimes
		B_{ab}^{(\chi)},
		\label{S-eq:generic_interaction}
	\end{align}
	and
	$
	\Omega_{ab}
	=
	\frac{E_a-E_b}{\hbar}.
	$
	The corresponding interaction-picture operator is
	\begin{align}
		\widetilde V_\chi(t)
		&=
		\sum_{a,b}
		e^{i\Omega_{ab}t}
		\ket a\bra b
		\otimes
		B_{ab}^{(\chi)}(t).
		\label{S-eq:generic_interaction_picture}
	\end{align}
	After absorbing environmental means into the reference Hamiltonian,
	$
	{\rm Tr}_{B_\chi}
	\left[
	B_{ab}^{(\chi)}
	\rho_{B_\chi}
	\right]
	=
	0.
	$
	The leading dissipative population kernel is second order  in the
	system-environment interaction~\cite{breuer2002theory},
	\begin{align}
		\mathcal K_\chi^{(2)}(t)\rho_S
		&=
		-\frac{1}{\hbar^2}
		{\rm Tr}_{B_\chi}
		\left[
		\widetilde V_\chi(t),
		\left[
		\widetilde V_\chi(0),
		\rho_S\otimes\rho_{B_\chi}
		\right]
		\right].
		\label{S-eq:K2}
	\end{align}
	The unevaluated environmental correlation function associated with the
	transition $\ket b\rightarrow\ket a$ is
	\begin{align}
		C_{a\leftarrow b}^{(\chi)}(t)
		&=
		{\rm Tr}_{B_\chi}
		\left[
		B_{ab}^{(\chi)}(t)
		B_{ba}^{(\chi)}(0)
		\rho_{B_\chi}
		\right].
		\label{S-eq:generic_C}
	\end{align}
	
	Projection onto populations gives
	
	\begin{align}
		\left.
		\dot P_a(t)
		\right|_{\chi}
		&=
		\sum_{b\neq a}
		\int_0^t
		d\tau
		\left[
		K_{a\leftarrow b}^{(\chi,2)}(\tau)
		P_b(t-\tau)
		-
		K_{b\leftarrow a}^{(\chi,2)}(\tau)
		P_a(t-\tau)
		\right],
		\\
		K_{a\leftarrow b}^{(\chi,2)}(t)
		&=
		\frac{2}{\hbar^2}
		{\rm Re}
		\left[
		e^{i\Omega_{ab}t}
		C_{a\leftarrow b}^{(\chi)}(t)
		\right].
		\label{S-eq:population_memory_kernel}
	\end{align}
	In the Markov regime ~\cite{breuer2002theory,Timm2008, harbola2006quantum},
	$
	\tau_{B_\chi}
	\ll
	\tau_{\rm pop},
	$
	the upper integration limit is extended to infinity and the general
	second-order rate becomes
	\begin{align}
		W_{a\leftarrow b}^{(\chi,2)}
		&=
		\frac{2}{\hbar^2}
		{\rm Re}
		\int_0^\infty
		dt\,
		e^{i\Omega_{ab}t}
		C_{a\leftarrow b}^{(\chi)}(t).
		\label{S-eq:W2_general}
	\end{align}
	Equation~\eqref{S-eq:W2_general} is only a structural result.  The
	physical meaning of a rate is determined by the operator and correlation
	function inserted into it.  Optical antenna correlations, protein
	energy-gap correlations, donor-acceptor spectral overlaps, and
	fermionic source-drain contractions therefore carry different
	information and cannot be interchanged.
	Cotunneling arises at fourth order in the tunnelling interaction, so that
	its leading irreducible contribution contains four applications of
	$V_{\rm tun}$~\cite{Averin1990,Konig1996,Timm2008},
	\begin{align}
		\mathcal K_{\rm cot}^{(4)}
		&=
		\mathcal O
		\left(
		V_{\rm tun}^4
		\right).
		\label{S-eq:K4_order}
	\end{align}
	
	The associated four-point reservoir object is
	
	\begin{align}
		C^{(4)}_{4321}
		(t,t_1,t_2,t_3)
		&=
		{\rm Tr}_{L,R}
		\left[
		F_4(t)
		F_3(t_1)
		F_2(t_2)
		F_1(t_3)
		\rho_L\rho_R
		\right].
		\label{S-eq:C4_general}
	\end{align}
	
	For noninteracting fermionic reservoirs, Wick's theorem gives
	~\cite{Harbola2008},
	\begin{align}
		\left\langle
		F_4F_3F_2F_1
		\right\rangle
		=
		\left\langle F_4F_3\right\rangle
		\left\langle F_2F_1\right\rangle
		-
		\left\langle F_4F_2\right\rangle
		\left\langle F_3F_1\right\rangle
		+
		\left\langle F_4F_1\right\rangle
		\left\langle F_3F_2\right\rangle.
		\label{S-eq:Wick}
	\end{align}
	
	The reducible contribution corresponding to two independent
	second-order sequential events is removed in constructing the irreducible
	fourth-order kernel~\cite{Konig1996,Timm2008},
	
	\begin{align}
		\mathcal K_{\rm cot}^{(4),{\rm irr}}
		&=
		\mathcal K_{\rm tun}^{(4)}
		-
		\mathcal K_{\rm red}^{(4)}.
		\label{S-eq:K4_irr}
	\end{align}
	
	Only after this general second- and fourth-order construction do we
	identify the individual PSIIRC rates.
	
	\section{Explicit evaluation of the PSIIRC rate families}
	\label{S-sec:rates}
	The baseline PSIIRC population model contains four rate families:
	the excitonic rates $r_{mn}$, the internal F\"orster rates $k_{ab}$, the sequential
	source-drain rates $\Gamma_{ab}$, and the optical pumping rates $\gamma_{ab}$. The cotunneling extension adds the fifth family
	$\omega_{ab}$. Throughout this SI the arrow fixes the
	direction explicitly, for example $k_{a\rightarrow b}$ denotes
	population transfer from $\ket a$ to $\ket b$.

	\subsection{Optical Antenna Pumping:
		\texorpdfstring{$W^{(2)}\rightarrow\gamma_{ab}, a,b \in g,X_n$}{W to gamma_{ab}}}
	\label{S-sec:optical}
	
	The optical excitation and depumping rates are obtained from the
	second-order correlation-function treatment of the effective bosonic
	antenna reservoir.~\cite{breuer2002theory,Nitzan2006}
	For the antenna reservoir,
	\begin{align}
		C_{\alpha}^{(+)}(\tau)
		&=
		\left\langle
		A_{\alpha}^{\dagger}(\tau)
		A_{\alpha}(0)
		\right\rangle_{\rm ss},
		\\
		C_{\alpha}^{(-)}(\tau)
		&=
		\left\langle
		A_{\alpha}(\tau)
		A_{\alpha}^{\dagger}(0)
		\right\rangle_{\rm ss}.
		\label{S-eq:antenna_correlations}
	\end{align}
	
	Using the nonequilibrium steady-state occupation $N_{\rm eff}$,
	
	\begin{align}
		C_{\alpha}^{(+)}(\tau)
		&=
		N_{\rm eff}
		e^{+i\Omega_\alpha\tau},
		\\
		C_{\alpha}^{(-)}(\tau)
		&=
		\left(
		N_{\rm eff}+1
		\right)
		e^{-i\Omega_\alpha\tau}.
		\label{S-eq:antenna_correlations_eval}
	\end{align}
	
	The one-sided Fourier transforms are
	
	\begin{align}
		\mathcal G_{\alpha}^{(\pm)}(\omega)
		&=
		\int_0^\infty
		d\tau\,
		e^{\pm i\omega\tau}
		C_{\alpha}^{(\pm)}(\tau).
		\label{S-eq:antenna_FT}
	\end{align}
	
	Using
	
	\begin{align}
		\int_0^\infty
		d\tau\,
		e^{ix\tau}
		&=
		\pi\delta(x)
		+
		i\mathcal P
		\left(
		\frac{1}{x}
		\right),
		\label{S-eq:one_sided_identity}
	\end{align}
	
	and neglecting the principal-value Lamb-shift contribution, the
	dissipative part samples resonant antenna modes.
	
	Replacing the discrete antenna modes by a continuum,
	
	\begin{align}
		\sum_\alpha
		\left|
		g_{n\alpha}
		\right|^2
		\delta
		\left(
		\omega-\Omega_\alpha
		\right)
		&\longrightarrow
		D(\omega)
		\left|
		g_n(\omega)
		\right|^2,
		\label{S-eq:antenna_continuum}
	\end{align}
	with
	$
	\left|
	g_n(\omega)
	\right|^2
	=
	\left|
	g_0(\omega)
	\right|^2
	\left|
	\mu_{gX_n}
	\right|^2,
	$
	gives the antenna spectral density
	
	\begin{align}
		J_{{\rm ant},n}(\omega)
		&=
		D(\omega)
		\left|
		g_0(\omega)
		\right|^2
		\left|
		\mu_{gX_n}
		\right|^2.
		\label{S-eq:J_ant}
	\end{align}
	The microscopic excitation and depumping rates become
	\begin{align}
		\gamma_{g\rightarrow X_n}
		&=
		\frac{2\pi}{\hbar}
		D(\Omega_{X_ng})
		\left|
		g_0(\Omega_{X_ng})
		\right|^2
		\left|
		\mu_{gX_n}
		\right|^2
		N_{\rm eff},
		\\
		\gamma_{X_n\rightarrow g}
		&=
		\frac{2\pi}{\hbar}
		D(\Omega_{X_ng})
		\left|
		g_0(\Omega_{X_ng})
		\right|^2
		\left|
		\mu_{gX_n}
		\right|^2
		\left(
		N_{\rm eff}+1
		\right).
		\label{S-eq:gamma_general}
	\end{align}
	
	Defining
	$
	\gamma_{\rm opt}(\omega)
	=
	\frac{2\pi}{\hbar}
	D(\omega)
	\left|
	g_0(\omega)
	\right|^2,
	$
	one obtains
	\begin{align}
		\gamma_{g\rightarrow X_n}
		&=
		\gamma_{\rm opt}
		\left|
		\mu_{gX_n}
		\right|^2
		N_{\rm eff},
		\\
		\gamma_{X_n\rightarrow g}
		&=
		\gamma_{\rm opt}
		\left|
		\mu_{gX_n}
		\right|^2
		\left(
		N_{\rm eff}+1
		\right).
		\label{S-eq:gamma_compact}
	\end{align}
	The correlation function in this derivation carries the occupation and
	spectral content of the nonequilibrium antenna reservoir.  It does not
	encode protein vibrational relaxation or electronic source-drain
	statistics.
	
	The excitonic transition dipole moments were evaluated by combining the
	site transition-dipole vector
	~\cite{Gelzinis2013} with the excitonic expansion coefficients reported
	in Supplementary Table S4 of Stones \textit{et al.}.
	~\cite{Stones2017} The excitonic transition dipole moments were
	therefore evaluated from
	$
	\boldsymbol{\mu}_{gX_n}
	=
	\sum_m
	c_m^{(n)}
	\boldsymbol{\mu}_m,
	\label{S-eq:exciton_dipole_numeric}
	$
	where $c_m^{(n)}$ is the contribution of pigment $m$ to the excitonic
	state $\ket{X_n}$, and $\boldsymbol{\mu}_m$ is the corresponding
	site transition-dipole vector.
	The excitonic transition dipole moments were evaluated from
	\begin{align}
		\boldsymbol{\mu}_{gX_n}
		&=
		\sum_m
		c_m^{(n)}
		\boldsymbol{\mu}_m,
		\label{S-eq:exciton_dipole_numeric}
	\end{align}
	where the excitonic coefficients $c_m^{(n)}$ are those already listed in
	Table~\ref{S-tab:exciton_parameters}. The Cartesian site transition-dipole
	vectors $\boldsymbol{\mu}_m$ were taken from Table~2 of Gelzinis
	\textit{et al.}.~\cite{Gelzinis2013} Using these quantities, the Cartesian
	components of each excitonic transition dipole were obtained directly from
	Eq.~\eqref{S-eq:exciton_dipole_numeric}. The corresponding transition-dipole
	strength was then calculated as
	\begin{align}
		\left|\mu_{gX_n}\right|^2
		&=
		\mu_{x,n}^{\,2}
		+
		\mu_{y,n}^{\,2}
		+
		\mu_{z,n}^{\,2}.
		\label{S-eq:exciton_dipole_strength}
	\end{align}
	
	For example, for the lowest excitonic state $X_1$, the coefficients listed
	in Table~\ref{S-tab:exciton_parameters} give
	\begin{align}
		\boldsymbol{\mu}_{gX_1}
		&=
		(-3.988592,\,
		2.292288,\,
		3.651045)\ {\rm D},
	\end{align}
	and therefore
	$
	\left|\mu_{gX_1}\right|^2
	=
	(-3.988592)^2
	+
	(2.292288)^2
	+
	(3.651045)^2=
	34.493578~{\rm D^2}.
	$
	
	Applying the same transformation to all six excitonic states gives the
	values summarized below.
	
	\begin{table}[H]
		\centering
		\begin{ruledtabular}
			\begin{tabular}{lrrrr}
				State &
				$\mu_x$ (D) &
				$\mu_y$ (D) &
				$\mu_z$ (D) &
				$\left|\mu_{gX_n}\right|^2$ (D$^2$)
				\\
				\hline
				$X_1$ & $-3.988592$ & $ 2.292288$ & $ 3.651045$ & $34.493578$ \\
				$X_2$ & $-3.644747$ & $-2.576942$ & $-1.860749$ & $23.387201$ \\
				$X_3$ & $ 0.180852$ & $ 0.341271$ & $-1.665148$ & $ 2.921892$ \\
				$X_4$ & $-3.936622$ & $ 1.213673$ & $ 0.283519$ & $17.050378$ \\
				$X_5$ & $ 3.771692$ & $-1.422807$ & $ 1.140639$ & $17.551100$ \\
				$X_6$ & $-0.939012$ & $-1.432151$ & $-1.492608$ & $ 5.160680$ \\
			\end{tabular}
		\end{ruledtabular}
		\caption{Excitonic transition-dipole vectors
			$\boldsymbol{\mu}_{gX_n}$ and the corresponding squared transition-dipole
			strengths $\left|\mu_{gX_n}\right|^2$ obtained from the excitonic
			coefficients in Table~\ref{S-tab:exciton_parameters} and the site
			transition-dipole vectors reported by Gelzinis \textit{et al.}.
			\cite{Gelzinis2013}}
		\label{S-tab:exciton_dipoles}
	\end{table}
	The quantities $\left|\mu_{gX_n}\right|^2$ in
	Table~\ref{S-tab:exciton_dipoles} are subsequently used to evaluate the
	state-resolved optical excitation and depumping rates in
	Eq.~\eqref{S-eq:gamma_compact}.
	For the numerical evaluation of the optical rates, the transition-dipole
	strengths listed in Table~\ref{S-tab:exciton_dipoles} were used together
	with an effective nonequilibrium antenna occupation of
	$N_{\rm eff}=50000$. The common effective optical coupling
	$\gamma_{\rm opt}$ was fixed by using the lowest excitonic transition
	$\ket g\leftrightarrow\ket{X_1}$ as the reference. Specifically, the
	excitation rate was set to
	$\gamma_{g\rightarrow X_1}=75~{\rm cm}^{-1}$, corresponding to the
	photoexcitation rate employed by Stones \textit{et al.}.
	~\cite{Stones2017} From Eq.~\eqref{S-eq:gamma_compact},
	
	\begin{align}
		\gamma_{\rm opt}
		&=
		\frac{
			\gamma_{g\rightarrow X_1}
		}{
			\left|\mu_{gX_1}\right|^2
			N_{\rm eff}
		}=
		\frac{
			75
		}{
			(34.493578)(50000)
		}=
		4.34864\times10^{-5}
		~{\rm cm}^{-1}{\rm D}^{-2}.
		\label{S-eq:gamma_opt_numeric}
	\end{align}
	The same value of $\gamma_{\rm opt}$ was then retained for all six
	excitonic transitions.  The excitation rates were evaluated as
	$
	\gamma_{g\rightarrow X_n}
	=
	\gamma_{\rm opt}
	\left|\mu_{gX_n}\right|^2
	N_{\rm eff},
	n=1,\ldots,6,
	$
	which gives
	\begin{align}
		\big(
		\gamma_{g\rightarrow X_1},
		\gamma_{g\rightarrow X_2},
		\gamma_{g\rightarrow X_3},
		\gamma_{g\rightarrow X_4},
		\gamma_{g\rightarrow X_5},
		\gamma_{g\rightarrow X_6}
		\big)
		&=
		\big(
		75.00000,\,
		50.85121,\,
		6.35312,\,
		37.07294,\,
		38.16167,\,
		11.22096
		\big)
		~{\rm cm}^{-1}.
		\label{S-eq:forward_optical_rates}
	\end{align}
	The corresponding depumping rates were obtained from the same effective
	coupling and transition-dipole strengths,
	$
	\gamma_{X_n\rightarrow g}
	=
	\gamma_{\rm opt}
	\left|\mu_{gX_n}\right|^2
	\left(
	N_{\rm eff}+1
	\right),
	$
	with $N_{\rm eff}+1=50001$.  The resulting rates are
	\begin{align}
		\big(
		\gamma_{X_1\rightarrow g},
		\gamma_{X_2\rightarrow g},
		\gamma_{X_3\rightarrow g},
		\gamma_{X_4\rightarrow g},
		\gamma_{X_5\rightarrow g},
		\gamma_{X_6\rightarrow g}
		\big)
		&=
		\big(
		75.00150,\,
		50.85222,\,
		6.35325,\,
		37.07368,\,
		38.16243,\,
		11.22118
		\big)
		~{\rm cm}^{-1}.
		\label{S-eq:reverse_optical_rates}
	\end{align}
	
	Thus, the $\ket g\leftrightarrow\ket{X_1}$ transition serves only to
	calibrate the common effective coupling $\gamma_{\rm opt}$.  Once this
	quantity is fixed, the relative strengths of all remaining excitation
	and depumping channels are determined by their corresponding excitonic
	transition-dipole strengths listed in
	Table~\ref{S-tab:exciton_dipoles}.

	\subsection{Excitonic Redistribution Rates:
		\texorpdfstring{$W^{(2)}\rightarrow r_{mn},\; m,n=1,\ldots,6$}{W to r}}
	\label{S-sec:exciton_rates}
	
	The excitonic redistribution rates act only within the manifold
	$\{\ket{X_1},\ldots,\ket{X_6}\}$. Throughout this section, $r_{mn}$ denotes population transfer from the initial excitonic state $\ket{X_m}$ to the final excitonic state $\ket{X_n}$, so that
	
	\begin{align}
		r_{mn}
		&\equiv
		W_{X_n\leftarrow X_m}^{({\rm exc},2)},
		\qquad
		m\neq n.
		\label{S-eq:r_definition}
	\end{align}
	
	The rates are evaluated from the second-order vibrational population kernel introduced above, using the modified-Redfield implementation appropriate to the adopted PSIIRC spectral density~\cite{Renger2001,Yang2002,Novoderezhkin2007,stones2017performance}. In this construction, the diagonal exciton--vibration coupling is incorporated through the state-dependent line-broadening functions, whereas the population-changing inter-exciton coupling is treated perturbatively.
	
	The vibrational correlation function associated with the transition
	$\ket{X_m}\rightarrow\ket{X_n}$ is
	
	\begin{align}
		C_{n\leftarrow m}^{({\rm vib})}(t)
		&=
		\left\langle
		B_{nm}(t)
		B_{mn}(0)
		\right\rangle_{\rm vib}.
		\label{S-eq:C_exciton}
	\end{align}
	
	For statistically independent fluctuations of the pigment sites,
	
	\begin{align}
		\left\langle
		B_i(t)B_j(0)
		\right\rangle
		&=
		\delta_{ij}C_i(t),
	\end{align}
	
	and transformation from the site basis to the exciton basis gives the corresponding bare Redfield contraction,
	
	\begin{align}
		C_{n\leftarrow m}^{({\rm vib})}(t)
		&=
		\sum_i
		\left|
		c_i^{(n)}
		\right|^2
		\left|
		c_i^{(m)}
		\right|^2
		C_i(t).
		\label{S-eq:C_exciton_eval}
	\end{align}
	
	Here, $c_i^{(m)}$ is the coefficient of pigment site $i$ in excitonic eigenstate $\ket{X_m}$. To include the state-dependent diagonal vibrational dressing required by modified-Redfield theory, we introduce the four-index excitonic overlap factors
	
	\begin{align}
		R_{abcd}
		&=
		\sum_i
		c_i^{(a)}
		c_i^{(b)}
		c_i^{(c)}
		c_i^{(d)},
		\\
		g_{abcd}(t)
		&=
		R_{abcd}g(t),
		\\
		\lambda_{abcd}
		&=
		R_{abcd}\lambda_{\rm site},
		\\
		\Lambda_{abcd}
		&=
		\frac{\lambda_{abcd}}{\hbar}.
		\label{S-eq:four_index}
	\end{align}
	
	The indices $a,b,c,d$ in Eq.~\eqref{S-eq:four_index} are auxiliary tensor indices and are distinguished from the transition labels $m,n$ used for the initial and final excitonic states.
	
	For the transition $\ket{X_m}\rightarrow\ket{X_n}$, the modified-Redfield correlation function entering the population kernel is written as
	
	\begin{align}
		C_{n\leftarrow m}^{({\rm MR})}(t)
		&=
		\hbar^2
		e^{-i\overline{\Lambda}_{nm}t}
		e^{-\overline g_{nm}(t)}
		\eta_{nm}(t),
		\label{S-eq:C_MR}
	\end{align}
	
	where
	
	\begin{align}
		\overline{\Lambda}_{nm}
		&=
		\Lambda_{nnnn}
		+
		\Lambda_{mmmm}
		-
		2\Lambda_{mmnn},
		\\
		\overline g_{nm}(t)
		&=
		g_{nnnn}(t)
		+
		g_{mmmm}(t)
		-
		2g_{mmnn}(t),
		\label{S-eq:MR_bar}
	\end{align}
	
	and
	
	\begin{align}
		\eta_{nm}(t)
		&=
		\ddot g_{mnmn}(t)
		-
		\left[
		\dot g_{mnmm}(t)
		-
		\dot g_{mnnn}(t)
		+
		2i\Lambda_{mnmm}
		\right]^2.
		\label{S-eq:MR_eta}
	\end{align}
	
	This line-broadening-dressed correlation function is the modified-Redfield form used for population redistribution among delocalized excitonic states in pigment--protein complexes~\cite{Renger2001,Yang2002,Novoderezhkin2007}. The exponential factor accounts for the diagonal vibrational dressing of the initial and final excitons, while $\eta_{nm}(t)$ contains the population-changing contribution generated by the off-diagonal exciton--bath coupling.
	
	Substitution of Eq.~\eqref{S-eq:C_MR} into the general second-order Nakajima--Zwanzig rate,
	Eq.~\eqref{S-eq:W2_general}, gives
	
	\begin{align}
		r_{mn}
		&\equiv
		W_{X_n\leftarrow X_m}^{({\rm exc},2)}
		\nonumber\\
		&=
		2\,{\rm Re}
		\int_0^\infty
		dt\,
		e^{i[
			\Omega_{X_nX_m}
			-
			\overline{\Lambda}_{nm}
			]t}
		e^{-\overline g_{nm}(t)}
		\eta_{nm}(t),
		\qquad
		m\neq n.
		\label{S-eq:r_exciton}
	\end{align}
	
	Thus, $r_{mn}$ is the population-transfer rate from $\ket{X_m}$ to $\ket{X_n}$. The reverse process is obtained independently as $r_{nm}=W_{X_m\leftarrow X_n}^{({\rm exc},2)}$ and is not assumed to be equal to $r_{mn}$.
	
	The 48 experimentally resolved vibrational modes entering the adopted
	PSIIRC spectral profile~\cite{Novoderezhkin2005,stones2017performance}
	enter Eq.~\eqref{S-eq:r_exciton} through the full line-broadening
	function $g(t)$ and its time derivatives. The resulting redistribution rates therefore retain the structured vibrational environment of PSIIRC rather than replacing it by a single phenomenological relaxation constant.
	
	The methodology is used in the regime in which the population-level
	master-equation treatment is consistent with previously benchmarked
	HEOM dynamics. For PSIIRC, a reduced hybrid spectral density containing
	the prominent $342~{\rm cm}^{-1}$ and $742~{\rm cm}^{-1}$ vibrational
	modes together with a Drude contribution was found to reproduce converged
	dynamics in comparisons between HEOM and master-equation calculations
	~\cite{stones2017performance}. The present calculation retains the full
	48-mode spectral information in the correlation and line-broadening
	functions while propagating the resulting dynamics at the population
	level.
	
	A time-nonlocal treatment could instead retain the finite memory
	kernel appearing in Eq.~\eqref{S-eq:population_memory_kernel}
	~\cite{breuer2002theory,Chen2015}. Such a calculation would require
	propagation of the memory kernel generated by the full structured
	spectral density and constitutes a different numerical treatment. Its
	principal advantage would be the explicit resolution of environmental
	memory effects. The present modified-Redfield construction does not
	assume that such effects are absent; rather, it provides the
	population-level Markovian rates used here to isolate the transport and
	energetic consequences of the PSIIRC kinetic network.
	\subsection{Internal Förster Transfer Rates:
		\texorpdfstring{$W^{(2)}\rightarrow k_{xy},x,y \in X_n,I_1,I_2,I_3,\alpha_1$}{W to k}}
	\label{S-sec:Forster_rates}
	
	The Förster rates connect weakly coupled dressed configurations in the
	exciton-charge-transfer and charge-transfer manifolds, following the
	standard weak-coupling treatment of vibrationally assisted energy and
	electron transfer.~\cite{Renger2001,Nitzan2006,may2023charge} For a
	particular pair $\ket x\rightarrow\ket y$,
	
	\begin{align}
		V_{\rm F}^{(xy)}
		&=
		V_{xy}
		\ket y\bra x
		+
		V_{ba}^{*}
		\ket x\bra y.
		\label{S-eq:V_F_pair}
	\end{align}
	
	The vibrationally dressed correlation function is
	
	\begin{align}
		C_{y\leftarrow x}^{({\rm F})}(t)
		&=
		\left|
		V_{yx}
		\right|^2
		e^{-i(\lambda_x+\lambda_y)t/\hbar}
		e^{-g_x(t)-g_y(t)}.
		\label{S-eq:C_Forster}
	\end{align}
	
	The NZ rate is
	
	\begin{align}
		k_{xy}
		&\equiv
		W_{y\leftarrow x}^{({\rm F},2)}
		\nonumber\\
		&=
		\frac{2\left|V_{yx}\right|^2}{\hbar^2}
		{\rm Re}
		\int_0^\infty
		dt\,
		e^{i\Omega_{yx}t}
		e^{-i(\lambda_x+\lambda_y)t/\hbar}
		e^{-g_x(t)-g_y(t)}.
		\label{S-eq:k_Forster_time}
	\end{align}
	The  quantity may be written as the overlap of the fluorescence
	profile of the donor state and the absorption profile of the acceptor
	.~\cite{Renger2001,Nitzan2006,may2023charge} We use
	\begin{align}
		F_x(\omega)
		&=
		2\,{\rm Re}
		\int_0^\infty
		dt\,
		e^{i\omega t}
		e^{-iE_xt/\hbar}
		e^{+i\lambda_xt/\hbar}
		e^{-g_x^{*}(t)},
		\\
		A_y(\omega)
		&=
		2\,{\rm Re}
		\int_0^\infty
		dt\,
		e^{i\omega t}
		e^{-iE_yt/\hbar}
		e^{-i\lambda_yt/\hbar}
		e^{-g_y(t)}.
		\label{S-eq:FA_lineshapes}
	\end{align}
	
	The spectral overlap is
	
	\begin{align}
		S_{x\rightarrow y}
		&=
		\frac{1}{2\pi\hbar^2}
		\int_{-\infty}^{\infty}
		d\omega\,
		F_x(\omega)
		A_y(\omega),
		\label{S-eq:spectral_overlap}
	\end{align}
	and consequently
	\begin{align}
		k_{xy}
		&=
		\left|
		V_{yx}
		\right|^2
		S_{x\rightarrow y}.
		\label{S-eq:k_Forster}
	\end{align}
	The correlation function in Eq.~\ref{S-eq:C_Forster} contains two distinct contributions that must remain separate. The weak electronic coupling $V_{yx}$ determines the electronic connectivity, whereas $g_x(t)$, $g_y(t)$, $\lambda_x$, and $\lambda_b$ determine the spectral overlap generated by nuclear broadening. The transfer rate therefore cannot be inferred from the energy gap alone. The retained Förster network follows the electronic connectivity of the adopted PSIIRC parametrization~\cite{stones2017performance} and includes $X_n\rightleftarrows I_1$, $X_n\rightleftarrows I_2$, $I_2\rightleftarrows I_3$, $I_1\rightleftarrows\alpha_1$, and $I_3\rightleftarrows\alpha_1$ for $n=1,\ldots,6$. No direct $X_n\leftrightarrow I_3$ transition is retained in this parametrization.~\cite{stones2017performance}
	
	\begin{align}
		k_{X_n I_3}
		&=
		k_{I_3\rightarrow X_n}
		=
		0.
		\label{S-eq:no_XI3}
	\end{align}
	For example,
	\begin{align}
		k_{X_nI_1}
		&=
		\left|
		V_{I_1X_n}
		\right|^2
		S_{X_n\rightarrow I_1},
		\\
		k_{X_n I_2}
		&=
		\left|
		V_{I_2X_n}
		\right|^2
		S_{X_n\rightarrow I_2},
		\\
		k_{I_2 I_3}
		&=
		\left|
		V_{I_3I_2}
		\right|^2
		S_{I_2\rightarrow I_3},
		\\
		k_{I_1\alpha_1}
		&=
		\left|
		V_{\alpha_1I_1}
		\right|^2
		S_{I_1\rightarrow\alpha_1},
		\\
		k_{I_3\alpha_1}
		&=
		\left|
		V_{\alpha_1I_3}
		\right|^2
		S_{I_3\rightarrow\alpha_1}.
		\label{S-eq:k_examples}
	\end{align}
	The  full 48-mode $g(t)$ used in the excitonic calculation is used to
	construct the donor fluorescence and acceptor absorption profiles.  No
	additional coarse-grained phenomenological linewidth is introduced at
	this stage.
	
	\subsection{Sequential Electronic Exchange/Transfer/Ejection Rates:
		\texorpdfstring{$W^{(2)}\rightarrow\Gamma_{a'b'}, a'b'\in \alpha_1,\beta_1,g $}{W to \Gamma_{ab}}}
	\label{S-sec:sequential_rates}
	
	Sequential one-electron exchange with the electronic reservoirs is treated
	to second order in the system-reservoir tunnelling interaction, following
	the standard master-equation description of weak-coupling electronic
	transport.~\cite{Harbola2006,Timm2008}
	The fermionic reservoir correlations are
	\begin{align}
		\left\langle
		c_{k\ell}^{\dagger}(t)c_{k'\ell}(0)
		\right\rangle
		&=
		\delta_{kk'}
		f_\ell(\epsilon_{k\ell})
		e^{+i\epsilon_{k\ell}t/\hbar},
		\\
		\left\langle
		c_{k\ell}(t)c_{k'\ell}^{\dagger}(0)
		\right\rangle
		&=
		\delta_{kk'}
		\left[
		1-f_\ell(\epsilon_{k\ell})
		\right]
		e^{-i\epsilon_{k\ell}t/\hbar}.
		\label{S-eq:fermionic_2pt}
	\end{align}
	The first correlation requires an initially occupied reservoir state,
	whereas the second requires an empty final reservoir state.  Their Fermi
	factors therefore follow from the operator ordering and are not inserted
	independently.
	The system creation operator is expanded in the many-body basis as
	\begin{align}
		d_\chi^\dagger
		&=
		\sum_{\substack{a',b'\\N_{a'}=N_{b'+1}}}
		D_{a'b'}^{(\chi,+)}
		\ket a'\bra b',
		\\
		D_{a'b'}^{(\chi,+)}
		&=
		\bra a'
		d_\chi^\dagger
		\ket b'.
		\label{S-eq:d_many-body}
	\end{align}
	For addition of one electron from reservoir $\ell$,
	\begin{align}
		C_{a'\leftarrow b'}^{(\ell,+)}(t)
		&=
		\sum_k
		\left|
		t_{k\chi}^{(\ell)}
		\right|^2
		\left|
		D_{a'b'}^{(\chi,+)}
		\right|^2
		f_\ell(\epsilon_{k\ell})
		e^{+i\epsilon_{k\ell}t/\hbar}.
		\label{S-eq:C_seq_add}
	\end{align}
	Substitution into Eq.~\eqref{S-eq:W2_general} gives
	\begin{align}
		W_{a'\leftarrow b'}^{(\ell,+)}
		&=
		\frac{2\pi}{\hbar}
		\sum_k
		\left|
		t_{k\chi}^{(\ell)}
		\right|^2
		\left|
		D_{a'b'}^{(\chi,+)}
		\right|^2
		f_\ell(\epsilon_{k\ell})
		\nonumber\\
		&\quad\times
		\delta
		\left(
		E_{a'}-E_{b'}-\epsilon_{k\ell}
		\right).
		\label{S-eq:W_seq_add}
	\end{align}
	Define the hybridization function without using the symbol $\Gamma$,
	\begin{align}
		\Delta_\ell(\epsilon)
		&=
		2\pi
		\sum_k
		\left|
		t_{k\chi}^{(\ell)}
		\right|^2
		\delta
		\left(
		\epsilon-\epsilon_{k\ell}
		\right).
		\label{S-eq:hybridization}
	\end{align}
	Then
	\begin{align}
		\Gamma_{b'\rightarrow a'}^{(\ell,+)}
		&=
		\frac{1}{\hbar}
		\Delta_\ell(E_{a'}-E_{'b})
		\left|
		D_{a'b'}^{(\chi,+)}
		\right|^2
		f_\ell(E_{a'}-E_{b'}).
		\label{S-eq:Gamma_add}
	\end{align}
	For electron removal from the system,
	
	\begin{align}
		\Gamma_{b'\rightarrow a'}^{(\ell,-)}
		&=
		\frac{1}{\hbar}
		\Delta_\ell(E_{b'}-E_{a'})
		\left|
		D_{a'b'}^{(\chi,-)}
		\right|^2
		\left[
		1-f_\ell(E_{b'}-E_{a'})
		\right].
		\label{S-eq:Gamma_remove}
	\end{align}
	
	The terminal sequential pathway retained in the PSIIRC junction is
	~\cite{stones2017performance,Sharma2025}
	$
	\alpha_1
	\longrightarrow
	\beta_1
	\longrightarrow
	g,
	\label{S-eq:terminal_path}
	$
	with rates
	$
	\Gamma_{\alpha_1\beta_1}
	=
	W_{\beta_1\leftarrow\alpha_1}^{({\rm seq},2)},
	\\
	\Gamma_{\beta_1 g}
	=
	W_{g\leftarrow\beta_1}^{({\rm seq},2)}.
	$
	The energy separations between $\alpha_1$ and $\beta_1$, and between
	$\beta_1$ and $g$, are sufficiently large in the adopted operating regime
	that the reverse electronic transfer is neglected,
	$
	\Gamma_{\beta_1\rightarrow\alpha_1}
	\simeq
	0,
	\Gamma_{g\rightarrow\beta_1}
	\simeq
	0.
	$
	This corresponds to the unidirectional, or effectively infinite-bias,
	transport limit commonly employed in weak-coupling junction master
	equations~\cite{Harbola2006,Timm2008} and adopted here for the terminal
	PSIIRC transport channel.~\cite{stones2017performance,Sharma2025}
	
	\subsection{Cotunneling Rates:
		\texorpdfstring{$W_{}^{({\rm cot},4)}
			\rightarrow\omega_{x'y',~x',y'\in~ I_1,I_3,\alpha_1,\alpha_2,g}$}{W to omega}}
	\label{S-sec:cot_rates}
	
	In the Nakajima-Zwanzig construction, the irreducible fourth-order contribution
	$\mathcal K_{\rm cot}^{(4),{\rm irr}}$ is generated by four applications of the tunneling interaction $V_{\rm tun}$. We now project this kernel onto the individual PSIIRC population transitions retained in the kinetic model. For an initial retained state $\ket{x'}$ and a final retained state $\ket{y'}$, the corresponding population-memory kernel is defined by
	
	\begin{align}
		K_{y'\leftarrow x'}^{({\rm cot},4)}(t)
		&=
		\bra{y'}
		{\rm Tr}_{L,R}
		\left[
		\mathcal K_{\rm cot}^{(4),{\rm irr}}(t)
		\left(
		\ket{x'}\bra{x'}
		\otimes
		\rho_L\rho_R
		\right)
		\right]
		\ket{y'}.
		\label{S-eq:K4_population}
	\end{align}
	In the  Markov regime used above for the second-order rates, the fourth-order population rate is obtained by integrating the corresponding irreducible memory kernel,
	\begin{align}
		W_{y'\leftarrow x'}^{({\rm cot},4)}
		&=
		\int_0^\infty
		dt\,
		K_{y'\leftarrow x'}^{({\rm cot},4)}(t).
		\label{S-eq:W4_from_K4}
	\end{align}
	
	The rates appearing in the PSIIRC kinetic equations are identified with these fourth-order population rates according to
	\begin{align}
		\omega_{x'y'}
		&\equiv
		W_{y'\leftarrow x'}^{({\rm cot},4)} ,
		\label{S-eq:omega_definition}
	\end{align}
	so that the ordering of the labels in $\omega_{x'y'}$ follows the convention used throughout the kinetic model, where the first label denotes the initial state and the second denotes the final state. In the present network,
	$
	x',y'
	\in
	\left\{
	I_1,I_3,\alpha_1,\alpha_2,g
	\right\}.
	$
	The state $I_2$ is not connected directly to $\alpha_2$ by a cotunneling jump in the retained kinetic scheme and therefore does not enter $\breve{\mathcal L}_{\omega}^{(4)}$.
	The four-point reservoir correlation function appearing in
	$\mathcal K_{\rm cot}^{(4),{\rm irr}}$ is reduced by Wick's theorem to products of two-point fermionic contractions. After subtraction of the reducible contribution describing two independent sequential tunneling events, the remaining terms contain two tunneling vertices on the forward amplitude and two on its complex-conjugate branch. The fourth-order rate can therefore be written in terms of an effective second-order virtual amplitude,
	\begin{align}
		W_{y'\leftarrow x'}^{({\rm cot},4)}
		&=
		\frac{2\pi\nu^2}{\hbar}
		\int
		d\varepsilon\,
		\mathcal F_{x'y'}(\varepsilon,U)
		\left|
		\mathcal A_{y'\leftarrow x'}^{(2)}
		(\varepsilon,U)
		\right|^2.
		\label{S-eq:W4_general}
	\end{align}
	Here, $\nu^2$ denotes the product of the electronic densities of states in the reduced wide-band parametrization, $\mathcal F_{x'y'}$ contains the fermionic occupation factors obtained from the Wick contractions, and $\mathcal A_{y'\leftarrow x'}^{(2)}$ contains the two one-electron tunneling vertices and the propagation through the virtual $(N+1)$ states. Equation~\eqref{S-eq:W4_general} is therefore the energy-domain representation of the population-projected fourth-order Nakajima-Zwanzig kernel rather than an independent phenomenological rate expression.
	
	The virtual $(N+1)$ manifold entering this amplitude is represented by the two quinone-dominated states $\ket{\alpha_1'}$ and $\ket{\alpha_2'}$ constructed in Sec.~\ref{S-sec:system}. We define the corresponding regularized propagators as
	\begin{align}
		\mathcal D_j(\varepsilon,U)
		&=
		\frac{1}{
			\varepsilon-E_{\alpha_j}^{\prime}+i\eta
		}
		-
		\frac{1}{
			\varepsilon-E_{\alpha_j}^{\prime}-U+i\eta
		},
		\qquad
		j=1,2.
		\label{S-eq:D_cot}
	\end{align}
	The first denominator describes propagation through the unshifted virtual one-electron state, whereas the second contains the interaction shift associated with the correlated two-electron sector. The regularizer $\eta>0$ prevents the numerical integration from sampling an infinitely sharp virtual pole. Unless a microscopic lifetime broadening is introduced independently, $\eta$ is interpreted as a regularization or phenomenological broadening parameter rather than as a measured population-decay rate.\cite{carmi01}
	
	For a general transition $\ket{x'}\rightarrow\ket{y'}$, the second-order virtual amplitude is
	\begin{align}
		\mathcal A_{y'\leftarrow x'}^{(2)}
		(\varepsilon,U)
		&=
		t_{y'\alpha_1'}
		t_{\alpha_1'x'}
		\mathcal D_1(\varepsilon,U)
		-
		t_{y'\alpha_2'}
		t_{\alpha_2'x'}
		\mathcal D_2(\varepsilon,U).
		\label{S-eq:A_general}
	\end{align}
	The individual factors in Eq.~\eqref{S-eq:A_general} originate directly from the two applications of $V_{\rm tun}$ on a given amplitude branch. Thus, $t_{\alpha_j'x'}$ describes the one-electron tunneling vertex connecting the retained initial state $\ket{x'}$ to the virtual state $\ket{\alpha_j'}$, while $t_{y'\alpha_j'}$ describes the second tunneling vertex connecting the virtual state to the retained final state $\ket{y'}$. Each tunneling amplitude inherits the appropriate source or drain coupling from $V_L$ or $V_R$ introduced above, while the reservoir label is suppressed here only to avoid overloading the notation. If the tunneling Hamiltonian is Hermitian, the reverse matrix elements satisfy the corresponding complex-conjugation relations.
	The relative sign between the two virtual branches follows the regularized cotunneling kernel used in the present PSIIRC parametrization.~\cite{Sharma2025} If the products of tunneling amplitudes associated with the two virtual channels are approximated by the same effective value,
	\begin{align}
		t_{y'\alpha_1'}
		t_{\alpha_1'x'}
		&\simeq
		t_{y'\alpha_2'}
		t_{\alpha_2'x'}
		\equiv
		\tau_{x'y'},
		\label{S-eq:tau_factorization}
	\end{align}
	Eq.~\eqref{S-eq:A_general} reduces to
	\begin{align}
		\mathcal A_{y'\leftarrow x'}^{(2)}
		(\varepsilon,U)
		&\simeq
		\tau_{x'y'}
		\mathcal B(\varepsilon,U),
		\label{S-eq:A_factorized}
	\end{align}
	where
	\begin{align}
		\mathcal B(\varepsilon,U)
		&=
		\frac{1}{
			\varepsilon-E_{\alpha_1}^{\prime}+i\eta
		}
		-
		\frac{1}{
			\varepsilon-E_{\alpha_1}^{\prime}-U+i\eta
		}
		-
		\frac{1}{
			\varepsilon-E_{\alpha_2}^{\prime}+i\eta
		}
		+
		\frac{1}{
			\varepsilon-E_{\alpha_2}^{\prime}-U+i\eta
		}.
		\label{S-eq:B_cot}
	\end{align}
	The compact quantity previously written schematically as
	$t_{x'y'}^{\,2}\mathcal B$ represents an effective product of two one-electron tunneling amplitudes. It should not be interpreted as the square of a direct tunneling matrix element between the retained initial and final states. Equation~\eqref{S-eq:A_general} displays the underlying state-resolved amplitudes explicitly.
	
	For the numerical implementation, the tunneling amplitudes entering the different cotunneling pathways were sampled over channel-dependent weak-coupling intervals. For the processes connecting $I_1$ and $I_3$ with the correlated $\alpha_2$ state, the corresponding lead-coupling amplitude was varied over $t_{\rm lead}=20$--$70~{\rm cm}^{-1}$, while the amplitude associated with the $\alpha_1\leftrightarrow\alpha_2$ pathway was varied over $t_{\alpha_1\alpha_2}=5$--$50~{\rm cm}^{-1}$. For the direct $\alpha_2\rightarrow g$ cotunneling channel, the amplitude was sampled over $t_{\alpha_2 g}=10$--$50~{\rm cm}^{-1}$ using 70 uniformly spaced values. These amplitudes are effective parameters of the reduced cotunneling description and are not identified with directly measured microscopic matrix elements. The channel-dependent intervals were chosen to probe the response of the different cotunneling pathways while retaining their perturbative character. Since each virtual amplitude contains two tunneling vertices and the corresponding population rate is fourth order in the tunneling interaction, $\omega_{x'y'}=\mathcal O(|t|^4)$, even a moderate variation of the tunneling amplitude generates a broad range of cotunneling rates. This parametrization therefore provides controlled access to the cotunneling-induced redistribution of the steady-state populations and the associated changes in passive-state ordering without imposing a single universal tunneling amplitude on all correlated pathways.
	The second electronic energy appearing after enforcement of energy conservation in the two-electron reservoir contraction is written in the numerical implementation as
	$
	E_f(\varepsilon,U)
	=
	-\varepsilon
	+
	E_{\alpha_1}^{\prime}
	+
	E_{\alpha_2}^{\prime}
	+
	U.
	$
	The formal reservoir derivation is expressed in terms of the source and drain chemical potentials. In the numerical reduced-state implementation, the reference entering the Fermi functions is assigned from the corresponding retained-state energy,
	$
	\mu_{x'}^{\rm eff}
	\equiv
	E_{x'},
	f_{x'}(\varepsilon)
	\equiv
	f
	\left(
	\varepsilon-E_{x'}
	\right).
	$
	This is a model-specific energy-reference prescription and does not imply that a macroscopic reservoir chemical potential is physically identical to a molecular many-body eigenenergy.
	For two occupied electronic lines referenced by retained states $x'$ and $y'$, the Wick-contracted occupation factor is written as
	\begin{align}
		\mathcal F_{x'y'}^{({\rm occ})}
		(\varepsilon,U)
		&=
		\frac{1}{2}
		\left[
		f
		\left(
		\varepsilon-E_{x'}
		\right)
		f
		\left(
		E_f(\varepsilon,U)-E_{y'}
		\right)
		\right.
		\left.
		+
		f
		\left(
		\varepsilon-E_{y'}
		\right)
		f
		\left(
		E_f(\varepsilon,U)-E_{x'}
		\right)
		\right].
		\label{S-eq:F_occ}
	\end{align}
	For two vacant final electronic lines,
	\begin{align}
		\mathcal F_{x'y'}^{({\rm vac})}
		(\varepsilon,U)
		&=
		\frac{1}{2}
		\left\{
		\left[
		1-
		f
		\left(
		\varepsilon-E_{x'}
		\right)
		\right]
		\left[
		1-
		f
		\left(
		E_f(\varepsilon,U)-E_{y'}
		\right)
		\right]
		\right.\left.
		+
		\left[
		1-
		f
		\left(
		\varepsilon-E_{y'}
		\right)
		\right]
		\left[
		1-
		f
		\left(
		E_f(\varepsilon,U)-E_{x'}
		\right)
		\right]
		\right\}.
		\label{S-eq:F_vac}
	\end{align}
	The two terms correspond to the two possible assignments of the electronic energies to the contracted fermionic lines. The identity
	$
	1-f(x)=f(-x)
	$
	may equivalently be used to rewrite the vacancy factors without changing the occupation requirement.
	The first class of cotunneling transitions couples the internal charge-transfer states $I_1$ and $I_3$ to the correlated state $\alpha_2$. For $z=1,3$, the virtual amplitude for the transition $I_z\rightarrow\alpha_2$ is
	\begin{align}
		\mathcal A_{\alpha_2\leftarrow I_z}^{(2)}
		(\varepsilon,U)
		&=
		t_{\alpha_2\alpha_1'}
		t_{\alpha_1'I_z}
		\mathcal D_1(\varepsilon,U)
		-
		t_{\alpha_2\alpha_2'}
		t_{\alpha_2'I_z}
		\mathcal D_2(\varepsilon,U),
		\qquad
		z=1,3.
		\label{S-eq:A_I_a2}
	\end{align}
	The corresponding fourth-order population rate is
	\begin{align}
		\omega_{I_z\alpha_2}
		&\equiv
		W_{\alpha_2\leftarrow I_z}^{({\rm cot},4)}=
		\frac{2\pi\nu^2}{\hbar}
		\int
		d\varepsilon\,
		\mathcal F_{I_1I_3}^{({\rm occ})}
		(\varepsilon,U)
		\left|
		\mathcal A_{\alpha_2\leftarrow I_z}^{(2)}
		(\varepsilon,U)
		\right|^2,
		\qquad
		z=1,3.
		\label{S-eq:omega_I_a2}
	\end{align}
	
	The appearance of $\mathcal F_{I_1I_3}^{({\rm occ})}$ reflects the two electronic lines used in the numerical cotunneling contraction, while it should not be confused with the initial and final labels of the individual population jump.
	For the reverse transition, the virtual amplitude is
	
	\begin{align}
		\mathcal A_{I_z\leftarrow\alpha_2}^{(2)}
		(\varepsilon,U)
		&=
		t_{I_z\alpha_1'}
		t_{\alpha_1'\alpha_2}
		\mathcal D_1(\varepsilon,U)
		-
		t_{I_z\alpha_2'}
		t_{\alpha_2'\alpha_2}
		\mathcal D_2(\varepsilon,U),
		\qquad
		z=1,3,
		\label{S-eq:A_a2_I}
	\end{align}
	and
	\begin{align}
		\omega_{\alpha_2 I_z}
		&\equiv
		W_{I_z\leftarrow\alpha_2}^{({\rm cot},4)}=
		\frac{2\pi\nu^2}{\hbar}
		\int
		d\varepsilon\,
		\mathcal F_{I_1I_3}^{({\rm vac})}
		(\varepsilon,U)
		\left|
		\mathcal A_{I_z\leftarrow\alpha_2}^{(2)}
		(\varepsilon,U)
		\right|^2,
		\qquad
		z=1,3.
		\label{S-eq:omega_a2_I}
	\end{align}
	
	The forward and backward amplitudes are related through the same tunneling Hamiltonian, but the corresponding rates differ because the occupied and vacant reservoir factors are different.
	For the $\alpha_1\leftrightarrow\alpha_2$ cotunneling pathway, the virtual amplitudes are
	\begin{align}
		\mathcal A_{\alpha_2\leftarrow\alpha_1}^{(2)}
		(\varepsilon,U)
		&=
		t_{\alpha_2\alpha_1'}
		t_{\alpha_1'\alpha_1}
		\mathcal D_1(\varepsilon,U)
		-
		t_{\alpha_2\alpha_2'}
		t_{\alpha_2'\alpha_1}
		\mathcal D_2(\varepsilon,U),
		\label{S-eq:A_a1_a2}
		\\
		\mathcal A_{\alpha_1\leftarrow\alpha_2}^{(2)}
		(\varepsilon,U)
		&=
		t_{\alpha_1\alpha_1'}
		t_{\alpha_1'\alpha_2}
		\mathcal D_1(\varepsilon,U)
		-
		t_{\alpha_1\alpha_2'}
		t_{\alpha_2'\alpha_2}
		\mathcal D_2(\varepsilon,U).
		\label{S-eq:A_a2_a1}
	\end{align}
	The associated fourth-order population rates are
	\begin{align}
		\omega_{\alpha_1\alpha_2}
		&\equiv
		W_{\alpha_2\leftarrow\alpha_1}^{({\rm cot},4)}
		=
		\frac{2\pi\nu^2}{\hbar}
		\int
		d\varepsilon\,
		\mathcal F_{\alpha_1\alpha_2}^{({\rm occ})}
		(\varepsilon,U)
		\left|
		\mathcal A_{\alpha_2\leftarrow\alpha_1}^{(2)}
		(\varepsilon,U)
		\right|^2,
		\label{S-eq:omega_a1_a2}
	\end{align}
	and
	\begin{align}
		\omega_{\alpha_2\alpha_1}
		&\equiv
		W_{\alpha_1\leftarrow\alpha_2}^{({\rm cot},4)}
		=
		\frac{2\pi\nu^2}{\hbar}
		\int
		d\varepsilon\,
		\mathcal F_{\alpha_1\alpha_2}^{({\rm vac})}
		(\varepsilon,U)
		\left|
		\mathcal A_{\alpha_1\leftarrow\alpha_2}^{(2)}
		(\varepsilon,U)
		\right|^2.
		\label{S-eq:omega_a2_a1}
	\end{align}
	
	The direct two-electron extraction channel $\alpha_2\rightarrow g$ is obtained from the same fourth-order construction. Its virtual amplitude is
	\begin{align}
		\mathcal A_{g\leftarrow\alpha_2}^{(2)}
		(\varepsilon,U)
		&=
		t_{g\alpha_1'}
		t_{\alpha_1'\alpha_2}
		\mathcal D_1(\varepsilon,U)
		-
		t_{g\alpha_2'}
		t_{\alpha_2'\alpha_2}
		\mathcal D_2(\varepsilon,U).
		\label{S-eq:A_a2_g}
	\end{align}
	For this channel the ground-state energy is used as the numerical Fermi reference,
	$
	\mu_g^{\rm eff}
	=
	E_g
	=
	0.
	$
	The corresponding vacancy factor reduces to
	\begin{align}
		\mathcal F_{gg}^{({\rm vac})}
		(\varepsilon,U)
		&=
		\left[
		1-
		f
		\left(
		\varepsilon-E_g
		\right)
		\right]
		\left[
		1-
		f
		\left(
		E_f(\varepsilon,U)-E_g
		\right)
		\right],
		\label{S-eq:F_gg_vac}
	\end{align}
	and the population rate is
	\begin{align}
		\omega_{\alpha_2 g}
		&\equiv
		W_{g\leftarrow\alpha_2}^{({\rm cot},4)}
		=
		\frac{2\pi\nu^2}{\hbar}
		\int
		d\varepsilon\,
		\mathcal F_{gg}^{({\rm vac})}
		(\varepsilon,U)
		\left|
		\mathcal A_{g\leftarrow\alpha_2}^{(2)}
		(\varepsilon,U)
		\right|^2.
		\label{S-eq:omega_a2_g}
	\end{align}
	The perturbative order of these rates now follows directly from the Nakajima-Zwanzig expansion. Each virtual amplitude contains two applications of the tunneling interaction,
	\begin{align}
		\mathcal A_{y'\leftarrow x'}^{(2)}
		&=
		\mathcal O
		\left(
		V_{\rm tun}^{2}
		\right)
		=
		\mathcal O
		\left(
		t^2
		\right),
	\end{align}
	whereas the population kernel contains the product of this amplitude with its complex conjugate,
	\begin{align}
		\omega_{x'y'}
		&=
		W_{y'\leftarrow x'}^{({\rm cot},4)}
		=
		\mathcal O
		\left(
		V_{\rm tun}^{4}
		\right)
		=
		\mathcal O
		\left(
		|t|^4
		\right).
		\label{S-eq:omega_order}
	\end{align}
	The perturbative cotunneling treatment is controlled away from a true virtual resonance. For the  virtual states, a sufficient condition is
	\begin{align}
		\left|
		\varepsilon-E_{\alpha_j}'
		\right|
		&\gg
		\max
		\left(
		\Delta_{\rm tun},
		k_{\rm B}T,
		\eta
		\right),
		\\
		\left|
		\varepsilon-E_{\alpha_j}'-U
		\right|
		&\gg
		\max
		\left(
		\Delta_{\rm tun},
		k_{\rm B}T,
		\eta
		\right),
		\qquad
		j=1,2,
		\label{S-eq:cot_validity}
	\end{align}
	where $\Delta_{\rm tun}$ denotes the characteristic tunneling-induced broadening. Near a true virtual resonance, the perturbative separation between sequential and irreducible cotunneling contributions is no longer controlled and a resummed treatment would be required.~\cite{Averin1990,Konig1996,Timm2008} The virtual states $\ket{\alpha_1'}$ and $\ket{\alpha_2'}$ therefore enter only as propagators inside $\mathcal A^{(2)}$ and are not additional stationary populations of the thirteen-state PSIIRC kinetic model.
	
	\subsection{Triplet-mediated Recombination}
	\label{S-sec:triplet_recombination}
	
	The 13-state photosystem II reaction center (PSIIRC) model does not contain explicit triplet populations. Nevertheless, triplet configurations associated with the D1 branch may provide an indirect recombination pathway from the radical-pair state $\ket{\alpha_1}$ to the neutral ground state $\ket{g}$. To estimate the magnitude of this omitted contribution without enlarging the kinetic state space, the triplet manifold is treated as an intermediate subspace and eliminated perturbatively through a Schrieffer-Wolff transformation.~\cite{Schrieffer1966,Bravyi2011,Bhattacharjee2023}
	
	We denote the projector onto the retained 13-state PSIIRC space by $\mathcal P_S=\sum_{a\in\mathcal B_S}\ket{a}\bra{a}$, where $\mathcal B_S$ is the retained system basis, and the projector onto the intermediate triplet sector by $\mathcal Q_T=\sum_\nu\ket{T_\nu}\bra{T_\nu}$. The states $\ket{T_\nu}$ are therefore not additional dynamical populations. The Hamiltonian relevant to this reduction is written as $H_{\rm SW}=H_S+H_T+V_{ST}$, where $H_S$ is the retained PSIIRC Hamiltonian, $H_T=\sum_\nu E_{T_\nu}\ket{T_\nu}\bra{T_\nu}$ describes the triplet sector, $E_{T_\nu}$ is the energy of triplet configuration $\ket{T_\nu}$, and $V_{ST}$ couples the retained and triplet subspaces.
	Eliminating the triplet sector generates an effective coupling between
	$\ket{\alpha_1}$ and $\ket{g}$ through a Schrieffer-Wolff
	transformation.~\cite{Schrieffer1966,Bravyi2011} For decay from $\ket{\alpha_1}$, the on-shell Schrieffer-Wolff coupling is $V_{g\alpha_1}^{(\mathrm{SW})}=\sum_\nu V_{gT_\nu}V_{T_\nu\alpha_1}/(E_{\alpha_1}-E_{T_\nu})$, where $V_{T_\nu\alpha_1}=\bra{T_\nu}V_{ST}\ket{\alpha_1}$ and $V_{gT_\nu}=\bra{g}V_{ST}\ket{T_\nu}$. When the individual triplet contributions are considered independently, the contribution from a single configuration is $V_{g\alpha_1}^{(\mathrm{SW},\nu)}=V_{gT_\nu}V_{T_\nu\alpha_1}/(E_{\alpha_1}-E_{T_\nu})$.
	
	This perturbative reduction is controlled when the couplings
	$\lvert V_{T_\nu\alpha_1}\rvert$ and $\lvert V_{gT_\nu}\rvert$
	remain much smaller than the energetic separation
	$\lvert E_{\alpha_1}-E_{T_\nu}\rvert$.
	~\cite{Schrieffer1966,Bravyi2011} For the triplet configurations considered here, this condition is satisfied at the electronic-Hamiltonian level. The closest charge-transfer triplet lies near $1.708~\mathrm{eV}$ and is separated from $\ket{\alpha_1}$ by approximately $196~\mathrm{cm}^{-1}$, whereas the corresponding coupling is estimated to be only $0.5$-$3~\mathrm{cm}^{-1}$. This condition establishes the validity of the perturbative elimination but should not be interpreted as proof that triplet populations can never arise in the complete microscopic dynamics. The nearby triplet energy lies within a scale comparable to thermal and low-frequency vibrational energies, so nuclear broadening and environmental coupling may still influence triplet formation. The present calculation is used only to estimate the effective $\ket{\alpha_1}\rightarrow\ket{g}$ correction associated with the triplet sector omitted from the working Liouvillian.
	
	The effective coupling gives the recombination correction
	$\kappa_{\alpha_1\rightarrow g}^{(\mathrm{SW})}
	=(2\pi/\hbar)\lvert V_{g\alpha_1}^{(\mathrm{SW})}\rvert^2
	\mathrm{FCWD}_{\alpha_1g}$ through Fermi's Golden Rule.
	~\cite{Nitzan2006} Here, $\hbar$ is the reduced Planck constant and $\mathrm{FCWD}_{\alpha_1g}$ is the Franck-Condon-weighted density of final nuclear states associated with the $\alpha_1\rightarrow g$ transition. In the classical Marcus limit~\cite{Marcus1956,Nitzan2006}, the nuclear response is described by an effective coordinate $Q$ with reorganization energy $\lambda_{\alpha_1g}=KQ_0^2/2$, where $K$ is the effective curvature and $Q_0$ is the displacement between the reactant and product free-energy surfaces. If $\Delta G_{\alpha_1g}$ denotes the corresponding free-energy change, the activation free energy becomes $\Delta G_{\alpha_1g}^{\ddagger}=(\Delta G_{\alpha_1g}+\lambda_{\alpha_1g})^2/(4\lambda_{\alpha_1g})$. The resulting Franck-Condon-weighted density is $\mathrm{FCWD}_{\alpha_1g}=(4\pi\lambda_{\alpha_1g}k_{\rm B}T)^{-1/2}\exp{- (\Delta G_{\alpha_1g}+\lambda_{\alpha_1g})^2/(4\lambda_{\alpha_1g}k_{\rm B}T)}$, where $k_{\rm B}$ is the Boltzmann constant and $T$ is the temperature.
	Three representative triplet configurations illustrate the relevant
	energetic and coupling scales. A localized chlorophyll triplet near
	$1.22~\mathrm{eV}$~\cite{Bhattacharjee2023} is assigned the weak-coupling
	range $\lvert V_{T\alpha_1}\rvert\simeq0.005$-$0.05~\mathrm{cm}^{-1}$.
	A charge-transfer triplet
	${}^{3}\mathrm{Chl}_{D1}^{+}\mathrm{Pheo}_{D1}^{-}$ near
	$1.708~\mathrm{eV}$ is assigned the stronger coupling range
	$\lvert V_{T\alpha_1}\rvert\simeq0.5$-$3~\mathrm{cm}^{-1}$, whereas the
	related ${}^{3}P_{D1}^{+}\mathrm{Pheo}_{D1}^{-}$ configuration is assigned
	the intermediate range $0.1$-$1~\mathrm{cm}^{-1}$. The triplet-to-ground
	coupling is likewise taken to remain weak for all three configurations,
	with $\lvert V_{gT}\rvert\simeq0.05$-$0.5~\mathrm{cm}^{-1}$.
	
	For the parameter range
	$\lambda_{\alpha_1g}=0.1$-$0.3~\mathrm{eV}$ and
	$\Delta G_{\alpha_1g}\simeq-1.73~\mathrm{eV}$, the condition
	$|\Delta G_{\alpha_1g}|>\lambda_{\alpha_1g}$ places the reaction in the Marcus inverted regime.~\cite{Marcus1956,Nitzan2006} The nuclear contribution is consequently strongly suppressed, making $\kappa_{\alpha_1\rightarrow g}^{(\mathrm{SW})}$ many orders of magnitude smaller than the charge-transfer and extraction rates retained in the PSIIRC kinetics. The purpose of this calculation is therefore not to introduce an additional triplet transport channel, but to verify that the effective recombination correction generated by the omitted triplet sector is negligible under the conditions considered. Accordingly, no triplet population and no corresponding correction are included in the final 13-state Liouvillian used for the numerical calculations.
	
	\label{S-sec:SW_triplet}
	
	\section{Population-only Liouvillian and range of validity}
	\label{S-sec:Liouvillian}
	
	The population vector is ordered as
	\begin{align}
		\bra{\hat\rho_q}
		&=
		\left\{
		\rho_g,
		\rho_{X_1},
		\rho_{X_2},
		\rho_{X_3},
		\rho_{X_4},
		\rho_{X_5},
		\rho_{X_6},
		\rho_{I_1},
		\rho_{I_2},
		\rho_{I_3},
		\rho_{\alpha_1},
		\rho_{\alpha_2},
		\rho_{\beta_1}
		\right\}.
		\label{S-eq:rho_vector}
	\end{align}
	
	The kinetic equation for the population vector is
	
	\begin{align}
		\ket{\dot{\hat\rho}_q}
		&=
		\breve{\mathcal L}
		\ket{\hat\rho_q},
		\label{S-eq:Liouville_equation}
	\end{align}
	with the total population Liouvillian decomposed according to the physical origin and perturbative order of the corresponding transitions,
	
	\begin{align}
		\breve{\mathcal L}
		&=
		\breve{\mathcal L}_{\gamma}^{(2)}
		+
		\breve{\mathcal L}_{r}^{(2)}
		+
		\breve{\mathcal L}_{k}^{(2)}
		+
		\breve{\mathcal L}_{\Gamma}^{(2)}
		+
		\breve{\mathcal L}_{\omega}^{(4)}.
		\label{S-eq:L_decomposition}
	\end{align}
	Here, $\breve{\mathcal L}_{\gamma}^{(2)}$ contains optical excitation and relaxation between $\ket{g}$ and the six excitonic states, $\breve{\mathcal L}_{r}^{(2)}$ contains vibrational redistribution within the excitonic manifold, $\breve{\mathcal L}_{k}^{(2)}$ contains the internal Förster-type transfer processes connecting the excitonic and charge-transfer states, $\breve{\mathcal L}_{\Gamma}^{(2)}$ contains sequential electron exchange with the electronic source and drain, and $\breve{\mathcal L}_{\omega}^{(4)}$ contains the cotunneling processes involving the correlated $\ket{\alpha_2}$ sector. The superscripts indicate the perturbative order in the corresponding microscopic coupling.
	
	Each allowed directed transition $b\rightarrow a$ can be represented by an elementary population-space jump Liouvillian $\breve{\mathcal J}_{a\leftarrow b}$. If $W_{a\leftarrow b}$ is the transition rate from state $b$ to state $a$, its matrix elements are
	
	\begin{align}
		\left(
		\breve{\mathcal J}_{a\leftarrow b}
		\right)_{ij}
		&=
		W_{a\leftarrow b}
		\left(
		\delta_{ia}
		-
		\delta_{ib}
		\right)
		\delta_{jb}.
		\label{S-eq:jump_liouvillian}
	\end{align}
	
	Thus, a single directed jump contributes $+W_{a\leftarrow b}\rho_b$ to the population of the destination state $a$ and $-W_{a\leftarrow b}\rho_b$ to the population of the initial state $b$. Each elementary jump therefore conserves the total probability independently. The population Liouvillian is the sum of all such directed jumps. Equivalently, for $a\neq b$,
	
	\begin{align}
		\left(
		\breve{\mathcal L}
		\right)_{ab}
		&=
		W_{a\leftarrow b},
	\end{align}
	whereas the diagonal elements collect all processes leaving state $a$,
	\begin{align}
		\left(
		\breve{\mathcal L}
		\right)_{aa}
		&=
		-
		\sum_{c\neq a}
		W_{c\leftarrow a}.
		\label{S-eq:L_elements}
	\end{align}
	In the rate notation used throughout this work, the first state label denotes the initial state and the second denotes the final state. For example, $\gamma_{gX_n}$ denotes the optical transition $g\rightarrow X_n$, $\Gamma_{\alpha_1\beta_1}$ denotes the sequential transition $\alpha_1\rightarrow\beta_1$, and $\omega_{\alpha_2g}$ denotes the cotunneling transition $\alpha_2\rightarrow g$. The corresponding elementary jump Liouvillians therefore enter as $\breve{\mathcal J}_{X_n\leftarrow g}$, $\breve{\mathcal J}_{\beta_1\leftarrow\alpha_1}$, and $\breve{\mathcal J}_{g\leftarrow\alpha_2}$, respectively.
	The optical contribution is consequently
	\begin{align}
		\breve{\mathcal L}_{\gamma}^{(2)}
		&=
		\sum_{n=1}^{6}
		\left[
		\breve{\mathcal J}^{(\gamma)}_{X_n\leftarrow g}
		+
		\breve{\mathcal J}^{(\gamma)}_{g\leftarrow X_n}
		\right],
		\label{S-eq:Lgamma_jumps}
	\end{align}
	with rates $\gamma_{gX_n}$ and $\gamma_{X_ng}$, respectively. The vibrational redistribution contribution is
	\begin{align}
		\breve{\mathcal L}_{r}^{(2)}
		&=
		\sum_{\substack{m,n=1\\m\neq n}}^{6}
		\breve{\mathcal J}^{(r)}_{X_n\leftarrow X_m},
		\label{S-eq:Lr_jumps}
	\end{align}
	where the jump $X_m\rightarrow X_n$ carries rate $r_{X_mX_n}$.
	The internal electron-transfer contribution contains the directed jumps along the Förster network,
	\begin{align}
		\breve{\mathcal L}_{k}^{(2)}
		&=
		\sum_{n=1}^{6}
		\left[
		\breve{\mathcal J}^{(k)}_{I_1\leftarrow X_n}
		+
		\breve{\mathcal J}^{(k)}_{X_n\leftarrow I_1}
		+
		\breve{\mathcal J}^{(k)}_{I_2\leftarrow X_n}
		+
		\breve{\mathcal J}^{(k)}_{X_n\leftarrow I_2}
		\right]
		+
		\breve{\mathcal J}^{(k)}_{I_3\leftarrow I_2}
		+
		\breve{\mathcal J}^{(k)}_{I_2\leftarrow I_3}
		+
		\breve{\mathcal J}^{(k)}_{\alpha_1\leftarrow I_1}
		+
		\breve{\mathcal J}^{(k)}_{I_1\leftarrow\alpha_1}
		+
		\breve{\mathcal J}^{(k)}_{\alpha_1\leftarrow I_3}
		+
		\breve{\mathcal J}^{(k)}_{I_3\leftarrow\alpha_1}.
		\label{S-eq:Lk_jumps}
	\end{align}
	The corresponding rates are $k_{X_nI_1}$, $k_{I_1X_n}$, $k_{X_nI_2}$, $k_{I_2X_n}$, $k_{I_2I_3}$, $k_{I_3I_2}$, $k_{I_1\alpha_1}$, $k_{\alpha_1I_1}$, $k_{I_3\alpha_1}$, and $k_{\alpha_1I_3}$.
	The sequential electronic contribution consists of the two directed extraction and reset processes,
	\begin{align}
		\breve{\mathcal L}_{\Gamma}^{(2)}
		&=
		\breve{\mathcal J}^{(\Gamma)}_{\beta_1\leftarrow\alpha_1}
		+
		\breve{\mathcal J}^{(\Gamma)}_{g\leftarrow\beta_1},
		\label{S-eq:Lgamma_seq_jumps}
	\end{align}
	with rates $\Gamma_{\alpha_1\beta_1}$ and $\Gamma_{\beta_1g}$.
	The fourth-order cotunneling contribution is
	\begin{align}
		\breve{\mathcal L}_{\omega}^{(4)}
		&=
		\breve{\mathcal J}^{(\omega)}_{\alpha_2\leftarrow I_1}
		+
		\breve{\mathcal J}^{(\omega)}_{I_1\leftarrow\alpha_2}
		+
		\breve{\mathcal J}^{(\omega)}_{\alpha_2\leftarrow I_3}
		+
		\breve{\mathcal J}^{(\omega)}_{I_3\leftarrow\alpha_2}
		+
		\breve{\mathcal J}^{(\omega)}_{\alpha_2\leftarrow\alpha_1}
		+
		\breve{\mathcal J}^{(\omega)}_{\alpha_1\leftarrow\alpha_2}
		+
		\breve{\mathcal J}^{(\omega)}_{g\leftarrow\alpha_2},
		\label{S-eq:Lomega_jumps}
	\end{align}
	with the respective rates $\omega_{I_1\alpha_2}$, $\omega_{\alpha_2I_1}$, $\omega_{I_3\alpha_2}$, $\omega_{\alpha_2I_3}$, $\omega_{\alpha_1\alpha_2}$, $\omega_{\alpha_2\alpha_1}$, and $\omega_{\alpha_2g}$. The first six terms describe exchange between the correlated $\alpha_2$ sector and the internal PSIIRC charge-transfer states, whereas the final jump $\breve{\mathcal J}^{(\omega)}_{g\leftarrow\alpha_2}$ represents the direct cotunneling exit from $\alpha_2$ to the ground-state sector. 
	The explicit jump decomposition makes clear that $\breve{\mathcal L}_{\gamma}^{(2)}$, $\breve{\mathcal L}_{r}^{(2)}$, $\breve{\mathcal L}_{k}^{(2)}$, $\breve{\mathcal L}_{\Gamma}^{(2)}$, and $\breve{\mathcal L}_{\omega}^{(4)}$ are not additional phenomenological generators introduced independently of the rate equations. Each is the sum of the elementary gain-loss jumps belonging to a particular microscopic process, and their sum reproduces exactly the population equations above. The full quantum master equation is 
	
	\vspace{1cm}
	
	\renewcommand{\arraystretch}{1.9}  
	\setlength{\arraycolsep}{4pt}      
	\[
	\begin{bmatrix}
		\dot{\rho}_{g}\\
		\dot{\rho}_{X_1}\\
		\dot{\rho}_{X_2}\\
		\dot{\rho}_{X_3}\\
		\dot{\rho}_{X_4}\\
		\dot{\rho}_{X_5}\\
		\dot{\rho}_{X_6}\\
		\dot{\rho}_{I_1}\\
		\dot{\rho}_{I_2}\\
		\dot{\rho}_{I_3}\\
		\dot{\rho}_{\alpha_1}\\
		\dot{\rho}_{\alpha_2}\\
		\dot{\rho}_{\beta_1}
	\end{bmatrix}
	=
	\left[
	\begin{array}{ccccccccccccc}
		
		\tilde{\mathcal L}_{gg} &
		\gamma_{1g} &
		\gamma_{2g} &
		\gamma_{3g} &
		\gamma_{4g} &
		\gamma_{5g} &
		\gamma_{6g} &
		0 &
		0 &
		0 &
		r_{\alpha_1 g}^{\mathrm{ISC}} &
		\omega_{\alpha_2 g} &
		\Gamma_{\beta_1 g}
		\\
		
		\gamma_{g1} &
		\tilde{\mathcal L}_{11} &
		r_{21} &
		r_{31} &
		r_{41} &
		r_{51} &
		r_{61} &
		k_{I_11} &
		k_{I_21} &
		0 &
		0 &
		0 &
		0
		\\
		
		\gamma_{g2} &
		r_{12} &
		\tilde{\mathcal L}_{22} &
		r_{32} &
		r_{42} &
		r_{52} &
		r_{62} &
		k_{I_12} &
		k_{I_22} &
		0 &
		0 &
		0 &
		0
		\\
		
		\gamma_{g3} &
		r_{13} &
		r_{23} &
		\tilde{\mathcal L}_{33} &
		r_{43} &
		r_{53} &
		r_{63} &
		k_{I_13} &
		k_{I_23} &
		0 &
		0 &
		0 &
		0
		\\
		
		\gamma_{g4} &
		r_{14} &
		r_{24} &
		r_{34} &
		\tilde{\mathcal L}_{44} &
		r_{54} &
		r_{64} &
		k_{I_14} &
		k_{I_24} &
		0 &
		0 &
		0 &
		0
		\\
		
		\gamma_{g5} &
		r_{15} &
		r_{25} &
		r_{35} &
		r_{45} &
		\tilde{\mathcal L}_{55} &
		r_{65} &
		k_{I_15} &
		k_{I_25} &
		0 &
		0 &
		0 &
		0
		\\
		
		\gamma_{g6} &
		r_{16} &
		r_{26} &
		r_{36} &
		r_{46} &
		r_{56} &
		\tilde{\mathcal L}_{66} &
		k_{I_16} &
		k_{I_26} &
		0 &
		0 &
		0 &
		0
		\\
		
		0 &
		k_{1I_1} &
		k_{2I_1} &
		k_{3I_1} &
		k_{4I_1} &
		k_{5I_1} &
		k_{6I_1} &
		\tilde{\mathcal L}_{I_1I_1} &
		0 &
		0 &
		k_{\alpha_1I_1} &
		\omega_{\alpha_2I_1} &
		0
		\\
		
		0 &
		k_{1I_2} &
		k_{2I_2} &
		k_{3I_2} &
		k_{4I_2} &
		k_{5I_2} &
		k_{6I_2} &
		0 &
		\tilde{\mathcal L}_{I_2I_2} &
		k_{I_3I_2} &
		0 &
		0 &
		0
		\\
		
		0 &
		0 &
		0 &
		0 &
		0 &
		0 &
		0 &
		0 &
		k_{I_2I_3} &
		\tilde{\mathcal L}_{I_3I_3} &
		k_{\alpha_1I_3} &
		\omega_{\alpha_2I_3} &
		0
		\\
		
		0 &
		0 &
		0 &
		0 &
		0 &
		0 &
		0 &
		k_{I_1\alpha_1} &
		0 &
		k_{I_3\alpha_1} &
		\tilde{\mathcal L}_{\alpha_1\alpha_1} &
		\omega_{\alpha_2\alpha_1} &
		0
		\\
		
		0 &
		0 &
		0 &
		0 &
		0 &
		0 &
		0 &
		\omega_{I_1\alpha_2} &
		0 &
		\omega_{I_3\alpha_2} &
		\omega_{\alpha_1\alpha_2} &
		\tilde{\mathcal L}_{\alpha_2\alpha_2} &
		0
		\\
		
		0 &
		0 &
		0 &
		0 &
		0 &
		0 &
		0 &
		0 &
		0 &
		0 &
		\Gamma_{\alpha_1\beta_1} &
		0 &
		\tilde{\mathcal L}_{\beta_1\beta_1}
		
	\end{array}
	\right]
	\begin{bmatrix}
		\rho_{g}\\
		\rho_{X_1}\\
		\rho_{X_2}\\
		\rho_{X_3}\\
		\rho_{X_4}\\
		\rho_{X_5}\\
		\rho_{X_6}\\
		\rho_{I_1}\\
		\rho_{I_2}\\
		\rho_{I_3}\\
		\rho_{\alpha_1}\\
		\rho_{\alpha_2}\\
		\rho_{\beta_1}
	\end{bmatrix}.
	\]
	\label{S-eq:liouvillian_matrix}
	Probability conservation follows from
	$
	\sum_a
	\left(
	\breve{\mathcal L}
	\right)_{ab}
	=
	0,
	\forall b,
	$
	and
	$
	\sum_a
	\dot\rho_a(t)
	=
	0.
	$
	The population and coherence sectors are decoupled in the kinetic
	description used here. The population-level treatment is appropriate for
	the stationary-state analysis after the short-time coherent dynamics have
	decayed. Ultrafast spectroscopy of the PSIIRC has identified electronic
	coherences with lifetimes of approximately $177$-$180~{\rm fs}$ and
	vibrational coherences persisting up to approximately $829~{\rm fs}$
	.~\cite{Jha2024} These observations establish that coherent dynamics is
	relevant on femtosecond timescales but does not persist indefinitely.
	The structured PSIIRC spectral density also contains prominent
	$342~{\rm cm}^{-1}$ and $742~{\rm cm}^{-1}$ vibrational modes, which have
	been identified as important components of the electron-transfer
	environment.~\cite{stones2017performance}  The calculated populations approach a stable
	steady state on a timescale of approximately $5~{\rm ps}$, whereas the
	coherences among $\ket{X_1}$-$\ket{X_6}$ decay to near zero within
	approximately $1~{\rm ps}$ in the parameter regime underlying the
	master-equation model. \cite{stones2017performance}  Electronic and vibronic
	coherences with reported lifetimes between $177$ and $829~{\rm fs}$
	likewise decay before the stationary populations are established.
	\cite{Jha2024}

	\FloatBarrier
	
	\begin{figure}[htbp]
		\centering
		\includegraphics[width=0.8\linewidth]{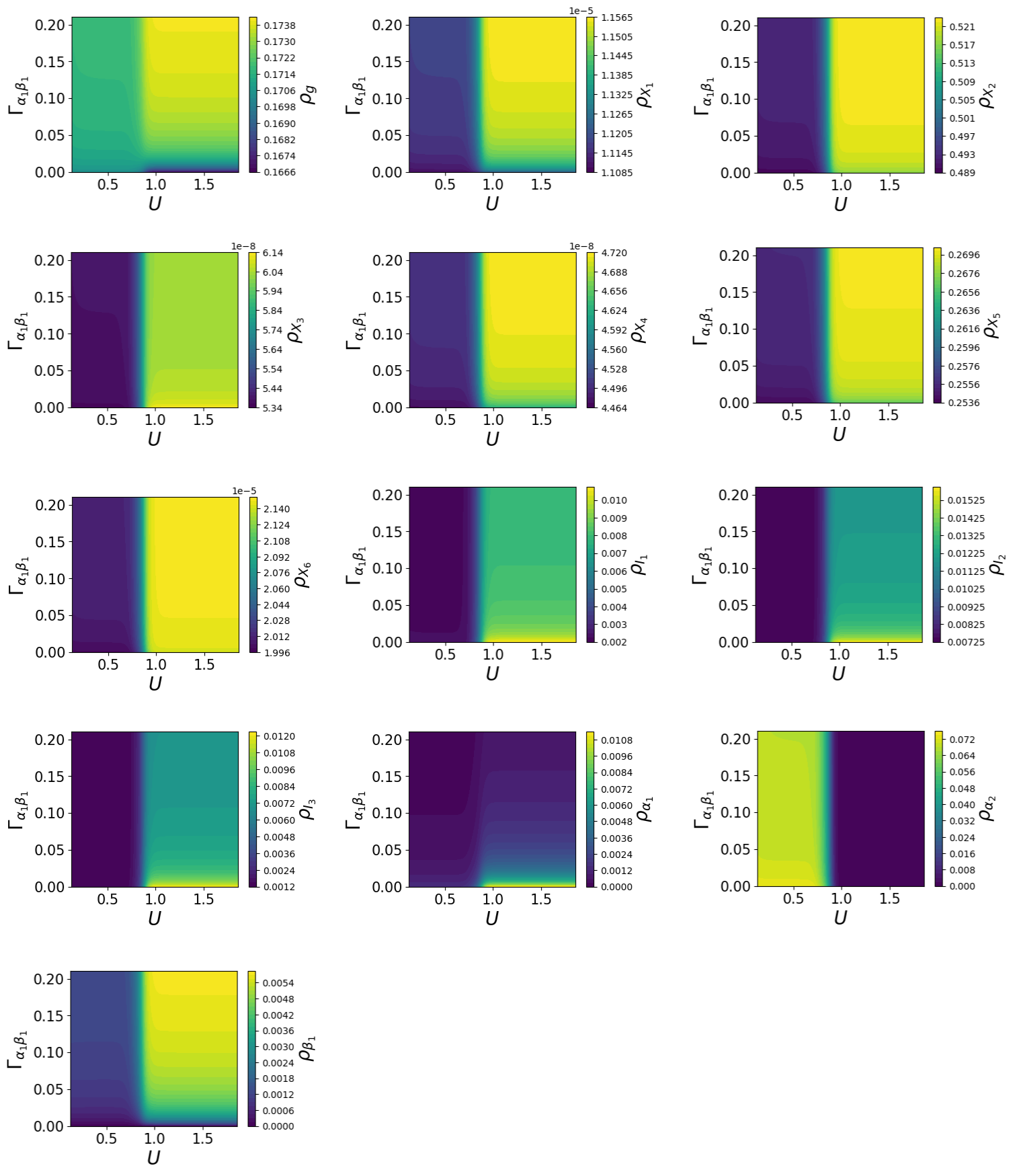}
		\caption{Steady-state population maps of the 13-state PSIIRC model as functions of the Coulomb interaction energy $U$ and the transition rate $\Gamma_{\alpha_1\beta_1}$. The panels show the populations $\rho_g$, $\rho_{X_1}$, $\rho_{X_2}$, $\rho_{X_3}$, $\rho_{X_4}$, $\rho_{X_5}$, $\rho_{X_6}$, $\rho_{I_1}$, $\rho_{I_2}$, $\rho_{I_3}$, $\rho_{\alpha_1}$, $\rho_{\alpha_2}$, and $\rho_{\beta_1}$, respectively. The color scale in each panel represents the magnitude of the corresponding steady-state population.}
		\label{fig:si1}
		\FloatBarrier
	\end{figure}
	\FloatBarrier
	\begin{figure}[htbp]
		\centering
		\includegraphics[width=0.6\linewidth]{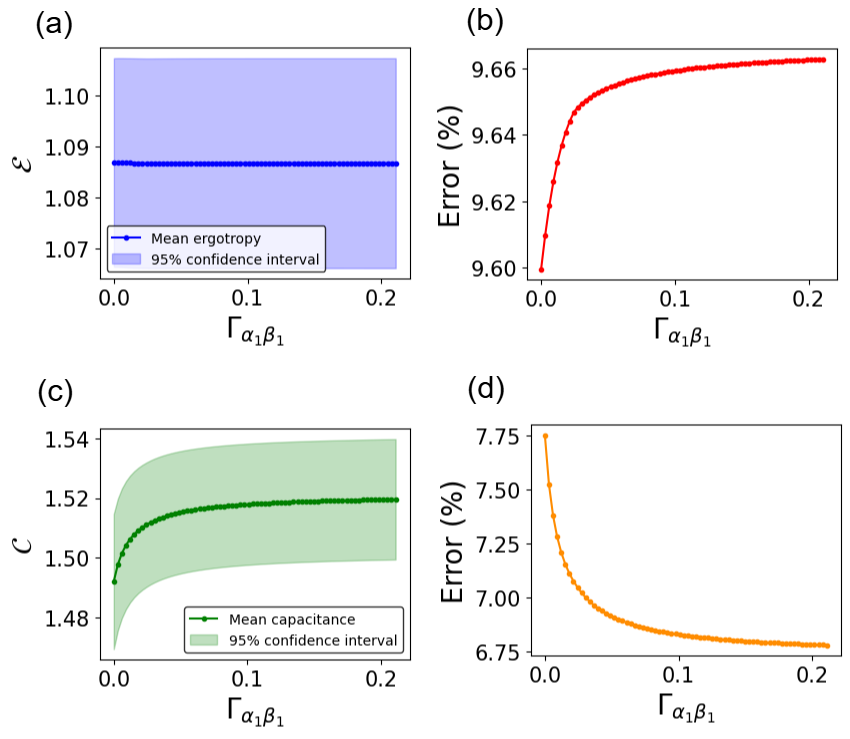}
		\caption{Uncertainty analysis of the energetic quantities as functions of the transition rate $\Gamma_{\alpha_1\beta_1}$. Panels (a) and (c) show the mean ergotropy $\mathcal{E}$ and mean energetic capacitance $\mathcal{C}$, respectively, with the shaded regions representing the corresponding $95\%$ confidence intervals. Panels (b) and (d) show the associated percentage errors in $\mathcal{E}$ and $\mathcal{C}$, respectively}
		\label{fig:si2}
	\end{figure}
	\FloatBarrier
	
	As a supplementary sensitivity test, separate from the nominal low- and high-$U$ response curves discussed in the main text, we examined the robustness of the energetic magnitudes against uncertainty in the system energies. We performed a Monte Carlo analysis using $100$ independent realizations, with a $5\%$ Gaussian variation applied to the physical state energies. For each realization, the energy-dependent kinetic rates and the corresponding nonequilibrium steady-state populations were evaluated over the complete $\Gamma_{\alpha_1\beta_1}$ sweep, from which the ergotropy $\mathcal{E}$ and energetic capacitance $\mathcal{C}$ were subsequently obtained. The solid curves in Figs.\ref{fig:si2}(a) and (c) represent the Monte Carlo ensemble means, while the shaded regions denote the corresponding $95\%$ confidence intervals, $\overline{Q}\pm1.96\sigma_Q/\sqrt{N}$, where $Q=\mathcal{E},\mathcal{C}$ and $N=100$. The relative uncertainty is quantified by $100\sigma_Q/|\overline{Q}|$ and is shown in Figs.\ref{fig:si2}(b) and (d).
	
	The mean ergotropy remains essentially insensitive to the extraction rate, varying only from approximately $1.0868$ to $1.0869$ eV over the entire $\Gamma_{\alpha_1\beta_1}$ range [Fig.\ref{fig:si2}(a)]. The corresponding relative uncertainty shows only a weak increase, from approximately $9.60\%$ to $9.66\%$ [Fig.\ref{fig:si2}(b)]. Thus, although energetic disorder produces a finite dispersion in $\mathcal{E}$, it does not appreciably modify its mean extraction-rate dependence. In contrast, the mean energetic capacitance increases from approximately $1.49$ eV at small $\Gamma_{\alpha_1\beta_1}$ to about $1.52$ eV and progressively approaches saturation [Fig.\ref{fig:si2}(c)]. Simultaneously, its relative uncertainty decreases from approximately $7.75\%$ to $6.78\%$ [Fig.\ref{fig:si2}(d)]. The opposite trends of the two relative uncertainties therefore show that increasing the extraction rate leaves the mean ergotropy nearly unchanged while progressively reducing the sensitivity of the energetic capacitance to variations in the underlying state energies.
	
	\begin{figure}[htbp]
		\centering
		\includegraphics[width=1.0\linewidth]{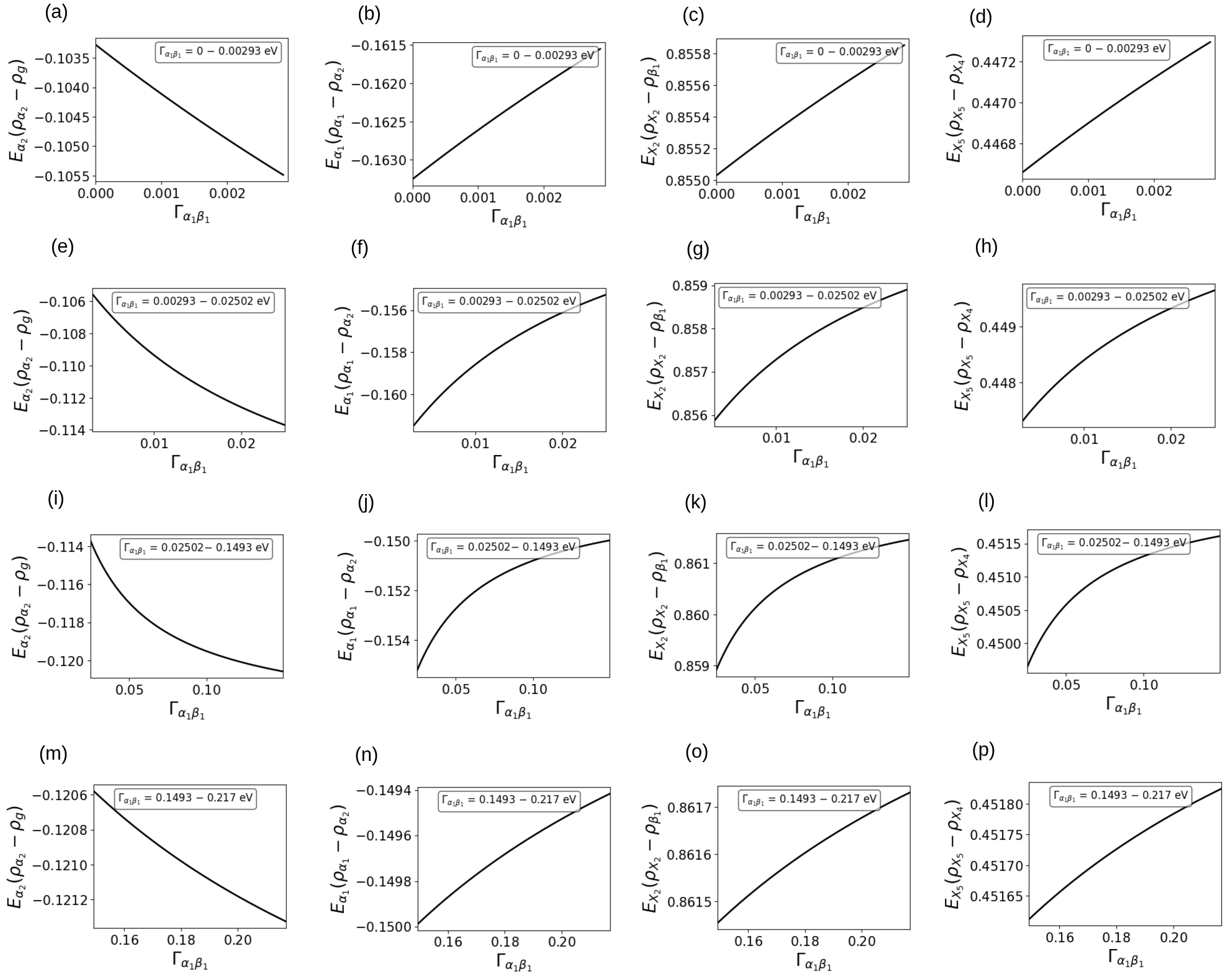}
		\caption{Dominant ergotropy contributions during the sequential electron-ejection-rate sweep $\Gamma_{\alpha_1\beta_1}$ at the low-interaction value $U=0.499~\mathrm{eV}$. The four columns show the contributions $E_{\alpha_2}(\rho_{\alpha_2}-\rho_g)$, $E_{\alpha_1}(\rho_{\alpha_1}-\rho_{\alpha_2})$, $E_{X_2}(\rho_{X_2}-\rho_{\beta_1})$, and $E_{X_5}(\rho_{X_5}-\rho_{X_4})$, respectively. Panels (a)-(d), (e)-(h), (i)-(l), and (m)-(p) correspond to the four passive-state intervals $\Gamma_{\alpha_1\beta_1}=0$-$0.00293~\mathrm{eV}$, $0.00293$-$0.02502~\mathrm{eV}$, $0.02502$-$0.1493~\mathrm{eV}$, and $0.1493$-$0.217~\mathrm{eV}$, respectively. The same four energy-weighted population differences remain dominant throughout the sweep despite the successive changes in passive-state ordering.}
		\label{fig:si4}
		
	\end{figure}

	\begin{figure}[htbp]
		\centering
		\includegraphics[width=0.8\linewidth]{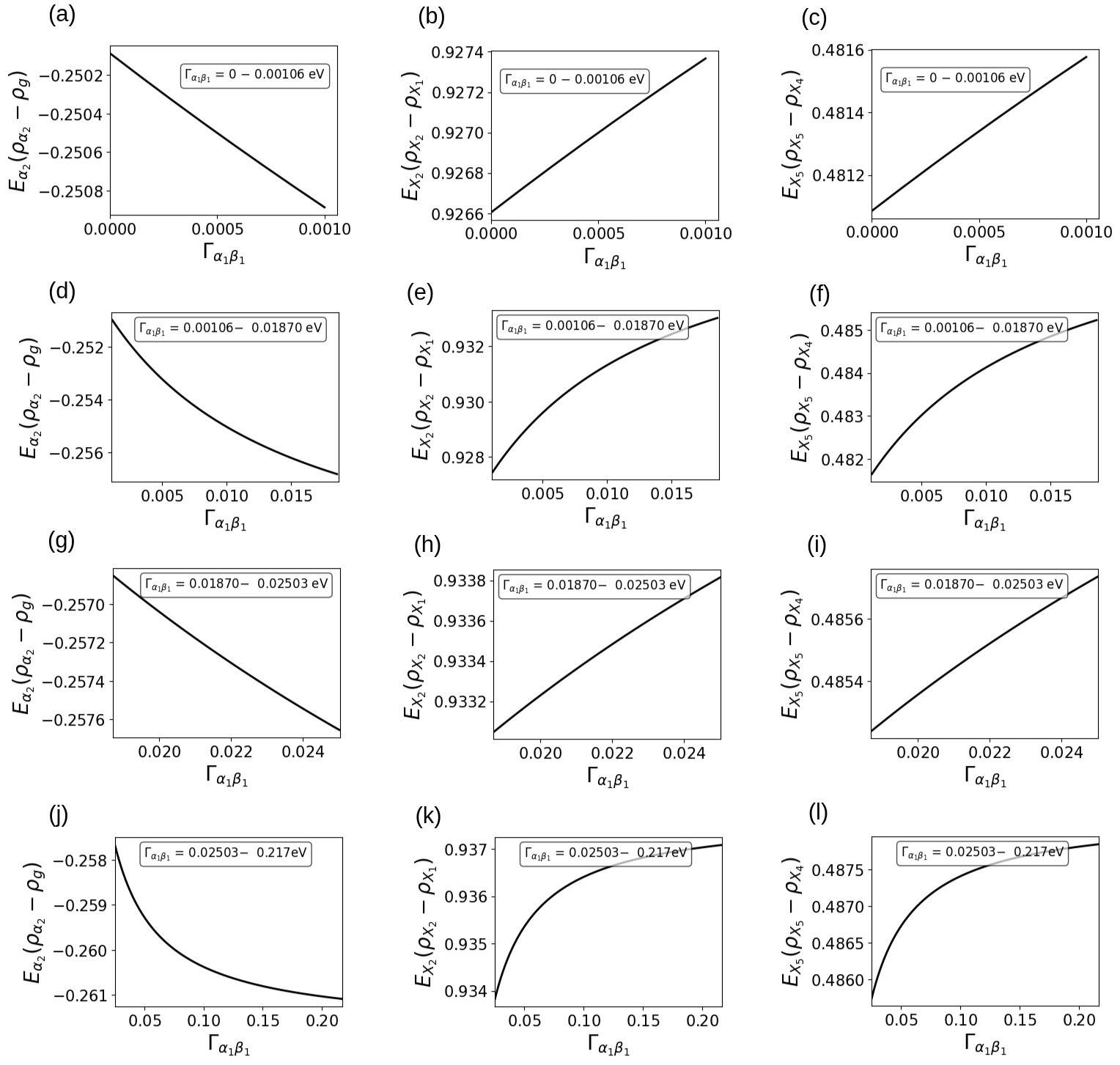}
		\caption{Dominant ergotropy contributions for the sequential electron-ejection-rate sweep $\Gamma_{\alpha_1\beta_1}$ at the high-interaction value $U=1.488~\mathrm{eV}$. The three columns show the contributions $E_{\alpha_2}(\rho_{\alpha_2}-\rho_g)$, $E_{X_2}(\rho_{X_2}-\rho_{X_1})$, and $E_{X_5}(\rho_{X_5}-\rho_{X_4})$, respectively. Panels (a)-(c), (d)-(f), (g)-(i), and (j)-(l) correspond to the four passive-state intervals $\Gamma_{\alpha_1\beta_1}=0$-$0.00106~\mathrm{eV}$, $0.00106$-$0.01870~\mathrm{eV}$, $0.01870$-$0.02503~\mathrm{eV}$, and $0.02503$-$0.217~\mathrm{eV}$, respectively. These three energy-weighted population differences remain the dominant contributions throughout the high-$U$ sweep despite the changes in passive-state ordering.}
		\label{fig:si5}
	\end{figure}
	\begin{figure}[htbp]
		\centering
		\includegraphics[width=1.0\linewidth]{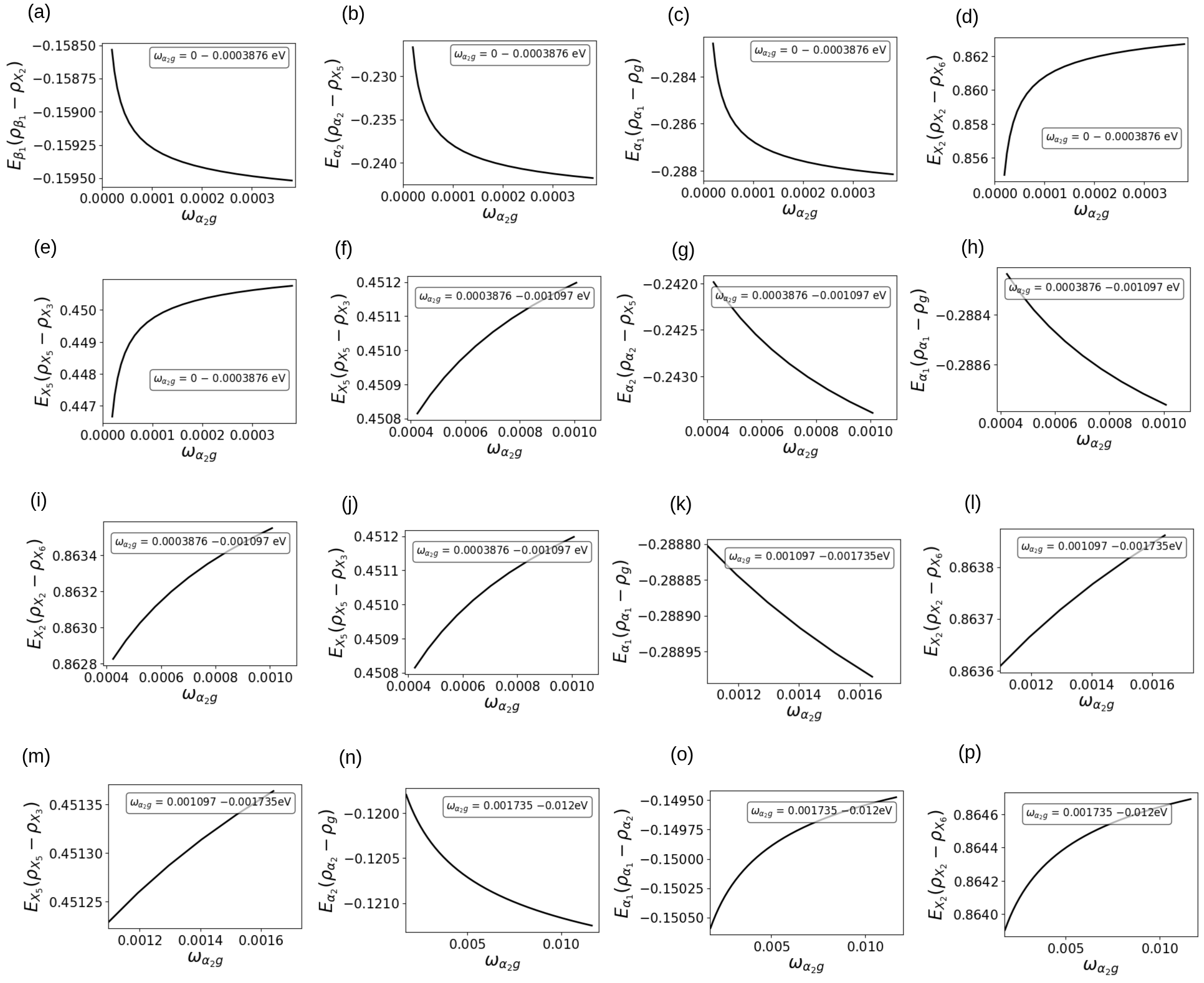}
		\caption{Dominant ergotropy contributions during the cotunneling-rate sweep $\omega_{\alpha_2 g}$ at the low-interaction value $U=0.496~\mathrm{eV}$. Panels (a)-(e) correspond to the interval $\omega_{\alpha_2 g}=0$-$0.0003876~\mathrm{eV}$, panels (f)-(j) to $\omega_{\alpha_2 g}=0.0003876$-$0.001097~\mathrm{eV}$, panels (k)-(m) to $\omega_{\alpha_2 g}=0.001097$-$0.001735~\mathrm{eV}$, and panels (n)-(p) to $\omega_{\alpha_2 g}=0.001735$-$0.012~\mathrm{eV}$. Within each interval, the plotted energy-weighted population differences represent the dominant contributions to the ergotropy for the corresponding passive-state configuration. The changes in the dominant contributions across the successive intervals reflect the passive-state reorganizations induced by increasing $\omega_{\alpha_2 g}$.}
		\label{fig:si6}
	\end{figure}
	
	\clearpage
	\renewcommand{\refname}{References}
	\bibliographystyle{aipnum4-2}
	\bibliography{Reference_JCP}

	\clearpage
\end{document}